\documentclass[prb,aps,shownopacs,twocolumn,superscriptaddress,10pt]{revtex4}
\usepackage{amsmath}
\usepackage{amssymb}
\usepackage{amsthm}
\usepackage{amsfonts}
\usepackage{listings}
\usepackage{enumerate}
\usepackage{multirow}
\usepackage{latexsym}
\usepackage{psfrag}
\usepackage{bm}
\usepackage{array}
\usepackage[all]{xy}
\usepackage{graphicx}
\usepackage{subfigure}
\usepackage{xcolor}
\usepackage[english]{babel}
\usepackage[autostyle]{csquotes}
\usepackage{chngcntr}
\usepackage{float}
\usepackage{lipsum}
\newcommand{\Tr}{\mathrm{Tr}}
\newcommand{\ch}{\mathcal{H}}

\usepackage{pdfpages}
\usepackage{tikz}
\usetikzlibrary{matrix,decorations.pathreplacing,calc}
\usepackage[pdftex,colorlinks=true]{hyperref}

\begin{document}    
\title{Structure of the Entanglement Entropy of (3+1)D Gapped Phases of Matter}

\author{Yunqin Zheng}
\affiliation{Physics Department, Princeton University, Princeton, New Jersey 08544, USA}

\author{Huan He}
\affiliation{Physics Department, Princeton University, Princeton, New Jersey 08544, USA}

\author{Barry Bradlyn}
\affiliation{Princeton Center for Theoretical Science, Princeton University, Princeton, New Jersey 08544, USA}

\author{Jennifer Cano}
\affiliation{Princeton Center for Theoretical Science, Princeton University, Princeton, New Jersey 08544, USA}

\author{Titus Neupert}
\affiliation{Department of Physics, University of Zurich, Winterthurerstrasse 190, 8057 Zurich, Switzerland}

\author{B. Andrei Bernevig}
\affiliation{Physics Department, Princeton University, Princeton, New Jersey 08544, USA}
\affiliation{Donostia International Physics Center, P. Manuel de Lardizabal 4, 20018 Donostia-San Sebastia ́n, Spain \thanks{On sabbatical}\footnotemark[1] }
\affiliation{Laboratoire Pierre Aigrain, Ecole Normale Sup ́erieure-PSL Research University, CNRS, Universit ́e Pierre et Marie Curie-Sorbonne Universit ́es,
	Universit ́e Paris Diderot-Sorbonne Paris Cit ́e, 24 rue Lhomond, 75231 Paris Cedex 05, France \footnotemark[1] }
\affiliation{Sorbonne Universit ́es, UPMC Univ Paris 06, UMR 7589, LPTHE, F-75005, Paris, France \footnotemark[1]}

\begin{abstract}
	We study the entanglement entropy of gapped phases of matter in three spatial dimensions. We focus in particular on size-independent contributions to the entropy across entanglement surfaces of arbitrary topologies. We show that for low energy fixed-point theories, the constant part of the entanglement entropy across any surface can be reduced to a linear combination of the entropies across a sphere and a torus. We first derive our results using strong sub-additivity inequalities along with assumptions about the entanglement entropy of fixed-point models, and identify the topological contribution by considering the renormalization group flow; in this way we give an explicit definition of topological entanglement entropy $S_{\mathrm{topo}}$ in (3+1)D, which sharpens previous results. We illustrate our results using several concrete examples and independent calculations, and show adding ``twist'' terms to the Lagrangian can change $S_{\mathrm{topo}}$ in (3+1)D. For the generalized Walker-Wang models, we find that the ground state degeneracy on a 3-torus is given by $\exp(-3S_{\mathrm{topo}}[T^2])$ in terms of the topological entanglement entropy across a 2-torus. We conjecture that a similar relationship holds for Abelian theories in $(d+1)$ dimensional spacetime, with the ground state degeneracy on the $d$-torus given by $\exp(-dS_{\mathrm{topo}}[T^{d-1}])$.
\end{abstract}

\date{\today}

\maketitle

\tableofcontents

\section{Introduction}
\label{introduction}

Classifying gapped phases of matter has recently emerged as one of the central themes of condensed matter physics\cite{wen2004quantum,bernevig2013topological,2012arXiv1203.4565S,zeng2015quantum,2015ARCMP...6..299S}. The ground states of two gapped Hamiltonians are in the same phase if they can be adiabatically connected to one another through local unitary transformations, without closing the energy gap\cite{wen2004quantum}. Prior to the discovery of topological order, the consensus in the physics community was that gapped phases could be classified by symmetry breaking order parameters\cite{landau1937theory,landau1950theory}. The discovery of topological order\cite{1981PhRvB..23.5632L,1982PhRvL..48.1559T,wen1990topological} revealed that two gapped systems can reside in distinct phases absent any global symmetries. The discovery of symmetry protected topological (SPT) order\cite{PhysRevLett.50.1153,1983PhLA...93..464H,kane2005quantum,bernevig2013topological,bernevig2006quantum,chen2013symmetry} further enriched the family of topological phases of matter: two systems with the same global symmetry can be in different phases even with trivial topological order. 

The classification of topological phases of matter has been studied systematically from many different angles. For noninteracting fermionic systems, phases have been classified according to time reversal symmetry, particle hole symmetry and chiral symmetry, summarized by the ten-fold way\cite{2010NJPh...12f5010R,2009AIPC.1134...22K}. Recently this classification was extended by considering crystal symmetries\cite{fu2011topological}, in particular nonsymmorphic symmetries\cite{2016Natur.532..189W,2016PhRvX...6b1008A}.  For interacting systems, multi-component Chern Simons theories\cite{PhysRevB.46.2290,PhysRevLett.65.1502,2012PhRvB..86l5119L,PhysRevB.93.155121,FrÃ¶hlich1997}, tensor category approaches\cite{2006AnPhy.321....2K,2015arXiv150605805B,2014arXiv1410.4540B,2008AnPhy.323.2709B}, various forms of boundary theories\cite{wen1995topological,2013PhRvX...3a1016V,Seiberg:2016rsg}, group cohomology constructions\cite{chen2013symmetry} and several additional methods\cite{kapustin2014anomalous,2017PhRvB..95c5131H,levin2012braiding} have been used to classify topological phases of matter.  

Given the ground state of a Hamiltonian, a variety of techniques have been developed to determine which phase it is in. One method exploits the anomalous boundary behavior of topological phases (such as nontrivial propagating modes if the boundary is gapless, or more exotic fractionalization if the boundary is gapped)\cite{Callan:1984sa,wen1995topological,fu2007topological,kane2005quantum,2013PhRvX...3a1016V,burnell2014exactly,PhysRevX.5.041013,2014PhRvB..89p5132C,PhysRevB.92.125111,2013PhRvB..88c5131M,2014arXiv1406.3032M,Seiberg:2016rsg} by studying systems with open boundary conditions. For topologically ordered phases, one can alternatively study the system on a closed manifold without boundaries, and examine the braiding and fusion properties of the gapped excitations, such as anyon excitations in (2+1)D and loop excitations in (3+1)D\cite{Moore:1991ks,PhysRevLett.53.722,levin2012braiding,wang2014braiding,wang2015non, 2016arXiv160205951W, 2016arXiv161209298P}. 

Additionally, the entanglement structure of the ground state can also reveal topological properties of the system. In particular, Kitaev and Preskill \cite{PhysRevLett.96.110404}, as well as Levin and Wen \cite{PhysRevLett.96.110405}, realized that in (2+1)D the existence of long range entanglement of the ground state, characterized by the topological entanglement entropy (TEE), indicates topological order. Among all approaches for probing topological order, studying the entanglement entropy is one of the more favorable \cite{isakov2011topological,jiang2012identifying,depenbrock2012nature}, because it depends on the ground state only and can be computed with periodic boundary conditions. There have been many attempts to generalize this construction to higher dimensions, in particular to better understand topological order in (3+1)D. The first attempt to study the TEE in (3+1)D was made in Ref.~\onlinecite{2008PhRvB..78o5120C}, where the authors computed the entanglement entropy (EE) for the (3+1)D toric code at finite temperature. In Ref.~\onlinecite{von2013three}, the (3+1)D entanglement entropy was computed for the semion model, which corresponds to the generalized Walker Wang (GWW) model of type $(n,p)=(2,1)$. (See Sec.~\ref{EntanglementEntropyOfTQFT} for the definition of the GWW models.) In Ref.~\onlinecite{bullivant2015entropic}, the authors discussed the tensor category representation of GWW models, and the entanglement entropy was computed in this framework. We note that these works only examine theories at exactly solvable fixed points. However, to isolate the topological part of the entanglement entropy, one needs to go beyond exactly solvable models; this is one of the motivations for the present work. The authors of Ref.~\onlinecite{grover2011entanglement}, for the first time, attempted to separate the topological and non-topological components of the entanglement entropy for a generic non-fixed-point system in (3+1)D. In particular, they realized that the constant (i.e., the contribution independent of the area of the entanglement surface) part of the entanglement entropy of a generic gapped system is not essentially topological, and contains a richer structure compared to that in (2+1)D. 

In this paper, based on previous works (especially Ref.~\onlinecite{grover2011entanglement}), we present a more detailed and complete analysis of the structure of the entanglement entropy (in particular the topological entanglement entropy) for gapped phases of matter in (3+1)D, whose low energy descriptions are topological quantum field theories (TQFT). We first make use of the strong sub-additivity (SSA) to constrain the general structure of the entanglement entropy for a TQFT. We find that the constant part of the entanglement entropy (in the ground state of a TQFT) across a general entanglement surface (which may contain multiple disconnected components) is a linear combination of the constant part of the EE across a sphere $S^2$ and that across a torus $T^2$, with the coefficients being topological invariants (Betti numbers) of the entanglement surface [see Eq.\eqref{TEEcentralresult}]. We further discuss the generalization of this result to generic non-fixed-point theories, where we study how the constant part of the entanglement entropy gets modified.  This allows us to isolate the topological entanglement entropy. We also provide explicit calculations of the entanglement entropy for a particular class of (3+1)D models, the GWW models. These calculations serve as an independent check of the result derived from the SSA inequalities, and also demonstrates that the EE can be modified by a topological twisting term in the action\footnote{For the GWW model in (3+1)D, the topological twisting term is the term depending on two form B-field only. For the Dijkgraaf-Witten models in any dimensions, the topological twisting term is the term depending on one form A-field only.}. This phenomena is new in (3+1)D as compared to (2+1)D, because the topological twisting term does not affect the TEE in (2+1)D. For example, the $\mathbb{Z}_2$ toric code and double semion theories, which differ by a topological twisting term, share the same TEE. Our approach has the advantage of simplicity: it starts from a simple-looking Lagrangian and does not require working with discrete lattice Hamiltonians. We conclude by conjecturing a formula for the TEE in terms of the ground state degeneracy for Abelian topological phases in general dimensions.  We give support to this conjecture by computing the entanglement entropy of BF theories in $(d+1)$ dimensions. 

The organization of this paper is as follows: In Sec.~\ref{GeneralStructure}, we present our approach to find a general formula for the constant part of the EE for TQFTs describing (3+1)D gapped phases of matter. The basic strategy is to use the SSA inequality to constrain the structure of the entanglement entropy. In the derivation, we assume a particular form of the entanglement entropy. In Sec.~\ref{EntanglementEntropyOfTQFT}, we justify this assumption through the study of the GWW models. We use a field theoretical approach, and compute the entanglement entropy of these models across general entanglement surfaces. We summarize our results in Sec.~\ref{DiscussionSummary}, and conclude with some open questions to be addressed in future work.

We present the details of our calculations in a series of appendices. In Appendix~\ref{DefinitionOfEntanglementEntropy} we review the definition of the entanglement entropy and the entanglement spectrum. In Appendix~\ref{Curvature} we review existing arguments about the local contributions to the entanglement entropy, which were first discussed in Ref.~\onlinecite{grover2011entanglement}. Appendices~\ref{app:recurrencederiv},~\ref{APPKPLWH2}, and~\ref{Appmoreon41} are dedicated to derivations of specific equations from the main text. In Appendix~\ref{TriangulationofTQFT} we review the basics of lattice formulation of TQFTs. In Appendix~\ref{Appclosedsurface} we explain why surfaces in the dual spacetime lattice are continuous and closed. In Appendix~\ref{AppMutualandSelfLinkingNumbers} we discuss the linking number integrals needed to formulate the GWW wave function. Finally, in Appendix~\ref{AppConjectureCaseStudy} we study BF theories in general $(d+1)$-dimensional spacetime, and give arguments for the validity of the conjecture that $\exp(-dS_{\mathrm{topo}}[T^{d-1}])$ gives the ground state degeneracy on the $d$-dimensional torus.

\section{Reduction formulas for Entanglement Entropy}
\label{GeneralStructure}

In this section, we study the general structure of the EE for gapped phases of matter in (3+1)D. The definitions of the entanglement entropy and the entanglement spectrum are reviewed in Appendix~\ref{DefinitionOfEntanglementEntropy}. We are inspired by the fact that for a (2+1)D system, the EE of the ground state of a local, gapped Hamiltonian obeys the area law. In particular, if we partition our system into two subregions, A and $\mathrm{A}^{\mathrm{c}}$, the EE of subregion A with the rest of the system $\mathrm{A}^{\mathrm{c}}$ takes the form
\begin{equation}
S(\mathrm{A})=\alpha l+\gamma+\mathcal{O}(1/l),
\end{equation} 
where $\alpha l$ is the area term, and $l$ is the length of the boundary of region A. Importantly the constant term  $- \gamma -$ is topological and thus dubbed  ``topological entanglement entropy" \cite{PhysRevLett.96.110404,PhysRevLett.96.110405}. We would like to understand whether an analogous formula holds for gapped phases of matter in (3+1)D. In particular, we ask how the constant part of the EE depends on the topological properties of both the Hamiltonian and the entanglement surface. 

Our approach to this question relies on the SSA inequality for the entanglement entropy. We also make certain locality assumptions about the form of the entropy, detailed in Appendix~\ref{Curvature}.
This allows us to derive an expression for the constant part of the EE of a subregion A for a TQFT, $S^{\mathrm{TQFT}}_{\mathrm{c}}(\mathrm{A})$, which depends on the topological properties (e.g. Betti numbers) of the entanglement surface $\partial \mathrm{A}\equiv \Sigma$.\footnote{In this paper, we will denote a generic entanglement surface as $\Sigma$.}

We start by reviewing some general facts about the EE and then use SSA inequalities to determine the formula for the EE across a general surface in Sec.~\ref{StrongSubAdditivity}. In Sec.~\ref{Discussions}, we discuss the implications of our EE formula, especially regarding models away from a renormalization group (RG) fixed point. Our approach is inspired by Ref.~\onlinecite{grover2011entanglement}.

\subsection{Strong Sub-Additivity}
\label{StrongSubAdditivity}

\subsubsection{Structure of the EE of Fixed Point TQFTs}
\label{StructureofEEofFixed}
As reviewed in Appendix~\ref{Curvature}, for a generic theory with an energy gap, the EE for a subregion A can be decomposed as
\begin{eqnarray}
S(\mathrm{A})&=&F_0 |\Sigma| +S_{\mathrm{topo}}(\mathrm{A})-4\pi F_2\chi(\Sigma)\nonumber\\&&+4F'_2\int_{\Sigma} d^2x \sqrt{h}H^2+\mathcal{O}(1/|\Sigma|),\label{variousterms}
\end{eqnarray}
where the coefficients $F_0, F_2$ and $F_2'$ are constants that depend on the system under study. The first term is the area law term, where $|\Sigma|$ is the area of the entanglement surface, $\Sigma$. The second term is the topological entanglement entropy, which is independent of the details of the entanglement surface and of the details of the Hamiltonian. The third term is proportional to the Euler characteristic $\chi(\Sigma)$ of the entanglement surface. Although it only depends on the topology of $\Sigma$, it is not universal, and we expect that the coefficient, $F_2$, will flow under the RG. The fourth term is proportional to the integral of the mean curvature, $H=(k_1+k_2)/2$, of $\Sigma$ (see Appendix~\ref{Curvature} for a derivation of the local contributions). It depends on the geometry (in contrast to the topology) of $\Sigma$, and its coefficient $F_2'$ also flows under the RG in general. The remaining terms are subleading in powers of the area $|\Sigma|$, and vanish when we take the size of the entanglement surface to infinity. One of the main goals of this paper is to understand the structure of the topological entanglement entropy, $S_{\mathrm{topo}}(\mathrm{A})$, and how it can be isolated from the Euler characteristic term and the mean curvature term.

In this section, unless otherwise stated, we consider (3+1)D TQFTs describing the low energy physics of a gapped topologically ordered phase. In this case the constant part of the EE depends only on the topology of the entanglement surface. The reason is the following: since a TQFT does not depend on the spacetime metric,  it is invariant under all diffeomorphisms, including dilatations as well as area-preserving diffeomorphisms. Hence, the term related to the mean curvature (which depends on the shape of $\Sigma$) should not appear. This implies that the coefficient $F_2'$ flows to zero at the fixed point. When we regularize the theory on the lattice, we explicitly break the scaling symmetry while maintaining the invariance under area preserving diffeomorphisms. Hence the area law term can survive, i.e. $F_0$ can flow to a non-vanishing value at the fixed point. (We relegate the explanation of this subtlety in Sec.~\ref{EEGWW3}.) Since the Euler characteristic is topological, $F_2$ can also flow to a non-vanishing value. In summary, the possible form of the EE for a low energy TQFT (when regularized on the lattice) is
\begin{equation}
S(\mathrm{A})=F_0 |\Sigma| + S_{\mathrm{topo}}(\mathrm{A})-4\pi F_2\chi(\Sigma)+\mathcal{O}(1/|\Sigma|).
\end{equation} 
For the sake of clarity, we denote the constant part of the EE for a generic theory as $S_{\mathrm{c}}(\mathrm{A})=S_{\mathrm{topo}}(\mathrm{A})-4\pi F_2\chi(\Sigma)\nonumber+4F'_2\int_{\Sigma} d^2x \sqrt{h}H^2$, and the constant part of the EE for a TQFT as $S_{\mathrm{c}}^{\mathrm{TQFT}}(\mathrm{A})=S_{\mathrm{topo}}(\mathrm{A})-4\pi F_2\chi(\Sigma)\nonumber$.  We point out that the value of $F_2$ for a general theory and for a TQFT are not the same, since its value flows under renormalization to the one in the TQFT, which will be specified in Sec.~\ref{AwayfromtheFixedPoint}. Furthermore, the area law part of the EE, $F_0 |\Sigma|$, is denoted as $S_{\mathrm{area}}(\mathrm{A})$.

For any quantum state, there are several information inequalities relating EEs between different subsystems that are universally valid\cite{Bao2015}, such as sub-additivity, strong sub-additivity, the Araki-Lieb inequality\cite{araki1970} and weak monotonicity\cite{Lieb1973}. Special quantum states, such as quantum error correcting codes\cite{devitt2013quantum} and holographic codes\cite{BaoNezami2015,Bao2015, 2015arXiv150206618L}, obey further independent information inequalities. The major constraint on the EE utilized in this paper is the strong sub-additivity inequality, which is typically used in quantum information theory. Explicitly, the SSA inequality is
\begin{equation}\label{SSA}
S(\mathrm{AB})+S(\mathrm{BC})\ge S(\mathrm{ABC})+S(\mathrm{B}),
\end{equation}
where the space is divided into four regions $\mathrm{A}, \mathrm{B}, \mathrm{C}$, and $(\mathrm{ABC})^{\mathrm{c}}$. Here, $(\mathrm{ABC})^{\mathrm{c}}$ is the complement of $\mathrm{ABC}\equiv\mathrm{A}\cup \mathrm{B}\cup \mathrm{C}$. SSA strongly constrains the structure of the constant part of $S(\mathrm{A})$, i.e., $S_{\mathrm{c}}(\mathrm{A})$, as we will see below.

\subsubsection{Reduction to the Constant Part of the EE}
\label{ReductiontoTopologicalEntropy}

The SSA is universal, and hence it is valid for any choice of the regions A, B and C. Here we will only need to consider the special cases with $\mathrm{A}\cap \mathrm{C}=\varnothing$. This configuration is chosen precisely to cancel the area law part of the EE on both sides of the SSA inequality, thus giving us information about the constant part $S_{\mathrm{c}}(\mathrm{A})$. Explicitly, when $\mathrm{A}\cap \mathrm{C}=\varnothing$, we have
\begin{equation}
S_{\mathrm{area}}(\mathrm{AB})+S_{\mathrm{area}}(\mathrm{BC})= S_{\mathrm{area}}(\mathrm{ABC})+S_{\mathrm{area}}(\mathrm{B}).
\end{equation}
Equation~\eqref{SSA} then implies
\begin{equation}\label{ScScScSc}
S_{\mathrm{c}}(\mathrm{AB})+S_{\mathrm{c}}(\mathrm{BC})\ge S_{\mathrm{c}}(\mathrm{ABC})+S_{\mathrm{c}}(\mathrm{B})	\;.
\end{equation}
When restricted to a TQFT, we have
\begin{equation}
S^{\mathrm{TQFT}}_{\mathrm{c}}(\mathrm{AB})+S^{\mathrm{TQFT}}_{\mathrm{c}}(\mathrm{BC})\ge S^{\mathrm{TQFT}}_{\mathrm{c}}(\mathrm{ABC})+S^{\mathrm{TQFT}}_{\mathrm{c}}(\mathrm{B})	\;.
\end{equation}

\subsubsection{Structure of $S_{\mathrm{c}}(\mathrm{A})$}
\label{Structure}

We need to parametrize $S^{\mathrm{TQFT}}_{\mathrm{c}}(\mathrm{A})$ in order to proceed. For a TQFT (where $F_2'=0$), we see that  $S^{\mathrm{TQFT}}_{\mathrm{c}}(\mathrm{A})=S_{\mathrm{topo}}(\mathrm{A})-4\pi F_2\chi(\Sigma)$ only depends on the topology of the entanglement surface $\Sigma$ through its Euler characteristic. Two-dimensional orientable surfaces are classified by a set of numbers $\lbrace n_0,n_1,n_2,\ldots \rbrace$, where $n_g$ is the number of disconnected components (parts) with genus $g$.\footnote{In this paper, the entanglement surfaces do not wrap around non-contractible cycles of the space. } We will show that this is an over-complete labeling for $S^{\mathrm{TQFT}}_{\mathrm{c}}(\mathrm{A})$, and that $S^{\mathrm{TQFT}}_{\mathrm{c}}(\mathrm{A})$ only depends on the zeroth and first Betti number\cite{nakahara2003geometry} of $\Sigma$ defined below in terms of $\{n_0,  n_1, n_2, \cdots\}$.

For the time being, we use the (over-)complete labeling scheme for $S^{\mathrm{TQFT}}_{\mathrm{c}}(\mathrm{A})$
\begin{equation}
S^{\mathrm{TQFT}}_{\mathrm{c}}[(0,n_0),(1,n_1),\cdots, (g,n_g),\cdots],
\end{equation}
where in each bracket, the first number denotes the genus, and the second number denotes the number of disconnected boundary components $\partial\mathrm{A}$ with the corresponding genus. The list ends precisely when $n_{g^*}\neq0$ and $n_{g}=0$ for any $g>g^*$. In other words, $S^{\mathrm{TQFT}}_{\mathrm{c}}[(0,n_0),(1,n_1),\dots, (g^*,n_{g^*})]$ is the constant part of the EE of the region with $n_0$ genus 0 boundaries, $n_1$ genus 1 boundaries, $\cdots$ and $n_{g^*}$ genus $g^*$ boundaries. We emphasize that the region A can have multiple disconnected boundary components. The set $\{n_g\}$ is related to the Betti numbers $b_i$ and the Euler characteristic $\chi$ through 
\begin{equation}
\sum_{g=0}^{g^*}n_g=b_0,	\;\;	\sum_{g=0}^{g^*}n_g(2-2g)=2b_0-b_1=\chi.
\end{equation}
These numbers  will be useful in the following calculations.

By applying the SSA inequality to a series of entanglement surfaces, we derive an expression for $S^{\mathrm{TQFT}}_c$ in terms of the Betti numbers $b_0$ and $b_1$, as well as the entropies $S^{\mathrm{TQFT}}_\mathrm{c}[T^2]$ and $S^{\mathrm{TQFT}}_\mathrm{c}[S^2]$ across the torus and sphere, respectively. Relegating the details of the derivation to Appendix~\ref{app:recurrencederiv}, we find: 
\begin{align}
S&_{\mathrm{c}}^{\mathrm{TQFT}}[(0,n_0),(1,n_1),\cdots, (g,n_g)]\nonumber\\
&=b_0 S_{\mathrm{c}}^{\mathrm{TQFT}}[T^2]+\frac{\chi}{2}\Big(S_{\mathrm{c}}^{\mathrm{TQFT}}[S^2]-S_{\mathrm{c}}^{\mathrm{TQFT}}[T^2]\Big). \label{TEEcentralresult}
\end{align}
Notice that Eq.~\eqref{TEEcentralresult} is consistent with the expectation that disconnected parts of the entanglement surface result in additive contributions due to the local nature of the mutual information.

\subsection{Topological Entanglement Entropy}
\label{Discussions}

Our first main result is Eq.~\eqref{TEEcentralresult}, which clarifies two points. First, as we mentioned in the introduction (and as was also discussed in Ref.~\onlinecite{grover2011entanglement}), given a general entanglement surface $[(0,n_0),(1,n_1),...,(g^*,n_{g^*})]$, we can reduce the computation of the constant part of the EE of a TQFT, $S_{\mathrm{c}}^{\mathrm{TQFT}}[(0,n_0),(1,n_1),...,(g^*,n_{g^*})]$, to that of  $S_{\mathrm{c}}^{\mathrm{TQFT}}[S^2]$ and $S_{\mathrm{c}}^{\mathrm{TQFT}}[T^2]$. Second, using Eq.~\eqref{TEEcentralresult}, we can identify the topological and universal part of $S_{\mathrm{c}}(\mathrm{A})$ for a generic theory beyond the TQFT fixed point. We now elaborate on these points.

\subsubsection{$S_{\mathrm{c}}^{\mathrm{TQFT}}[S^2]$ and $S_{\mathrm{c}}^{\mathrm{TQFT}}[T^2]$}

For a TQFT, Eq.~\eqref{TEEcentralresult} proves that the constant part of the EE across a general surface can be reduced to a linear combination of the constant part of the EE across $S^2$ and $T^2$. Whether $S_{\mathrm{c}}^{\mathrm{TQFT}}[S^2]$ and $S_{\mathrm{c}}^{\mathrm{TQFT}}[T^2]$ are independent of each other depends on the type of TQFT. As we show in Sec.~\ref{EntanglementEntropyOfTQFT}, for a BF theory [see Eq.~\eqref{BFtheory}] in (3+1)D, $S_{\mathrm{c}}^{\mathrm{TQFT}}[S^2]=S_{\mathrm{c}}^{\mathrm{TQFT}}[T^2]$. For the GWW models [see Eq.~\eqref{GWW}] in (3+1)D, we show in Sec.~\ref{EntanglementEntropyOfTQFT} that $S_{\mathrm{c}}^{\mathrm{TQFT}}[S^2]$ and $S_{\mathrm{c}}^{\mathrm{TQFT}}[T^2]$ are different in general. Thus, Eq.~\eqref{TEEcentralresult} is the simplest expression that is universally valid for any TQFT. 

\subsubsection{Away from the Fixed Point}
\label{AwayfromtheFixedPoint}

In Sec.~\ref{StructureofEEofFixed} and Appendix~\ref{Curvature}, we revisited the arguments presented in Ref.~\onlinecite{grover2011entanglement} that the constant part of the EE for a theory away from the fixed point  is generically not topological. The structure of the EE of a generic theory was shown in Eq.~\eqref{variousterms}. Combining Eq.~\eqref{variousterms} and Eq.~\eqref{TEEcentralresult}, we now extract more information about the structure of the EE.

First, we argued in Sec.~\ref{StructureofEEofFixed} that
\begin{eqnarray}
 F_2'\to 0,
\end{eqnarray}
when the theory is renormalized to a TQFT fixed point. 

Second, by setting $F_2'=0$ in Eq.~\eqref{variousterms} and comparing the TEE and the coefficient of the Euler characteristic $\chi$ in Eq.~\eqref{variousterms} and Eq.~\eqref{TEEcentralresult}, we find that
\begin{equation}\label{DefineTopoEE}
\begin{split}
&S_{\mathrm{topo}}[(0,n_0),\cdots,(g^*,n_{g^*})]
\\
&= b_0 S_{\mathrm{c}}^{\mathrm{TQFT}}[T^2]= \bigg( \sum_{i=0}^{g^*}n_i \bigg) S_{\mathrm{c}}^{\mathrm{TQFT}}[T^2],
\end{split}
\end{equation}
and
\begin{eqnarray}\label{F2}
	F_2\to -\frac{1}{8\pi}\Big(S_{\mathrm{c}}^{\mathrm{TQFT}}[S^2]-S_{\mathrm{c}}^{\mathrm{TQFT}}[T^2]\Big).
\end{eqnarray}
Equation \eqref{DefineTopoEE} suggests that the TEE across an arbitrary entanglement surface (for a generic theory) is proportional to $S_{\mathrm{c}}^{\mathrm{TQFT}}[T^2]$; in particular, the TEE across $T^2$ (for a generic theory) equals $S_{\mathrm{c}}^{\mathrm{TQFT}}[T^2]$,  i.e., $S_{\mathrm{topo}}[T^2]=S_{\mathrm{c}}^{\mathrm{TQFT}}[T^2]$. Equation \eqref{F2} shows that while $F_2$ can flow when the theory is renormalized, it converges to a nontrivial value $-\frac{1}{8\pi}\big(S_{\mathrm{c}}^{\mathrm{TQFT}}[S^2]-S_{\mathrm{c}}^{\mathrm{TQFT}}[T^2]\big)$ at the RG fixed point. Our identification of the TEE Eq.~\eqref{DefineTopoEE} further elaborates on the result from Ref.~\onlinecite{grover2011entanglement}, which showed that the TEE across a genus $g$ entanglement surface $\Sigma_g$ is $S_{\mathrm{topo}}[\Sigma_g]=gS_{\mathrm{topo}}[T^2]-(g-1)S_{\mathrm{topo}}[S^2]$. Our result Eq.~\eqref{DefineTopoEE} suggests that $S_{\mathrm{topo}}[S^2]=S_{\mathrm{topo}}[T^2]$ and therefore further simplifies the result of Ref.~\onlinecite{grover2011entanglement} to $S_{\mathrm{topo}}[\Sigma_g]=S_{\mathrm{topo}}[T^2]$ for any $g$. Our identification of the TEE also works for entanglement surfaces with multiple disconnected components.

\subsubsection{Extracting the TEE}
\label{TowardExtractingTEE}

\begin{figure}[t]
	\includegraphics[width=0.5\textwidth]{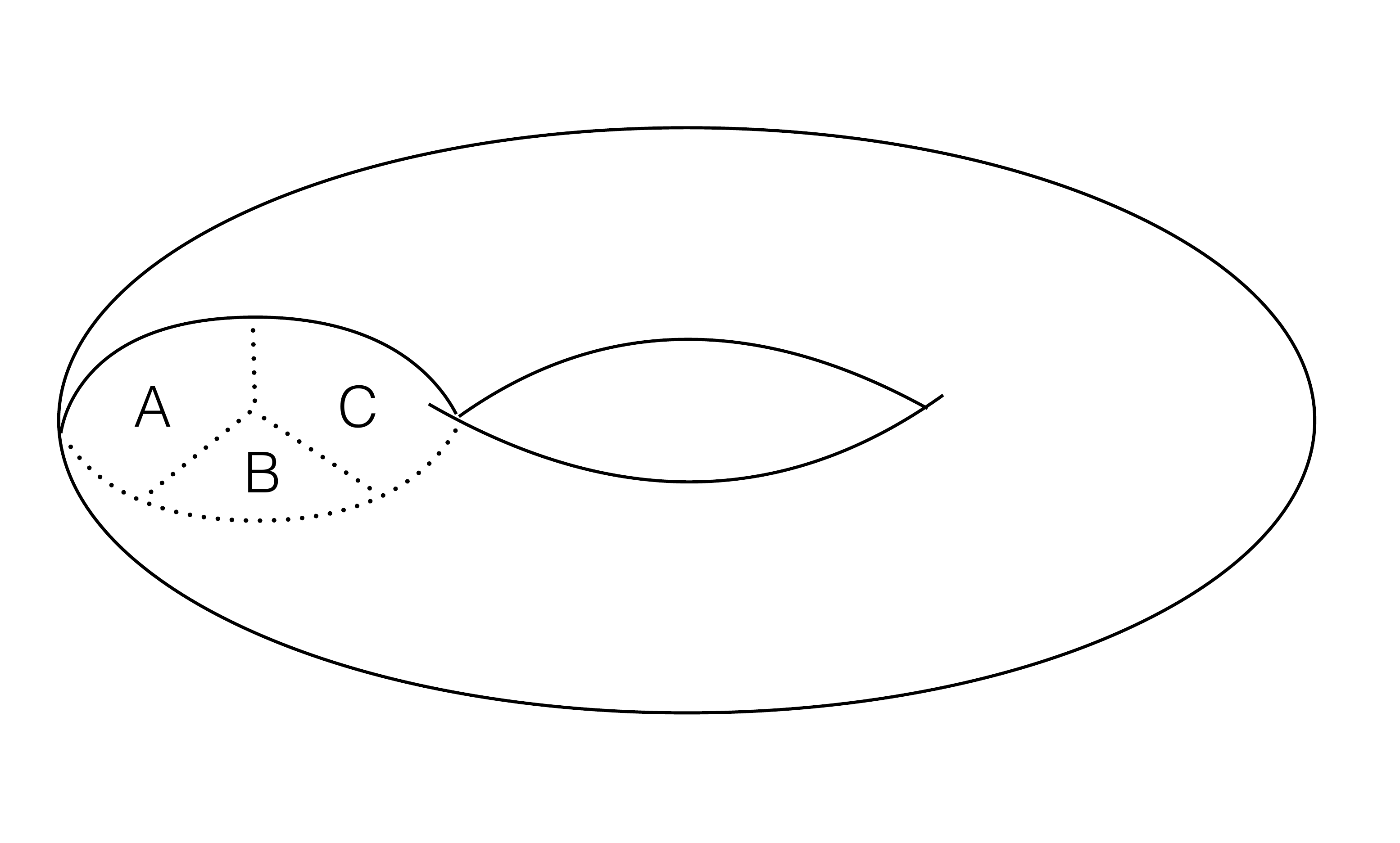}
	\centering
	\caption{KPLW prescription of entanglement surface $T^2$. The space inside the two torus is divided into three regions, A, B and C, each being a solid torus. }
	\label{KPT2}
\end{figure}

Equation \eqref{DefineTopoEE} suggests an ``algorithm" to compute the TEE for a generic theory: 1) take a ground state wavefunction $|\psi\rangle$ for a generic system; 2) renormalize $|\psi\rangle$ to the fixed point; 3) compute the entanglement entropy for an entanglement surface $T^2$, $S^{\mathrm{TQFT}}[T^2]$. The constant part $S_{\mathrm{c}}^{\mathrm{TQFT}}[T^2]$ is the TEE across $T^2$. Notice that this is consistent with our definition $S_{\mathrm{c}}^{\mathrm{TQFT}}[T^2]=S_{\mathrm{topo}}[T^2]-4\pi F_2 \chi(T^2) $ since $\chi(T^2)=0$. The TEE across an arbitrary surface immediately follows from Eq.~\eqref{DefineTopoEE}.

In this section, we will explain a more practical algorithm for extracting the TEE (across $T^2$) which is applicable to the groundstate wavefunction of any generic theory, and does not require renormalization to the TQFT fixed point. Our algorithm (which is termed the KPLW prescription) builds upon the study of the topological entanglement entropy in (2+1)D systems initiated by Kitaev, Preskill, Levin and Wen\cite{PhysRevLett.96.110404,PhysRevLett.96.110405}(KPLW) and the proposal in Ref.~\onlinecite{grover2011entanglement} in (3+1)D. We compute a particular combination of the EE of different regions, which we call $S_{\mathrm{KPLW}}[T^2]$, and demonstrate that this combination equals $S_{\mathrm{topo}}[T^2]$. The same KPLW prescription was studied in Ref.~\onlinecite{grover2011entanglement}, but here we provide a rigorous proof of the equivalence between the entanglement entropy from the KPLW prescription Eq.~\eqref{KPLWT2} and the TEE $S_{\mathrm{topo}}[T^2]$, as we derive in Eq.~\eqref{KPLWTEE}. Via Eq.~\eqref{DefineTopoEE}, we can then obtain the TEE across a general surface.

We generalize the KPLW prescription to (3+1)D by considering the configuration of the entanglement regions shown in Fig.~\ref{KPT2} and computing the combination of EEs
\begin{eqnarray}
S_{\mathrm{KPLW}}[T^2]&\equiv&S(\mathrm{A})+S(\mathrm{B})+S(\mathrm{C})-S(\mathrm{AB})\nonumber\\&&-S(\mathrm{AC})-S(\mathrm{BC})+S(\mathrm{ABC}).\label{KPLWT2}
\end{eqnarray}
Following similar arguments in Ref.~\onlinecite{PhysRevLett.96.110404}, it can be shown that $S_{\mathrm{KPLW}}[T^2]$ satisfies two properties:
\begin{enumerate}
	\item $S_{\mathrm{KPLW}}[T^2]$ is insensitive to local deformations of the entanglement surface.
	\item $S_{\mathrm{KPLW}}[T^2]$ is insensitive to local perturbations of the Hamiltonian.
\end{enumerate}

We first argue that the property 1 holds. If we locally deform the common boundary of region A and B (but away from the common boundary of region A, B and C, which is a line), the deformation of $S_{\mathrm{KPLW}}[T^2]$ is
\begin{equation}
\begin{split}
\Delta S_{\mathrm{KPLW}}[T^2]=&\,[\Delta S(\mathrm{A})-\Delta S(\mathrm{AC})]\\
&+[\Delta S(\mathrm{B})-\Delta S(\mathrm{BC})].
\end{split}
\end{equation}
Because the deformation is far away from region C (farther than the correlation length $\xi\simeq 1/m$, where $m$ is the energy gap), $\Delta S(\mathrm{A})-\Delta S(\mathrm{AC})=0$, and similarly  $\Delta S(\mathrm{B})-\Delta S(\mathrm{BC})=0$. Hence $S_{\mathrm{KPLW}}[T^2]$ is unchanged under the deformation of common boundary of A and B, away from the line which represents the common boundary of A, B and C. If we now locally deform the common boundary of regions A, B and C \footnote{We should distinguish between the common boundary of A, B and C, which is a line $\mathrm{A}\cap \mathrm{B}\cap \mathrm{C}$, and the boundary of region ABC, which is a surface} (the line $\mathrm{A}\cap \mathrm{B}\cap \mathrm{C}$),
\begin{equation}
\begin{split}
\Delta S_{\mathrm{KPLW}}[T^2]
\equiv&\,\Delta S(\mathrm{A})+\Delta  S(\mathrm{B})+\Delta S(\mathrm{C})-\Delta S(\mathrm{AB})\\
&-\Delta S(\mathrm{AC})-\Delta S(\mathrm{BC})\\
=&\,[\Delta S(\mathrm{DBC})-\Delta S(\mathrm{BC})]+[\Delta S(\mathrm{DAC})\\
&-\Delta S(\mathrm{AC})]+[\Delta S(\mathrm{DAB})-\Delta S(\mathrm{AB})],
\end{split}
\end{equation}
where region D is the complement of the region ABC, i.e., $\mathrm{D}=(\mathrm{ABC})^{\mathrm{c}}$, and we have used  $\mathrm{A}^{\mathrm{c}}=\mathrm{DBC}$ and $S(\mathrm{A})=S(\mathrm{A}^{\mathrm{c}})$. Since the deformation is far from region D (farther than the correlation length $\xi$) as it is acting only on the line $\mathrm{A}\cap \mathrm{B}\cap \mathrm{C}$, each of three square brackets vanishes separately. Hence $S_{\mathrm{KPLW}}[T^2]$ is unchanged under the deformation of the common boundary line of A, B and C. In summary $\Delta S_{\mathrm{KPLW}}[T^2]=0$ under an arbitrary  deformation of the entanglement surface. Therefore property 1 holds. 

We now argue that property 2 holds. As suggested in Refs.~\onlinecite{PhysRevLett.96.110404,PhysRevLett.96.110405}, when we locally perturb the Hamiltonian far inside one region\footnote{Quantitatively, the shortest distance $d$ between the position of the local deformation and the entanglement surface should be much longer than the correlation length $\xi\simeq 1/m$, i.e., $d\gg \xi$.}, for instance region A, the finiteness of the correlation length $\xi$ guarantees that the perturbation does not affect the reduced density matrix for the region $\mathrm{A}^{\mathrm{c}}$. Therefore the entanglement entropy $S(\mathrm{A})=S(\mathrm{A}^{\mathrm{c}})$ is unchanged. If a perturbation of the Hamiltonian occurs on the common boundary of multiple regions, for example region A and B, one can deform the entanglement surface using property 1 such that the perturbation is non-vanishing in one region only. This shows that $S_{\mathrm{KPLW}}[T^2]$ is invariant under local deformations of the Hamiltonian which does not close the gap (i.e., those which leave $\xi<\infty$), and property 2 holds. In summary $S_{\mathrm{KPLW}}[T^2]$ is a topological and universal quantity. 

Lastly we show that the combination $S_{\mathrm{KPLW}}[T^2]$ equals the TEE, $S_{\mathrm{topo}}[T^2]$, i.e., 
\begin{eqnarray}
S_{\mathrm{KPLW}}[T^2]=S_{\mathrm{topo}}[T^2]\label{KPLWTEE},
\end{eqnarray}
where $S_{\mathrm{topo}}[T^2]$ is defined in Eq.~\eqref{DefineTopoEE}. We insert the expansion of the EE \eqref{variousterms} in the definition of $S_{\mathrm{KPLW}}[T^2]$. First, it is straightforward to check that the KPLW combination of the area law terms cancel. Second, the KPLW combination of the Euler characteristic terms vanish since each region in the KPLW combination is topologically a $T^2$, and $\chi(T^2)=0$. Third, as we prove in Appendix~\ref{APPKPLWH2}, the KPLW combination of the mean curvature terms vanishes as well, i.e, 
\begin{eqnarray}
4F'_2\int_{\substack{\partial \mathrm{A}+\partial \mathrm{B}+\partial \mathrm{C}\\-\partial \mathrm{AB}-\partial \mathrm{AC}\\-\partial \mathrm{BC}+\partial \mathrm{ABC}}}d^2x \sqrt{h}H^2=0.\label{KPLWH22}
\end{eqnarray}
This was assumed implicitly in Ref. \onlinecite{grover2011entanglement}, but we demonstrate it explicitly here so as to close the loop in the argument. 

Finally, the KPLW combination simplifies to $S_{\mathrm{topo}}[T^2]$: it is given by the sum of the TEE across the four tori $\partial\mathrm{A}, \partial\mathrm{B}, \partial\mathrm{C}$ and $\partial \mathrm{ABC}$, minus the TEE across the three tori $\partial\mathrm{AB}, \partial\mathrm{AC}$ and $\partial\mathrm{BC}$. Therefore, Eq.~\eqref{KPLWTEE} holds. In summary, we have demonstrated that the KPLW prescription, Eq.~\eqref{KPLWT2}, gives a concrete method to extract the TEE for a generic (non-fixed-point) theory.

\section{Application: Entanglement Entropy of Generalized Walker-Wang Theories}
\label{EntanglementEntropyOfTQFT}

In this section, we construct lattice ground state wave functions for a class of TQFTs known as the generalized Walker-Wang (GWW) models, whose actions are given by Eq.~\eqref{GWW} below. We then compute the EE across various two dimensional entanglement surfaces. The calculations in this section are independent of the SSA inequality used in Sec.~\ref{GeneralStructure}. The calculations in this section provide support for our assumptions about the entanglement entropy for fixed-point models, and suggest a conjecture about higher dimensional topological phases. 

The GWW models are described by a TQFT with the action\cite{Walker2012,Gaiotto2015,seiberg2014coupling}
\begin{equation}\label{GWW}
\mathcal{S}_{\mathrm{GWW}} = \int \frac{n}{2\pi} B\wedge dA + \frac{np}{4\pi} B \wedge B,	\;\; n,p\in\mathbb{Z}.
\end{equation}
The Walker-Wang models correspond to the special cases $p=0$ and $p=1$. In Eq.~\eqref{GWW} $B$ is a 2-form $U(1)$ gauge field and $A$ is a 1-form $U(1)$ gauge field. (When we formulate the theory on a lattice, they will be $\mathbb{Z}_n$ valued. See Appendix~\ref{TriangulationofTQFT} for details.)  The gauge transformations of the gauge fields are
\begin{eqnarray}\label{gaugetransformation}
\begin{split}
&A\to A+dg-p\lambda,\\
&B\to B+d\lambda,
\end{split}
\end{eqnarray}
where $\lambda$ is a $u(1)$ valued 1-form gauge field (where $u(1)$ is the Lie algebra of $U(1)$) with gauge transformation $\lambda\to \lambda+df$ (where $f$ is a scalar satisfying $f\simeq f+2\pi$), and $g$ is a compact scalar (i.e., $g\simeq g+2\pi$). The gauge invariant surface and line operators are respectively
\begin{eqnarray}
\begin{split}
&\exp\Big(i k\oint_{\Sigma_1} B\Big), ~k\in \{0,1, ..., n-1\},\\
&\exp\Big(i l \oint_{\gamma}A+ilp\int_{\Sigma_2} B\Big), ~l\in \{0,1, ..., n-1\},\label{SurfaceLineOp}
\end{split}
\end{eqnarray}
where $\Sigma_1$ is a closed two dimensional surface, $\gamma$ is a closed one dimensional loop and $\Sigma_2$ is an open two dimensional surface whose boundary is $\gamma$. The gauge invariance follows from the compactification of the scalar $g$ and the standard Dirac flux quantization condition of $U(1)$ gauge field $\lambda$: $\oint_{\gamma} dg\in 2\pi \mathbb{Z}$ and $\oint_{\Sigma_1}d\lambda\in 2\pi \mathbb{Z}$.\footnote{The Dirac flux quantization of the $U(1)$ gauge field $\lambda$ can be derived as follows: $\oint_{\Sigma_1}d\lambda=\int_{\Sigma_1^+}d\lambda^+-\int_{\Sigma_1^-}d\lambda^- =\int_{\partial \Sigma_1^+}\lambda^+-\int_{\partial \Sigma_1^-}\lambda^-$, where $\Sigma_1^+\cup \Sigma_1^-=\Sigma_1$ and the minus sign of the $\Sigma_1^-$ term is due to orientation. We use $\lambda^+$ and $\lambda^-$ to emphasis that the gauge field are evaluated in $\Sigma_1^+$ and $\Sigma_1^-$ respectively. The $U(1)$ gauge symmetry implies that $\lambda^+-\lambda^-$ on the common boundary $\partial \Sigma_1^+=\partial \Sigma_1^-=\Sigma_1^+\cap \Sigma_1^-$ does not have to vanish, but it can be a pure gauge $df$. Therefore, $\oint_{\Sigma_1}d\lambda=\oint_{\Sigma_1^+\cap \Sigma_1^-}df\in 2\pi \mathbb{Z}$. This proves the Dirac flux quantization. }  We will use canonical quantization to explain that $\exp(i n\oint_{\Sigma_1} B)$ and $\exp(i n \oint_{\gamma}A+inp\int_{\Sigma_2} B)$ are trivial operators in App.~\ref{TriangulationofTQFT}. 

\subsection{Wave Function of GWW Models}

\subsubsection{BF Theory: $(n,0)$}

\begin{figure}[t]
	\includegraphics[width=0.5\textwidth]{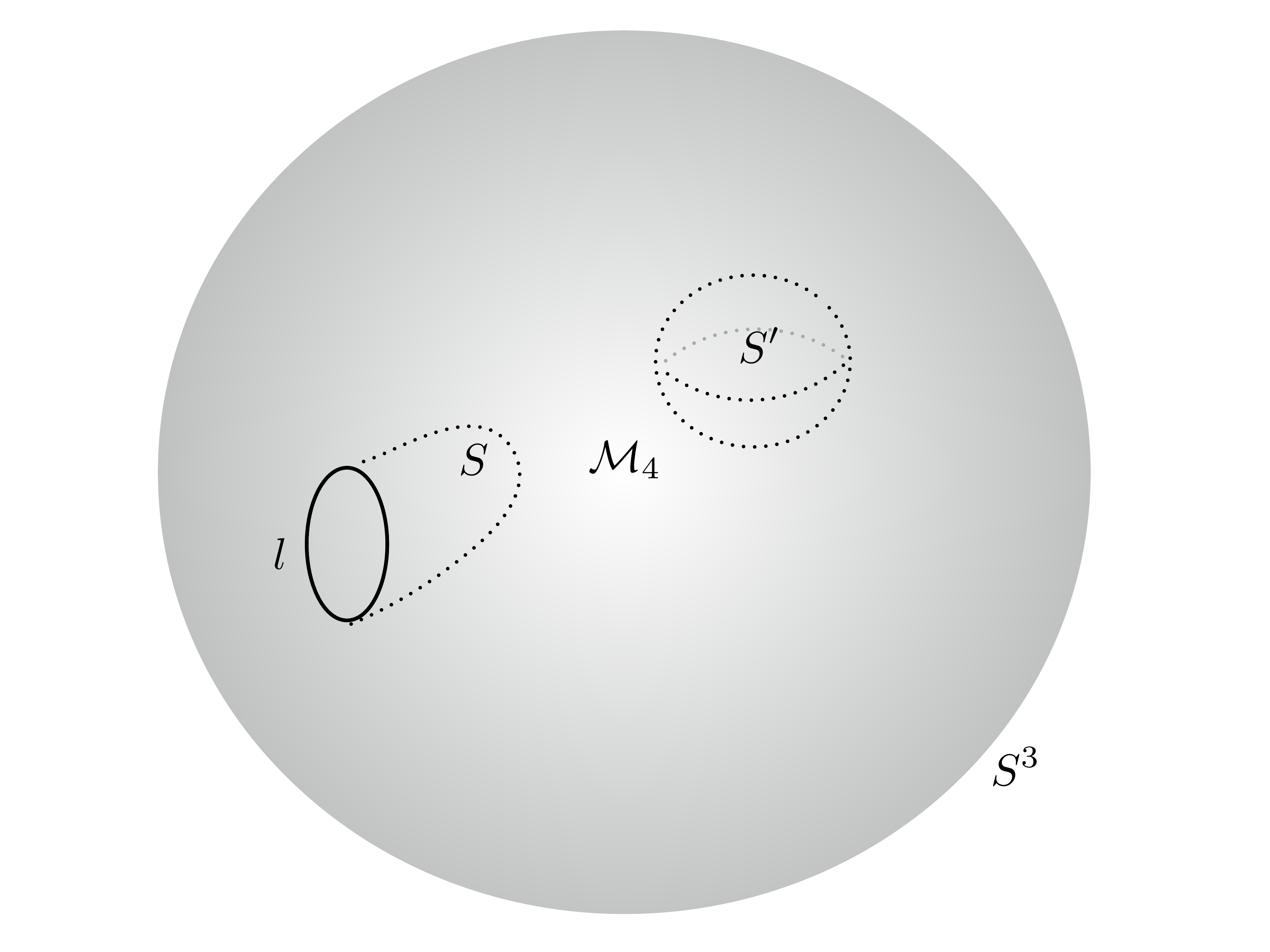}
	\centering
	\caption{A schematic figure of the topology of spacetime $\mathcal{M}_4$ and space $S^3$. Inside $S^3$, we schematically draw a loop $l$ representing the loop  configurations $\mathcal{C}$ of the $B$ field in the dual lattice. The dashed surface $S$ bounding the loop $l$ extends into the spacetime bulk $\mathcal{M}_4$, representing the $B$ field configuration in the dual lattice of spacetime. $S'$ represents the $B$ field configurations that form closed surfaces away from the boundary of the spacetime $\partial \mathcal{M}_4$. The boundary condition in the path integral Eq.~\eqref{WavefunctionBF} is specified by a fixed $B$ configuration $\mathcal{C}$ on $S^3$. The path integral should integrate over all the configurations in the spacetime bulk $\mathcal{M}_4$ with the boundary configuration $\mathcal{C}$ on $S^3$ fixed.}
	\label{M4S3}
\end{figure}

For simplicity, we first discuss the special case when $p=0$, which is referred to as a BF theory. The action is 
\begin{equation}\label{BFtheory}
\mathcal{S}_{\mathrm{BF}} = \int_{\mathcal{M}_4} \frac{n}{2\pi} B \wedge dA,
\end{equation}
where $A$ is a 1-form gauge field and $B$ is a 2-form gauge field. The theory is defined on a spacetime which is topologically a four ball, $\mathcal{M}_4\simeq\mathcal{B}^4$, whose boundary $S^3$ is a spatial slice, as shown in Fig.~\ref{M4S3}. In the following, we formulate the theory on a triangulated spacetime lattice. The 1-form gauge field $A$ corresponds to 1-cochains $A(ij)\in \frac{2\pi}{n}\mathbb{Z}_{n}$ living on 1-simplices $(ij)$. The 2-form gauge field $B$ corresponds to 2-cochains $B(ijk)\in \frac{2\pi}{n}\mathbb{Z}_{n}$ living on 2-simplices $(ijk)$\footnote{We use $i,j,k$ to label vertices, and $(ij),(ijk)$ to label 1-simplices and 2-simplices with the specified vertices.}. We define the Hilbert space to be $\mathcal{H}=\otimes_{(ijk)}H_{(ijk)}$, where $H_{(ijk)}$ is a local Hilbert space on the 2-simplex $(ijk)$ spanned by the basis $|B(ijk)\rangle=|2\pi q/n\rangle, q\in \mathbb{Z}_n$.\footnote{Note that the Hilbert space on each 1-simplex is defined independently, and does not have to satisfy the closed loop (Gauss law) constraint Eq.~\eqref{constraintBijk}.} More details about the lattice formulation of the TQFT are given in Appendix~\ref{TriangulationofTQFT}.

We now discuss the ground state wave function for this theory. The ground state wave function is defined on the boundary of the open spacetime manifold $S^3=\partial\mathcal{M}_4$ as\cite{witten1989,Elitzur:1989nr}
\begin{equation}\label{WavefunctionBF}
|\psi\rangle=\mathfrak{C}\sum_{\mathcal{C},\mathcal{C}'}\int\limits_{\mathcal{C}'|_{\partial\mathcal{M}_4}} \mathcal{D}A\int\limits_{\mathcal{C}|_{\partial\mathcal{M}_4}}\mathcal{D}B \exp\Big(i\frac{n}{2\pi} \int\limits_{\mathcal{M}_4} B\wedge dA \Big)|\mathcal{C}\rangle,
\end{equation}
where $\mathcal{C}'$ and $\mathcal{C}$ indicate the boundary configurations for the $A$ and $B$ fields respectively, i.e., the value of $A$ and $B$ fields on $\partial \mathcal{M}_4$. We integrate over all $A$ and $B$ subject to the boundary conditions $\mathcal{C}'$ and $\mathcal{C}$. $\mathfrak{C}$ is a normalization factor. Because $A$ and $B$ are canonically conjugate, the states are specified by the configuration of $B$ only; $|\mathcal{C}\rangle$ is a specific state corresponding to the particular $B$ field configuration $\mathcal{C}$ on $\partial\mathcal{M}_4$. The summation over $\mathcal{C}$ ranges over all possible configurations of $B$-cochain with weights determined by the path integral. $\mathcal{C}|_{\partial\mathcal{M}_4}$ means the path integral is subject to the fixed boundary conditions $\mathcal{C}$ on $\partial \mathcal{M}_4$, and similarly for $\mathcal{C}'|_{\partial\mathcal{M}_4}$. If we take the spacetime $\mathcal{M}_4$ to be a closed manifold, Eq.~\eqref{WavefunctionBF} reduces to the partition function over $\mathcal{M}_4$. Because the spacetime is topologically a 4-ball $\mathcal{B}^4$, there is only one ground state associated with the boundary $S^3$.\footnote{Topologically degenerate ground states are the representation of line and surface operators which wrap around the nontrivial spatial cycles. Since there are no nontrivial 1-cycles and 2-cycles in the spatial manifold $S^3$ that line and surface operators can wrap around, the ground state is topologically non-degenerate.}

We first work out the wavefunction for the BF theory with $n=2$ explicitly as a generalizable example. We use $B$ field values as a basis to express $|\mathcal{C}\rangle$. Integrating out $A$ (notice that we both integrate over the configurations of the $A$-field with fixed boundary configurations and also sum over the boundary configurations, i.e., $\sum_{\mathcal{C}'}\int_{\mathcal{C}'|_{\partial\mathcal{M}_4}} \mathcal{D}A$, which is tantamount to integrating over all configurations of $A$), we get the constraint $\delta(dB)$, 
\begin{equation}\label{BFWF}
|\psi\rangle= \mathfrak{C}\sum_{\mathcal{C}} \int\limits_{\mathcal{C}|_{\partial\mathcal{M}_4}} \mathcal{D}B \delta\big(dB\big) |\mathcal{C}\rangle.
\end{equation}
where the delta function $\delta(dB)$ constrains $dB(ijkl)=0\mod 2\pi$ on each tetrahedron $(ijkl)$ in $\mathcal{M}_4$. Concretely,
\begin{equation}\label{constraintBijk}
\begin{split}
dB(ijkl)&= B(jkl)-B(ikl)+B(ijl)-B(ijk)\\
&= 0 \mod 2\pi.
\end{split}
\end{equation}
Any $B$ configuration satisfying this constraint is said to be flat (see Appendix~\ref{TriangulationofTQFT} for details). 
Since $B(ijk)\in \{0, \pi\}, \forall i,j,k$ for the $n=2$ theory, Eq.~\eqref{constraintBijk} means that for each tetrahedron, there are an even number of 2-simplices where $B(ijk)=\pi\mod 2\pi$, and an even number of 2-simplices with $B(ijk)=0\mod 2\pi$. We refer to the $\pi$ 2-simplices as occupied and to the $0$ 2-simplices  as unoccupied. 

\begin{figure}[t]
	\includegraphics[width=0.5\textwidth]{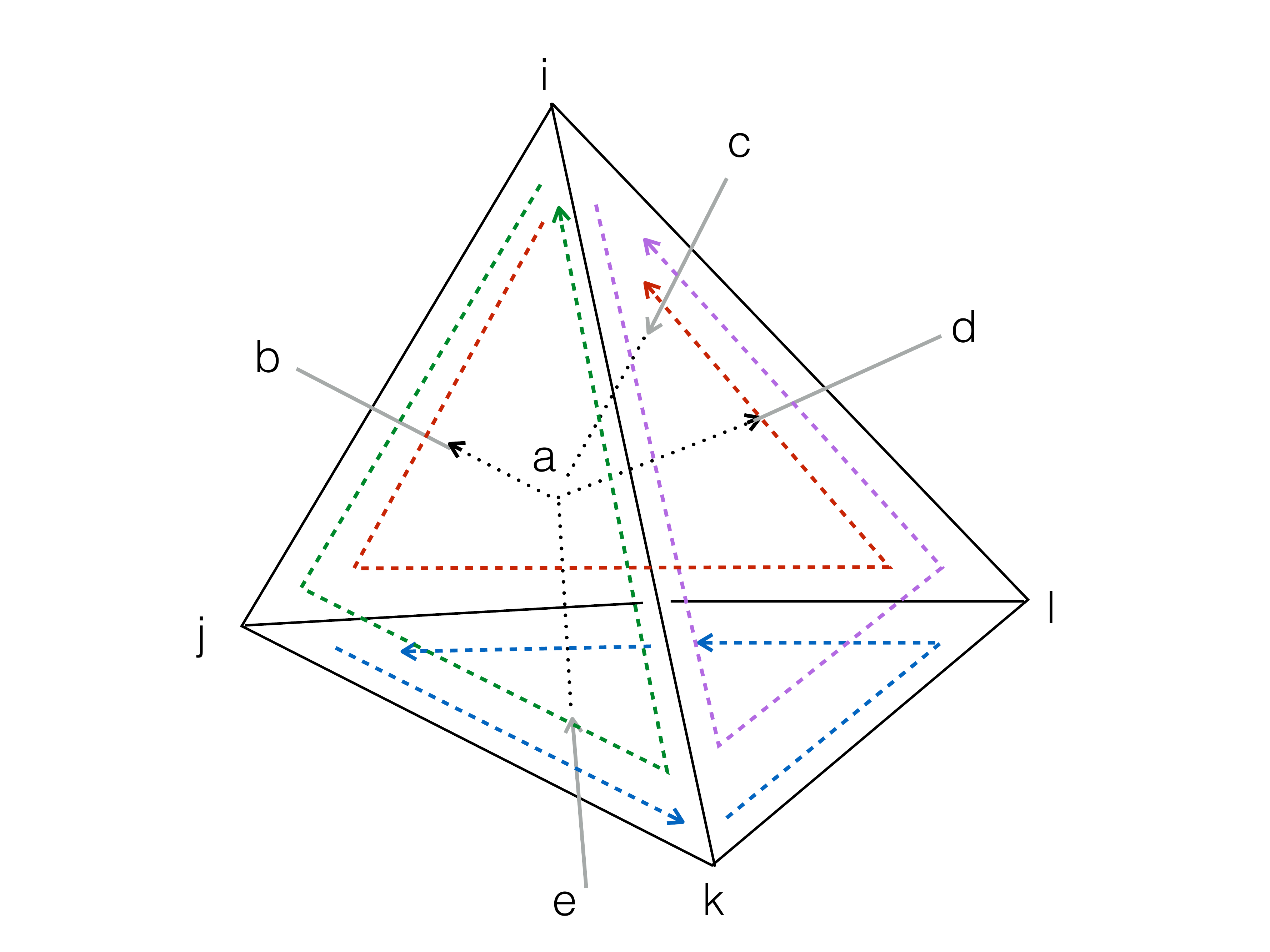}
	\centering
	\caption{A tetrahedron is drawn with solid lines, and its dual is drawn in dash and gray lines. The 2-simplex $(ijk)$ in the original lattice is dual to the 1-simplex $(ab)$ in the dual lattice. Similarily, $(ikl)$ is dual to $(ad)$, $(ijl)$ is dual to $(ca)$ and  $(jkl)$ is dual to $(ea)$. The colored dash arrows indicate the orientations of the four 2-simplices, where $(ijk)$ and $(ikl)$ share the same orientation, and $(ijl)$ and $(jkl)$ share the opposite orientation. The orientations of the dual-lattice 1-simplices are also indicated by the arrows on the grey/dashed lines. }
	\label{DualLattice}
\end{figure}

It is more transparent to consider the configurations in the dual lattice of the spatial slice $S^3$. (In the next paragraph, we will discuss the dual lattice configurations in the spacetime $\mathcal{M}_4$.) As an example, the dual lattice of a tetrahedron is shown in Fig.~\ref{DualLattice}. The 2-simplices in the original lattice are mapped to 1-simplices in the dual lattice.\footnote{The dual lattice of a triangulation is not necessarily a triangulation. For example, the dual lattice of a triangular lattice in two dimensions is a honeycomb lattice. Therefore, it is inappropriate to talk about cochains and simplices in the dual lattice of a triangulation. However, we will still use such notions for simplicity as long as the context is clear. In the dual lattice, we use ``1-simplex'' to denote a link, and ``1-cochain'' to denote a discretized 1-form on the link.} A 2-cochain $B(ijk)$ defined on a 2-simplex in the original lattice is mapped to a 1-cochain $\tilde{B}(ab)$ defined on an 1-simplex in the dual lattice. If $B(ijk)=\pi$, then we define the corresponding $\tilde{B}(ab)=\pi$ in the dual lattice. In the dual lattice, Eq.~\eqref{constraintBijk} means that there are an even number of occupied bonds (1-simplices) associated with each vertex, as well as an even number of unoccupied bonds. If we glue different tetrahedra together, we find that the occupied bonds in the dual lattice form loops. Pictorially, this is reminiscent of the wave function of the toric code model in one lower dimension\cite{2009arXiv0904.2771K,2003AnPhy.303....2K,2006AnPhy.321....2K}. 

In the $(3+1)$D spacetime $\mathcal{M}_4$ [rather than the $3$D space $S^3$], 2-simplices are dual to the $(4-2)=2$-simplices [rather than the 1-simplices] in the dual lattice. Equation~\eqref{constraintBijk} means the occupied 2-simplices form continuous surfaces in the dual spacetime lattice. (Continuous means that the simplices in the dual lattice connect via edges, rather than via vertices. We discuss the continuity of the dual lattice surfaces in Appendix~\ref{Appclosedsurface}.) If these surfaces are inside the bulk of the spacetime and do not touch $\partial \mathcal{M}_4$ (such as $S'$ in Fig.~\ref{M4S3}), they are continuous and closed surfaces; if the surfaces intersect with the spatial slice $\partial \mathcal{M}_4$ (such as $S$ in Fig.~\ref{M4S3}), the intersections are closed loops in $\partial \mathcal{M}_4$.

For the BF theory with a general coefficient $n$, the wavefunction is also a superposition of loop configurations. The only difference is that the loops are formed by 1-simplices in the dual lattice with $\tilde{B}=\frac{2\pi}{n}$. When there is a loop formed by 1-simplices with $\tilde{B}=\frac{2\pi l}{n}$  in the dual lattice, we regard the loop as composed of $l$ overlapping loops formed by the same 1-simplices with $\tilde{B}=\frac{2\pi }{n}$. We emphasize that the loop configuration is enforced by the flatness condition Eq.~\eqref{constraintBijk}. For $n>2$, we need to specify the orientations of the simplices and keep tract of the signs in Eq.~\eqref{constraintBijk}. The orientation of each simplex is specified in Fig.~\ref{DualLattice}, where the orientations of $(jkl)$ and $(ijl)$ are pointing into the tetrahedron, while the orientation of $(ikl)$ and $(ijk)$ are pointing out of the tetrahedron. For example, if the values of the $B$-cochains are $B=2\pi q_1/n, 2\pi q_2/n, 2\pi q_3/n, 0$ with $q_1-q_2+q_3=0$ on the 2-simplices $(jkl), (ikl), (ijl), (ijk)$ respectively, the dual of $(jkl)$ and $(ikl)$ (i.e., $(ea)$ and $(ad)$) belong to one loop in the dual lattice, while the dual of $(ijl)$ and $(ikl)$ (i.e., $(ca)$ and $(ad)$) belong to another loop in the dual lattice. Note that the two loops share the same dual lattice bond $(ad)$ where the value of the $B$-cochain is the sum of the $B$ values from the two loops $B(ad)=2\pi (q_1+q_3)/n=2\pi q_2/n$. The gauge transformation, $B(ijk)\to B(ijk)+\lambda(jk)-\lambda(ik)+\lambda(ij)$, preserves Eq.~\eqref{constraintBijk}. Hence, although it deforms the position of loops, it never turns closed loops into open lines. Open lines in the dual lattice violate the flatness condition Eq.~\eqref{constraintBijk}, and so do not contribute to the wave function Eq.~\eqref{BFWF}. Summing over the configurations $\mathcal{C}$ ensures gauge invariance of the wave function. Notice that Eq.~\eqref{BFWF} implies that the weights associated with different loop configurations $\mathcal{C}$ are equal, similar to the toric code. Thus we see that Eq.~\eqref{BFWF} reduces to
\begin{equation}
|\psi\rangle=\mathfrak{C}\sum_{\mathcal{C}\in\mathcal{L}}|\mathcal{C}\rangle,\label{WavefunctionBFTheory}
\end{equation}
where the sum is taken over the set $\mathcal{L}$ of all possible loop configurations $\mathcal{C}$ at the spatial slice $S^3=\partial\mathcal{M}_4$. This is termed ``loop condensation", since the wave function is the equal weight superposition of all loop configurations in the dual lattice. 

\subsubsection{General Case: $(n,p)$}

In this section, we consider GWW models with nontrivial $p$ described by the action in Eq.~\eqref{GWW}, where $A$ is still a 1-form and $B$ a 2-form. Canonical quantization of the GWW theories implies that $B\in \frac{2\pi}{n}\mathbb{Z}_n$ on the lattice (see Appendix~\ref{TriangulationofTQFT} for more details).

In order to find the ground state wave function, we still use $B$ as the basis to label the configurations $\mathcal{C}$ and the corresponding states $|\mathcal{C}\rangle$ on the spatial slice.  The wave function is formally given by 
\begin{equation}\label{Wavefunction}
\begin{split}
|\psi\rangle
=&\mathfrak{C}\sum_{\mathcal{C},\mathcal{C}'}\int\limits_{\mathcal{C}'|_{\partial\mathcal{M}_4}} \mathcal{D}A\int\limits_{\mathcal{C}|_{\partial\mathcal{M}_4}}\mathcal{D}B	\\
& \exp\Big(i \frac{n}{2\pi}\int\limits_{\mathcal{M}_4} B \wedge dA+i\frac{np}{4\pi}\int\limits_{\mathcal{M}_4} B\wedge B \Big)|\mathcal{C}\rangle.
\end{split}
\end{equation}
For simplicity, we consider the case $n=2, p=1$ in the following. As in the BF theory, we first integrate out the $A$ fields, yielding
\begin{equation}
|\psi\rangle=\mathfrak{C}\sum_{\mathcal{C}}\int\limits_{\mathcal{C}|_{\partial\mathcal{M}_4}} \mathcal{D}B\, \delta\big(dB\big)\,\exp\Big(i\frac{2}{4\pi}\int\limits_{\mathcal{M}_4} B\wedge B \Big)|\mathcal{C}\rangle.
\end{equation}
The difference between this wave function and that of the BF theory, Eq.~\eqref{BFWF}, is that when the flatness condition $\delta(dB)$ is satisfied, the states with different configurations $\mathcal{C}$ are associated with different weights. The weights are determined by the integral
\begin{equation}\label{linking}
\begin{split}
&\exp\Big(i\frac{2}{4\pi}\int_{\mathcal{M}_4} B\wedge B\Big),
\end{split}
\end{equation}
where $B$ must satisfy the flatness condition $dB=0$ with the boundary condition labeled by $\mathcal{C}$. 

We proceed to evaluate the integral in Eq.~\eqref{linking}. Notice that the flatness condition, Eq.~\eqref{constraintBijk}, implies that the 2-simplices at which $B=\pi$ form two-dimensional spacetime surfaces in the dual lattice of $\mathcal{M}_4$ whose boundaries on the spatial slice $S^3$ are closed loops belonging to $\mathcal{C}$. Relegating the details of the derivation to Appendix~\ref{AppMutualandSelfLinkingNumbers}, we show that when $B=\pi$ only at two  dual lattice surfaces $S_1, S_2$, whose boundaries are  dual lattice loops $l_1=\partial S_1, ~l_2=\partial S_2$ in $\mathcal{C}$,  it follows that
\begin{equation}\label{BwedgeB}
\begin{split}
&\exp\Big(i\frac{2}{4\pi}\int_{\mathcal{M}_4} B\wedge B \Big)\\
=&\exp\Big(i\pi \mathrm{link}(l_1,l_2)+i\frac{\pi}{2}\mathrm{link}(l_1,l_1)+i\frac{\pi}{2}\mathrm{link}(l_2,l_2)\Big).
\end{split}
\end{equation}
The first term is associated with the mutual linking number, $\mathrm{link}(l_1,l_2)$, between different loops, while the second and the third terms are associated with the self-linking number, $\mathrm{link}(l_i,l_i)$, of one loop, $l_i$, with itself, defined in Appendix~\ref{AppMutualandSelfLinkingNumbers}. Equation \eqref{BwedgeB} can be generalized to configurations with many loops, and the weights of different configurations are determined by the linking numbers of the loops. In summary, the ground state wave function for the $(n,p)=(2,1)$ theory is:
\begin{equation}\label{WFGWW1}
|\psi\rangle
=\mathfrak{C}\sum_{\mathcal{C}\in\mathcal{L}}(-1)^{\#(\mathrm{Mutual~links})}i^{\#(\mathrm{Self~links})}|\mathcal{C}\rangle.
\end{equation}

For general $(n,p)$, a similar argument can be made. $B$ can now take $n$ different values $\textstyle \frac{2\pi k}{n}, k=0,1,\cdots, n-1$ on each 2-simplex in the lattice, or on each 1-simplex in the dual lattice. Due to the constraint of Eq.~\eqref{constraintBijk}, the 1-simplices where $B=2\pi/n$ form loops in the dual lattice. Similar to the discussion of the case $p=0$ and general $n$, two dual-lattice loops can touch in one tetrahedron. We also regard a loop with $B=2\pi q/n$ to be $q$ overlapping loops with $B=2\pi /n$.  If there are $q_1$ loops with $B=2\pi/n$ that are overlapping on $l_1$ (which is equivalent to one loop with $B=2\pi q_1/n$ on $l_1$) and $q_2$ loops with $B=2\pi/n$ that are overlapping on $l_2$ (which is equivalent to one loop with $B=2\pi q_2/n$ on $l_2$), then
\begin{equation}
\begin{split}
&\exp\Big(i\frac{np}{4\pi}\int\limits_{\mathcal{M}_4} B\wedge B \Big)	\\
=&\exp\Big[2i\frac{np(2\pi)^2q_1q_2}{4\pi n^2} \mathrm{link}(l_1,l_2)+i\frac{np(2\pi)^2q_1^2}{4\pi n^2}\mathrm{link}(l_1,l_1)	\\
&+i\frac{np(2\pi)^2q_2^2}{4\pi n^2}\mathrm{link}(l_2,l_2)\Big]	\\
=& \exp\Big[i\frac{2\pi pq_1q_2}{n}\mathrm{link}(l_1,l_2)+i\frac{\pi pq_1^2}{ n}\mathrm{link}(l_1,l_1)	\\
&+i\frac{\pi pq_2^2}{n}\mathrm{link}(l_2,l_2)\Big]\label{BFBBalphabeta}.
\end{split}
\end{equation}
Therefore after evaluating these weights, the wave function Eq.~\eqref{Wavefunction} reduces to
\begin{equation}
|\psi\rangle=\mathfrak{C}\sum_{\mathcal{C}\in \mathcal{L}} 
\mathrm{e}^{i\frac{2\pi p}{n} \#(\mathrm{Mutual~links})} 
\mathrm{e}^{i\frac{\pi p}{n}\#(\mathrm{Self~links})}
|\mathcal{C}\rangle,\label{WaveFunctionnp}
\end{equation}
where the mutual-linking and self-linking numbers are counted with multiplicities $q_1$ and $q_2$ as given in Eq.~\eqref{BFBBalphabeta}. The sum over $\mathcal{C}\in \mathcal{L}$ contains configurations with all possible $q_1$ and $q_2$.

\subsection{Entanglement Entropy of GWW Models}

In this section, we show that the constant part of the EE of GWW theories depends on the topology of the entanglement surface in a nontrivial way. In particular, $S_{\mathrm{c}}[S^2]\neq S_{\mathrm{c}}[T^2]$ in general. 
Hence, $S_{\mathrm{c}}[S^2]$ and $S_{\mathrm{c}}[T^2]$ are truly independent quantities. 

This section is divided into two parts: In Sec.~\ref{EEGWW1}, we calculate the EE for GWW models with arbitrary $(n,p)$ across the entanglement surface $T^2$. In Sec.~\ref{EEGWW3}, we compute the EE for GWW models across closed surfaces with arbitrary genus and an arbitrary number of disconnected components. These independent calculations confirm Eq.~\eqref{TEEcentralresult}.

\subsubsection{EE for the Torus, $n=2, p=1$}
\label{EEGWW1}

In this subsection, we compute the EE of GWW models across $\Sigma=T^2$. For simplicity, we first consider the case $n=2, p=1$, and then generalize to models with arbitrary $n$ and $p$. 

\begin{figure}[t]
	\includegraphics[width=0.46\textwidth]{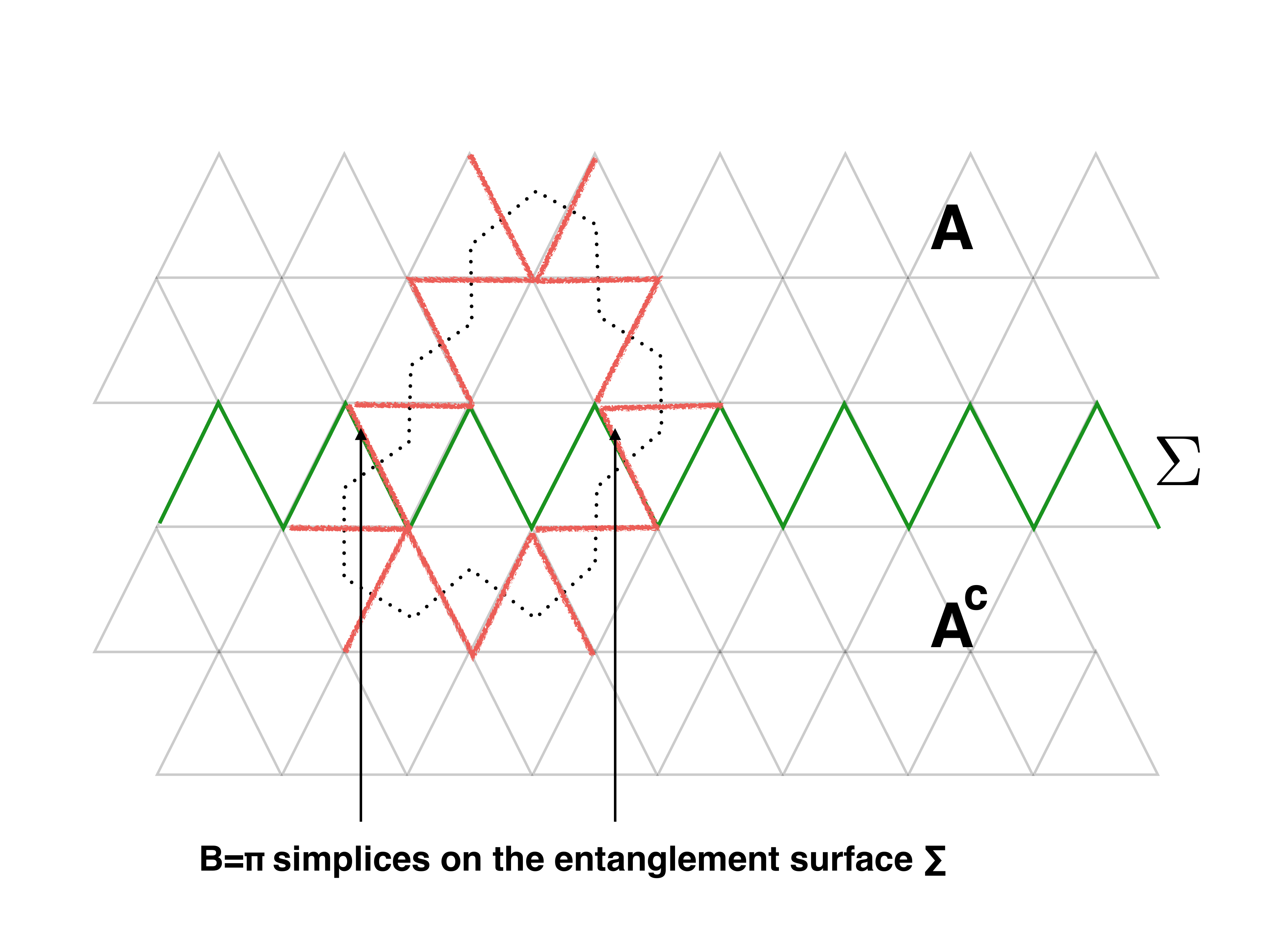}
	\centering
	\caption{An example of the lattice structure of an entanglement cut in $(2+1)$D. The green simplices form the entanglement cut $\Sigma$, which partitions the lattice into region A and region $\mathrm{A}^\mathrm{c}$. We include $\Sigma$ as part of region A. $B=\pi$ on the red simplices, while $B=0$ elsewhere. The dotted loop is the dual lattice configuration of the red simplices.  In this example, the configuration $\mathcal{C}_\mathrm{E}$ contains two $B=\pi$ 1-simplices at the entanglement cut $\Sigma$, which are the fourth and eighth 1-simplices of $\Sigma$ (counting from the left side) as shown in the figure.}
	\label{entanglementsurface}
\end{figure}

\begin{figure}[H]
	\includegraphics[width=0.46\textwidth, height=6cm]{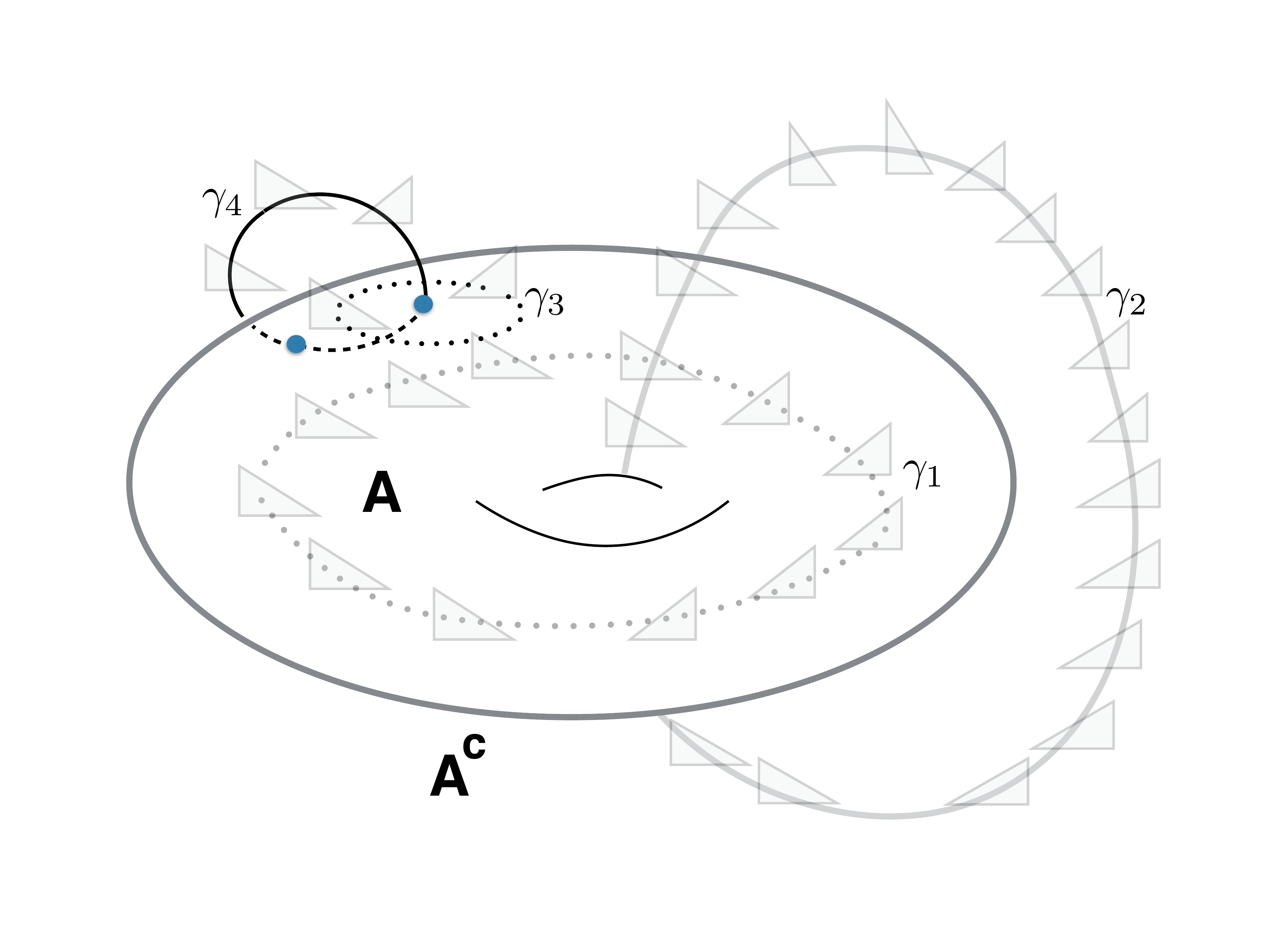}
	\centering
	\caption{A particular spatial configuration with one loop $\gamma_1$ (dashed line) threading through the hole (the hole itself belongs to region $\mathrm{A}^{\mathrm{c}}$) inside the region A and one loop $\gamma_2$ (grey line) threading through the hole inside the region $\mathrm{A}^\mathrm{c}$. $\gamma_3$ and $\gamma_4$ are two linked contractible loops, where $\gamma_3$ locates inside region A, and $\gamma_4$ locates both in region A and $\mathrm{A}^{\mathrm{c}}$. The two blue points are the intersection of $l_4$ with $\Sigma$. The simplices (gray triangles) are living in the real lattice where $B=\pi$. The lines perpendicular to the simplices are living in the dual lattice where $\widetilde{B}=\pi$ and they form loops in the dual lattice. This configuration corresponds to $\alpha=\mathrm{o}, \beta=\mathrm{o}$.}
	\label{solidtorus}
\end{figure}

We start with the wave function obtained in the last section, Eq.~\eqref{WFGWW1}:
\begin{equation}
|\psi\rangle=\mathfrak{C}\sum_{\mathcal{C}}(-1)^{\#(\mathrm{Mutual~links})}i^{\#(\mathrm{Self~links})}|\mathcal{C}\rangle. \label{eq:WWWF}
\end{equation} 
We choose the subregion A to be a solid torus whose surface is $T^2$, and $\mathrm{A}^{\mathrm{c}}$ to be the complement of A. We illustrate the microscopic structure of the spatial partitioning in Fig.~\ref{entanglementsurface} via a lower-dimensional example. The entanglement surface $\Sigma$ is chosen to be a smooth surface in the real spatial lattice (green simplices in Fig.~\ref{entanglementsurface}).  The real space simplices that form the entanglement surface $\Sigma$ are counted as part of region A. \footnote{There are other choices of spatial partitioning. For example, we can count the real simplices that form the entanglement surface as part of region $\mathrm{A}^\mathrm{c}$. We will consider only consider the partitioning mentioned in the main text for definiteness. } We will find the Schmidt decomposition of the wavefunction corresponding to this spatial partitioning in order to calculate the EE.  To do so, we first parametrize the configurations $\mathcal{C}$ appearing in Eq.~(\ref{eq:WWWF}) as:
\begin{equation}
\mathcal{C}\mapsto\{\mathcal{C}_\mathrm{E},(a,\alpha),(b,\beta)\},
\end{equation}

\noindent which we now explain. $\mathcal{C}_\mathrm{E}$ labels the real space $B$-cochain configuration at the entanglement surface $\Sigma$. (In Fig.~\ref{entanglementsurface}, the fourth and the eighth green 1-simplices (counting from the left side) are occupied on the entanglement surface $\Sigma$, which also belong to region A according to our partition.) We denote by $N_\mathrm{A}(\mathcal{C}_\mathrm{E})$ the number of configurations in the region A (but not including $\Sigma$) consistent with the choice of $\mathcal{C}_\mathrm{E}$. We label such configurations by $(a,\alpha)$, where $\alpha$ is the parity (even $\mathrm{e}$ or odd $\mathrm{o}$) of the number of occupied loops winding around the nontrivial spatial cycle inside the region A in the dual lattice, and the configurations of either parity are enumerated by $a=1,\dots,N_\mathrm{A}(\mathcal{C}_\mathrm{E})/2$.\footnote{We can establish a one-to-one correspondence between the configurations of loops in the even parity sector and the odd parity sector. If we start with a configuration in the even parity sector in which $k$ dual lattice loops wrap around the non-contractible cycle in region A, we can obtain a configuration in the odd parity sector by adding a single loop wrapping the non-contractible cycle so that there are $(k+1)$ non-contractible dual lattice loops in total. Similarly, we can start with the odd parity sector and obtain the even parity sector. This demonstrates that the number of configurations in the even parity sector is equal to that of the odd parity sector. Therefore, we denote the number of configurations in both sectors by $N_\mathrm{A}(\mathcal{C}_\mathrm{E})/2$. This argument can be generalized to the case of general $n$.} Similarly, $(b,\beta)$  labels the $N_\mathrm{\mathrm{A}^{\mathrm{c}}}(\mathcal{C}_\mathrm{E})$ configurations in region $\mathrm{A}^\mathrm{c}$. Figure~\ref{solidtorus} presents a particular configuration where, besides two contractible dual-lattice loops, there is one dual lattice loop wrapping the non-contractible cycle in the dual lattice of region A and one dual lattice loop wrapping the non-contractible cycle in the dual lattice of region $\mathrm{A}^\mathrm{c}$, which corresponds to $\alpha=\mathrm{o}$ and $\beta=\mathrm{o}$. Note that two non-contractible cycles are in different regions A and $\mathrm{A}^{\mathrm{c}}$.  To be illustrative, we also draw 2-simplices in the real lattice where $B=\pi$ whose dual configurations form loops in the space. Hence the summation over $\mathcal{C}$ splits as:
\begin{equation}
\sum_{\mathcal{C}}~=~\sum_{\mathcal{C}_\mathrm{E}}~~~\sum_{a=1}^{N_{\mathrm{A}}(\mathcal{C}_\mathrm{E})/2}~\sum_{b=1}^{N_{\mathrm{A}^{\mathrm{c}}}(\mathcal{C}_\mathrm{E})/2}~~\sum_{\alpha=\mathrm{e},\mathrm{o}}~~~\sum_{\beta=\mathrm{e},\mathrm{o}}.
\end{equation}
For convenience we also introduce the notation
\begin{widetext}
	\begin{equation}\label{llss}
	\begin{split}
	l_{a,\mathrm{e}}^{\mathcal{C}_\mathrm{E}}=&(-1)^{\#(\mathrm{Mutual~links~with~fixed~\mathcal{C}_\mathrm{E}~configuration~of~region~A~in~even~sector})},	\\
	l_{a,\mathrm{o}}^{\mathcal{C}_\mathrm{E}}=&(-1)^{\#(\mathrm{Mutual~links~with~fixed~\mathcal{C}_\mathrm{E}~configuration~of~region~A~in~odd~sector})},	\\
	s_{a,\mathrm{e}}^{\mathcal{C}_\mathrm{E}}=&i^{\#(\mathrm{Self~links~with~fixed~\mathcal{C}_\mathrm{E}~configuration~of~region~A~in~even~sector})},	\\
	s_{a,\mathrm{o}}^{\mathcal{C}_\mathrm{E}}=&i^{\#(\mathrm{Self~links~with~fixed~\mathcal{C}_\mathrm{E}~configuration~of~region~A~in~odd~sector})},
	\end{split}
	\end{equation}
\end{widetext}
where even/odd sector refers to the set of states with an even/odd number of loops in the dual lattice threading the non-contractible cycle in region A. Similar definitions apply to region $\mathrm{A}^\mathrm{c}$. See Fig.~\ref{solidtorus} for an illustration. We further define $|\tilde{\mathrm{A}}_a^{\mathcal{C}_\mathrm{E}}\rangle_{\alpha}$ to be a state associated with one particular configuration in region A, which is labeled by $\{\mathcal{C}_\mathrm{E}, a, \alpha\}$, and define $|\tilde{\mathrm{A}^{\mathrm{c}}}_b^{\mathcal{C}_\mathrm{E}}\rangle_{\beta}$ likewise in region $\mathrm{A}^{\mathrm{c}}$.  There is a subtlety: we also need to specify the mutual-linking/self-linking number of loops which cross the entanglement surface. We specify that when two loops (among which at least one of them crosses the entanglement surface) are linked, such as $\gamma_3$ and $\gamma_4$ in Fig.~\ref{solidtorus}, the mutual-linking number is counted as part of the A side, i.e., $l_{a,\mathrm{e}}^{\mathcal{C}_\mathrm{E}}$ and $l_{a,\mathrm{o}}^{\mathcal{C}_\mathrm{E}}$. Additionally, when a loop crosses the entanglement surface, the self-linking number of the loop is counted as part of the A side, i.e., $s_{a,\mathrm{e}}^{\mathcal{C}_\mathrm{E}}$ and $s_{a,\mathrm{o}}^{\mathcal{C}_\mathrm{E}}$. We are able to make such a choice because there is a phase ambiguity in the Schmidt decomposition, and phases can be shuffled between A and $\mathrm{A}^{\mathrm{c}}$ by redefining the basis $|\tilde{\mathrm{A}}_a^{\mathcal{C}_\mathrm{E}}\rangle_{\mathrm{e}/\mathrm{o}}$ and $|\tilde{\mathrm{A}^{\mathrm{c}}}_b^{\mathcal{C}_\mathrm{E}}\rangle_{\mathrm{e}/\mathrm{o}}$. (For example, we can define another set of states via $|\hat{\mathrm{A}}_a^{\mathcal{C}_\mathrm{E}}\rangle_{\mathrm{e}/\mathrm{o}}=s_{a,\mathrm{e}/\mathrm{o}}^{\mathcal{C}_\mathrm{E}-1}|\tilde{\mathrm{A}}_a^{\mathcal{C}_\mathrm{E}}\rangle_{\mathrm{e}/\mathrm{o}}$, and $ |\hat{\mathrm{A}^{\mathrm{c}}}_b^{\mathcal{C}_\mathrm{E}}\rangle_{\mathrm{e}/\mathrm{o}}=s_{a,\mathrm{e}/\mathrm{o}}^{\mathcal{C}_\mathrm{E}}|\tilde{\mathrm{A}^{\mathrm{c}}}_b^{\mathcal{C}_\mathrm{E}}\rangle_{\mathrm{e}/\mathrm{o}}$.) As we will see, the reduced density matrix Eq.~\eqref{21reduceddensitymatrix} does not depend on the choice of phase assignment. Combining the above, 
we get
\begin{equation}\label{n2p1entanglementcutwavefunction}
\begin{split}
|\psi\rangle
=&\mathfrak{C} \sum_{\mathcal{C}_\mathrm{E}}\sum_{a=1}^{N_{\mathrm{A}}(\mathcal{C}_\mathrm{E})/2}\sum_{b=1}^{N_{\mathrm{A}^{\mathrm{c}}}(\mathcal{C}_\mathrm{E})/2}\sum_{\alpha=\mathrm{e}/\mathrm{o}}\sum_{\beta=\mathrm{e}/\mathrm{o}}	\\
&(-1)^{\alpha\beta}l_{a,\alpha}^{\mathcal{C}_\mathrm{E}}l_{b,\beta}^{\mathcal{C}_\mathrm{E}}s_{a,\alpha}^{\mathcal{C}_\mathrm{E}}s_{b,\beta}^{\mathcal{C}_\mathrm{E}}|\tilde{\mathrm{A}}_a^{\mathcal{C}_\mathrm{E}}\rangle_{\alpha}|\tilde{\mathrm{A}^{\mathrm{c}}}_b^{\mathcal{C}_\mathrm{E}}\rangle_{\beta}.
\end{split}
\end{equation}
The factor $(-1)^{\alpha\beta}$, which equals $-1$ when $\alpha=\beta=\mathrm{o}$ and $1$ otherwise, reflects the mutual-linking between the non-contractible loops in region A (such as $\gamma_1$ in Fig.~\ref{solidtorus}) and the non-contractible loops in region $\mathrm{A}^\mathrm{c}$ (such as $\gamma_2$ in Fig.~\ref{solidtorus}). Figure \ref{solidtorus} shows a special configuration where there is one non-contractible loop in region A and one non-contractible loop in region $\mathrm{A}^\mathrm{c}$.

From this we easily obtain the reduced density matrix for region A by tracing over the Hilbert space in region $\mathrm{A}^\mathrm{c}$,
\begin{equation}\label{21reduceddensitymatrix}
\begin{split}
\rho_\mathrm{A}=&|\mathfrak{C}|^2\sum_{\mathcal{C}_\mathrm{E}}\frac{N_{\mathrm{A}^{\mathrm{c}}}(\mathcal{C}_\mathrm{E})}{2}\sum_{a,\tilde{a}=1}^{N_{\mathrm{A}}(\mathcal{C}_\mathrm{E})/2}\sum_{\alpha, \tilde{\alpha}, \gamma=\mathrm{e}, \mathrm{o}}\\&(-1)^{(\alpha-\tilde{\alpha})\gamma}	|\mathrm{A}_a^{\mathcal{C}_\mathrm{E}}\rangle_{\alpha}\langle \mathrm{A}_{\tilde{a}}^{\mathcal{C}_\mathrm{E}}|_{\tilde{\alpha}}\\
=&|\mathfrak{C}|^2\sum_{\mathcal{C}_\mathrm{E}}N_{\mathrm{A}^{\mathrm{c}}}(\mathcal{C}_\mathrm{E})\sum_{a,\tilde{a}=1}^{N_{\mathrm{A}}(\mathcal{C}_\mathrm{E})/2}	\\
&\Big(|\mathrm{A}_a^{\mathcal{C}_\mathrm{E}}\rangle_\mathrm{e}\langle \mathrm{A}_{\tilde{a}}^{\mathcal{C}_\mathrm{E}}|_\mathrm{e}+|\mathrm{A}_a^{\mathcal{C}_\mathrm{E}}\rangle_\mathrm{o}\langle \mathrm{A}_{\tilde{a}}^{\mathcal{C}_\mathrm{E}}|_\mathrm{o}\Big),
\end{split}
\end{equation}
where we have performed unitary transformations on the bases $|\tilde{\mathrm{A}}_a^{\mathcal{C}_\mathrm{E}}\rangle_{\mathrm{e}/\mathrm{o}}$ and $|\tilde{\mathrm{A}^{\mathrm{c}}}_b^{\mathcal{C}_\mathrm{E}}\rangle_{\mathrm{e}/\mathrm{o}}$ to absorb the mutual-linking and self-linking factors within region A and region $\mathrm{A}^\mathrm{c}$ respectively. The transformed bases are denoted $|\mathrm{A}_a^{\mathcal{C}_\mathrm{E}}\rangle_{\alpha}=l_{a,\alpha}^{\mathcal{C}_\mathrm{E}}s_{a,\alpha}^{\mathcal{C}_\mathrm{E}}|\tilde{\mathrm{A}}_a^{\mathcal{C}_\mathrm{E}}\rangle_{\alpha}$ and $|\mathrm{A}_b^{\mathrm{c}\mathcal{C}_\mathrm{E}}\rangle_{\beta}=l_{b,\beta}^{\mathcal{C}_\mathrm{E}}s_{b,\beta}^{\mathcal{C}_\mathrm{E}}|\tilde{\mathrm{A}^{\mathrm{c}}}_b^{\mathcal{C}_\mathrm{E}}\rangle_{\beta}$. 

Furthermore, the constraint 
\begin{equation}\label{normalization}
\Tr_{\ch_\mathrm{A}}(\rho_\mathrm{A})=|\mathfrak{C}|^2\sum_{\mathcal{C}_\mathrm{E}}N_{\mathrm{A}^{\mathrm{c}}}(\mathcal{C}_\mathrm{E})N_{\mathrm{A}}(\mathcal{C}_\mathrm{E})=1
\end{equation}
fixes the normalization constant $\mathfrak{C}$.
For each fixed configuration $\mathcal{C}_\mathrm{E}$ on the entanglement surface, the product of the number of configurations in the region A and the number of configurations in region $\mathrm{A}^\mathrm{c}$, i.e.,  $N_{\mathrm{A}^{\mathrm{c}}}(\mathcal{C}_\mathrm{E})N_{\mathrm{A}}(\mathcal{C}_\mathrm{E})$, is independent of $\mathcal{C}_\mathrm{E}$ (see Appendix~\ref{Appmoreon41} for details). Thus, to compute $\mathfrak{C}$ we need only to count the number of different choices of $\mathcal{C}_\mathrm{E}$. There are in total $2^{|\Sigma|-1}$ different boundary configurations, where the $1$ comes from the constraint that closed dual lattice loops always intersect the entanglement surface twice (hence the number of occupied 1-simplices on $\Sigma$ is even), and $|\Sigma|$ is the number of 2-simplices on the entanglement surface. Since $|\mathfrak{C}|^2N_{\mathrm{A}^{\mathrm{c}}}(\mathcal{C}_\mathrm{E})N_{\mathrm{A}}(\mathcal{C}_\mathrm{E})$ is independent of $\mathcal{C}_\mathrm{E}$, and there are $2^{|\Sigma|-1}$ choices of $\mathcal{C}_\mathrm{E}$,
\begin{equation}\label{1over2Sigma}
{|\mathfrak{C}|^2N_{\mathrm{A}^{\mathrm{c}}}(\mathcal{C}_\mathrm{E})N_{\mathrm{A}}(\mathcal{C}_\mathrm{E})=\frac{1}{2^{|\Sigma|-1}}.}
\end{equation}
We give a more detailed derivation of this formula in Appendix~\ref{Appmoreon41}.

From the reduced density matrix $\rho_{\mathrm{A}}$, we can calculate the entanglement entropy of the ground state $|\psi\rangle$ associated with the torus entanglement surface by the replica trick,
\begin{equation}
\begin{split}
S(\mathrm{A})&=-\Tr_{\ch_\mathrm{A}} \rho_{\mathrm{A}} \log \rho_{\mathrm{A}}=-\frac{d}{dN} \bigg(\frac{\Tr_{\ch_\mathrm{A}}\rho_{\mathrm{A}}^N}{(\Tr_{\ch_\mathrm{A}}\rho_{\mathrm{A}})^N}\bigg)\bigg|_{N=1}
\end{split}
\end{equation}
Using Eq.~\eqref{21reduceddensitymatrix}, 
\begin{equation}
\begin{split}
&\Tr_{\ch_\mathrm{A}}\rho_{\mathrm{A}}^N=|\mathfrak{C}|^{2N}\sum_{\mathcal{C}_{\mathrm{E}_0}}\sum_{a_0=1}^{N_{\mathrm{A}_0}/2}\sum_{\alpha_0=\mathrm{e,o}}\langle \mathrm{A}_{a_0}^{\mathcal{C}_{\mathrm{E}_0}}|_{\alpha_0}\prod_{I=1}^N\bigg( \sum_{\mathcal{C}_{\mathrm{E}_I}}\\
&\sum_{a_I, \tilde{a}_I=1}^{N_{\mathrm{A}_I}/2}\sum_{\alpha_I=\mathrm{e,o}}N_{\mathrm{A}^{\mathrm{c}}}(\mathcal{C}_{\mathrm{E}_I})|\mathrm{A}_{a_I}^{\mathcal{C}_{\mathrm{E}_I}}\rangle_{\alpha_I}\langle \mathrm{A}_{\tilde{a}_I}^{\mathcal{C}_{\mathrm{E}_I}}|_{\alpha_I}\bigg)|\mathrm{A}_{a_0}^{\mathcal{C}_{\mathrm{E}_0}}\rangle_{\alpha_0}\\
&= |\mathfrak{C}|^{2N} \sum_{\mathcal{C}_{\mathrm{E}_0}, a_0, \alpha_0}\prod_{I=1}^N \bigg(\sum_{\mathcal{C}_{\mathrm{E}_I}, a_I, \tilde{a}_I, \alpha_I}N_{\mathrm{A}^{\mathrm{c}}}(\mathcal{C}_{\mathrm{E}_I})\bigg)\delta_{\mathcal{C}_{\mathrm{E}_0} \mathcal{C}_{\mathrm{E}_1}}\\
&\delta_{\mathcal{C}_{\mathrm{E}_1} \mathcal{C}_{\mathrm{E}_2}}\cdots \delta_{\mathcal{C}_{\mathrm{E}_N}\mathcal{C}_{\mathrm{E}_0}}\times \delta_{a_0a_1}\delta_{\tilde{a}_1a_2}\delta_{\tilde{a}_2 a_3}\cdots \delta_{\tilde{a}_{N-1}a_{N}} \delta_{\tilde{a}_Na_0} \\
&\times \delta_{\alpha_0\alpha_1}\delta_{\alpha_1\alpha_2}\cdots \delta_{\alpha_N\alpha_0}\\
&=|\mathfrak{C}|^{2N} \sum_{\mathcal{C}_{\mathrm{E}_0}}N_{\mathrm{A}^{\mathrm{c}}}(\mathcal{C}_{\mathrm{E}_0})^N\sum_{\alpha_0=\mathrm{o,e}}\sum_{a_1=1}^{N_{\mathrm{A}}(\mathcal{C}_{\mathrm{E}_0}/2)}\cdots\sum_{a_N=1}^{N_{\mathrm{A}}(\mathcal{C}_{\mathrm{E}_0}/2)} 1\\
&=|\mathfrak{C}|^{2N} \sum_{\mathcal{C}_{\mathrm{E}_0}}2N_{\mathrm{A}^{\mathrm{c}}}(\mathcal{C}_{\mathrm{E}_0})^N\bigg(\frac{N_{\mathrm{A}}(\mathcal{C}_{\mathrm{E}_0})}{2}\bigg)^N=2^{-|\Sigma|(N-1)}.
\end{split}
\end{equation}
In the first equation, we expand the trace over the Hilbert space in region A. In the second equation, we use the orthogonal condition $\langle A_{a}^{\mathcal{C}_{\mathrm{E}}}|_{\alpha}|A_{a'}^{\mathcal{C}'_{\mathrm{E}}}\rangle_{\alpha'}=\delta_{\mathcal{C}_{\mathrm{E}}\mathcal{C}'_{\mathrm{E}}}\delta_{aa'}\delta_{\alpha\alpha'}$. In the third equation, we simplify the formula using the delta functions $\mathcal{C}_{\mathrm{E}_0}=\mathcal{C}_{\mathrm{E}_1}=\cdots =\mathcal{C}_{\mathrm{E}_N}$, $\alpha_0=\alpha_1=\cdots =\alpha_N$, and eliminate $\{a_0, \tilde{a}_I\}$ by $\{a_I\}$. In the last equation, we used Eq.~\eqref{1over2Sigma}. Moreover, notice that $\Tr_{\ch_\mathrm{A}}\rho_{\mathrm{A}}=1$, we obtain the entanglement entropy 
\begin{eqnarray}
S(A)=-\frac{d}{dN}2^{-|\Sigma|(N-1)}|_{N=1}=|\Sigma|\log 2.
\end{eqnarray}
Since $|\Sigma|$ is the number of 2-simplices on $\Sigma$, which is proportional to the area of $\Sigma$, hence it is the area law term. Since there is no constant term, the topological entanglement entropy is trivial, reflecting the absence of topological order in this model. 

\subsubsection{EE for the Torus: general $(n,p)$}

We carry out the analogous calculations for a general GWW theory with arbitrary coefficients $n$ and $p$. We start by writing down the ground state wave function, 
\begin{equation}
\begin{split}
|\psi\rangle=&\mathfrak{C}\sum_{\mathcal{C}_\mathrm{E}}\sum_{a=1}^{N_{\mathrm{A}}(\mathcal{C}_\mathrm{E})/n}\sum_{b=1}^{N_{\mathrm{A}^{\mathrm{c}}}(\mathcal{C}_\mathrm{E})/n}\sum_{\alpha, \beta=0}^{n-1}\\
& 
e^{\frac{i 2\pi p\alpha\beta}{n}} l_{a,\alpha}^{\mathcal{C}_\mathrm{E}}l_{b,\beta}^{\mathcal{C}_\mathrm{E}}s_{a,\alpha}^{\mathcal{C}_\mathrm{E}}s_{b,\beta}^{\mathcal{C}_\mathrm{E}}|\tilde{\mathrm{A}}_a^{\mathcal{C}_\mathrm{E}}\rangle_{\alpha}|\tilde{\mathrm{A}^{\mathrm{c}}}_b^{\mathcal{C}_\mathrm{E}}\rangle_{\beta},
\end{split}
\end{equation}
where $l_{a,\alpha}^{\mathcal{C}_\mathrm{E}},l_{b,\beta}^{\mathcal{C}_\mathrm{E}},s_{a,\alpha}^{\mathcal{C}_\mathrm{E}},s_{b,\beta}^{\mathcal{C}_\mathrm{E}}$ are straightforward generalizations of Eq.~\eqref{llss} to the cases with arbitrary coefficients $p$ and $n$, c.f. Eq.~\eqref{WaveFunctionnp}. The reduced density matrix is
\begin{align}\label{Reduceddensitymatrixnp1}
\rho_\mathrm{A}=|\mathfrak{C}|^2 \sum_{\mathcal{C}_\mathrm{E}}\frac{N_{\mathrm{A}^{\mathrm{c}}}(\mathcal{C}_\mathrm{E})}{n}\sum_{a,\tilde{a}=1}^{N_{\mathrm{A}}(\mathcal{C}_\mathrm{E})/n}\sum_{\alpha,\tilde{\alpha},\gamma=0}^{n-1}\nonumber
\\
e^{\frac{i2\pi p (\alpha-\tilde{\alpha})\gamma}{n}}|\mathrm{A}_a^{\mathcal{C}_\mathrm{E}}\rangle_{\alpha}\langle \mathrm{A}_{\tilde{a}}^{\mathcal{C}_\mathrm{E}}|_{\tilde{\alpha}},
\end{align}
where we again performed the unitary transformations to absorb the self-linking and mutual-linking factors, and denote the resulting new basis as $|\mathrm{A}_a^{\mathcal{C}_\mathrm{E}}\rangle_{\alpha}$ and $|\mathrm{A}_b^{\mathrm{c}\mathcal{C}_\mathrm{E}}\rangle_{\beta}$.

For the same reason as in Eq.~\eqref{1over2Sigma}, 
\begin{equation}\label{AAA}
|\mathfrak{C}|^2N_{\mathrm{A}^{\mathrm{c}}}(\mathcal{C}_\mathrm{E})N_{\mathrm{A}}(\mathcal{C}_\mathrm{E})=\frac{1}{n^{|\Sigma|-1}},
\end{equation}
where $|\Sigma|$ is the number of 2-simplices on the entanglement surface. 

In order to compute the entanglement entropy 
\begin{equation}
S_{\mathrm{A}}=-\mathrm{Tr}_{\mathcal{H}_{\mathrm{A}}} \,\rho_\mathrm{A}\log \rho_\mathrm{A},
\label{eq: EE definition}
\end{equation}
we first calculate the entanglement spectrum, i.e., we diagonalize $\rho_\mathrm{A}$. As a first step, we carry out the sum over $\gamma$ in Eq.~\eqref{Reduceddensitymatrixnp1}. We note that the sum is nonvanishing only if $p (\alpha-\tilde{\alpha})/n$ is an integer, in which case the sum takes the value $n$. Thus, 
\begin{equation}\label{sumovergamma}
\sum_{\gamma=0}^{n-1}e^{\frac{i2\pi p (\alpha-\tilde{\alpha})\gamma}{n}}
=
n~\delta\bigg(\alpha-\tilde{\alpha}=0\mod \frac{n}{\gcd(n,p)}\bigg).
\end{equation}
We find
\begin{widetext}
	\begin{subequations}
		\begin{eqnarray}
		\rho_\mathrm{A}
		&=& |\mathfrak{C}|^2
		\sum_{\mathcal{C}_\mathrm{E}}N_{\mathrm{A}^{\mathrm{c}}}(\mathcal{C}_\mathrm{E})
		\sum_{a,\tilde{a}=1}^{N_{\mathrm{A}}(\mathcal{C}_\mathrm{E})/n}
		\sum_{\alpha,\tilde{\alpha}}^{n-1}\delta\bigg(\alpha-\tilde{\alpha}=0\mod \frac{n}{\gcd(n,p)}\bigg)|\mathrm{A}_a^{\mathcal{C}_\mathrm{E}}\rangle_{\alpha}\langle \mathrm{A}_{\tilde{a}}^{\mathcal{C}_\mathrm{E}}|_{\tilde{\alpha}}\label{eq:rhoAlaststep}\\
		&=&\sum_{\mathcal{C}_\mathrm{E},a,\alpha,\tilde{a},\tilde{\alpha}}\left[\rho^{\mathcal{C}_\mathrm{E}}_\mathrm{A}\right]_{a,\alpha;\tilde{a}\tilde{\alpha}}|\mathrm{A}_a^{\mathcal{C}_\mathrm{E}}\rangle_{\alpha}\langle \mathrm{A}_{\tilde{a}}^{\mathcal{C}_\mathrm{E}}|_{\tilde{\alpha}},
		\end{eqnarray}
		where $\left[\rho^{\mathcal{C}_\mathrm{E}}_\mathrm{A}\right]_{a,\alpha;\tilde{a}\tilde{\alpha}}$ are matrix elements given by
		\begin{equation}
		\left[\rho^{\mathcal{C}_\mathrm{E}}_\mathrm{A}\right]_{a,\alpha;\tilde{a}\tilde{\alpha}}=|\mathfrak{C}|^2N_{\mathrm{A}^{\mathrm{c}}}(\mathcal{C}_\mathrm{E})\left[\openone_{\frac{n}{\mathrm{gcd}(n,p)}}\otimes J_{\mathrm{gcd}(n,p)}\right]_{\alpha\tilde{\alpha}}\otimes \left[J_{\frac{N_A(\mathcal{C}_\mathrm{E})}{n}}\right]_{a\tilde{a}}.
		\end{equation}
		Here, $\openone_m$ is the $m\times m$ identity matrix, and $J_{l}$ is an $l\times l$ matrix of ones (which has one nonzero eigenvalue equal to $l$). The first term in this expression originates from the periodic delta function in Eq.~\eqref{eq:rhoAlaststep}, and the second term comes from the sum over $a,\tilde{a}$ in the outer product.
	\end{subequations}
	Noting that each $J_m$ is a rank one matrix with nonzero eigenvalue $m$, we see immediately that $\rho_A^{\mathcal{C}_\mathrm{E}}$ can be put in diagonal form
	\begin{equation}\label{rhoAgeneralpn}
	\rho_A^{\mathcal{C}_\mathrm{E}}=|\mathfrak{C}|^2N_{\mathrm{A}^{\mathrm{c}}}(\mathcal{C}_\mathrm{E})\frac{N_\mathrm{A}(\mathcal{C}_\mathrm{E})}{n}\mathrm{gcd}(n,p)(\openone_{\frac{n}{\mathrm{gcd}(n,p)}}\oplus\mathbf{0}_{N_{\mathrm{A}}(\mathcal{C}_\mathrm{E})-n/\mathrm{gcd}(n,p)}).
	\end{equation}
\end{widetext}
The matrix in Eq.~\eqref{rhoAgeneralpn} is
\begin{eqnarray}
\begin{tikzpicture}[decoration={brace,amplitude=5pt},baseline=(current bounding box.west)]
\matrix (magic) [matrix of math nodes,left delimiter=(,right delimiter=)] {
	1 \\
	& 1 \\
	& & \ddots \\
	& & & 1 \\
	& & & & 0 \\
	& & & & & 0 \\
	& & & & & & \ddots \\
	& & & & & & & 0\\
	& & & & & & & &0\\
	& & & & & & & & &0\\
};
\draw[decorate] (magic-1-1.north) -- (magic-4-4.north east) node[above=5pt,midway,sloped] {$\frac{n}{\gcd(n,p)}$ $1$'s};
\draw[decorate] (magic-5-5.north east) -- (magic-8-8.north east) node[above=5pt,midway,sloped] {$N_{\mathrm{A}}(\mathcal{C}_{\mathrm{E}})-\frac{n}{\gcd(n,p)}$  $0$'s};
\end{tikzpicture}
\end{eqnarray}

Finally, using Eq.~\eqref{AAA}, we find that the nonzero entanglement eigenvalues are given by
\begin{equation}
e^{-\xi_{\mathcal{C}_\mathrm{E},r}}=\frac{\mathrm{gcd}(n,p)}{n^{|\Sigma|}},
\end{equation}
where $r=1,\cdots,n^{|\Sigma|}/\mathrm{gcd}(n,p)$.
With this spectrum, it is straightforward to evaluate Eq.~\eqref{eq: EE definition} to obtain the entanglement entropy as
\begin{equation}
S(\mathrm{A})= |\Sigma|\log n-\log \gcd(n,p).
\end{equation}
The first term is proportional to the area of the entanglement surface. The second constant term is the TEE \footnote{The constant part of the EE is $S^{\mathrm{TQFT}}_{\mathrm{c}}(A)=-\log \gcd(n,p)$. According to the discussion in Sec.~\ref{GeneralStructure}, because the entanglement surface is $T^2$, whose Euler characteristic vanishes,  $S_{\mathrm{topo}}(A)\equiv S_{\mathrm{topo}}[T^2]=S^{\mathrm{TQFT}}_{\mathrm{c}}(A)=-\log \gcd(n,p)$.}:
\begin{equation}
S^{\mathrm{TQFT}}_{\mathrm{c}}(\mathrm{A})=S_{\mathrm{topo}}(\mathrm{A})=-\log \gcd(n,p).
\end{equation}
We see that the TEE depends nontrivially on the parameters $n$ and $p$. If $n$ and $p$ are coprime, i.e., $\gcd(n,p)=1$, the TEE vanishes. If $p=0$, using the definition $\gcd(n,0)=n$, the constant part of the EE reduces to $-\log n$. Alternatively, we can also compute the EE of the BF theory using the wave function  Eq.~\eqref{WavefunctionBFTheory}, and we find the constant part to be $-\log n$.

Note that this result is consistent with Refs.~\onlinecite{Gaiotto2015} and \onlinecite{2016arXiv161201418W} where the ground state degeneracy (GSD) on $T^3$ was computed to be $\gcd(n,p)^3$. The ground state degeneracy suggests that the GWW models can be topologically ordered, which, in our context, is reflected by the nonzero TEE, $-\log \gcd(n,p)$. When $\gcd(n,p)=1$, the ground state on $T^3$ is non-degenerate, and the TEE vanishes. In particular, for the case of the Walker-Wang model $n=2$, $p=1$, we obtain
\begin{equation}\label{EEofS2}
S(\mathrm{A})=|\Sigma|\log 2,
\end{equation}
and there is no topological order. We notice the relation between the GSD on $T^3$ and the TEE across the torus $T^2$,
\begin{eqnarray}
\exp(-3S_{\mathrm{topo}}[T^2])=\mathrm{GSD}[T^3], 
\end{eqnarray} 
which should be compared to the similar relation, $\exp(-2S_{\mathrm{topo}}[T^1])=\mathrm{GSD}[T^2]$, for the (2+1)D Abelian theories. 

For an Abelian theory in $(d+1)$D, our computation leads us to conjecture that
\begin{eqnarray}\label{conjecture}
\exp(-dS_{\mathrm{topo}}[T^{d-1}])=\mathrm{GSD}[T^d].
\end{eqnarray} 
For $(d+1)$D BF theory with level $n$, we have computed both the TEE and the $\mathrm{GSD}[T^d]$, and we found $S_{\mathrm{topo}}[T^{d-1}]=-\log n$ and $\mathrm{GSD}=n^d$. This is consistent with our conjecture. (See Appendix~\ref{AppConjectureCaseStudy} for details.) We conjecture that this relationship is true for more general theories such as Dijkgraaf-Witten models, and higher dimensional Chern-Simons theories as well. For a generic $(2+1)$ dimensional nonabelian Chern-Simons theory, Eq.~\eqref{conjecture} may not hold. For example, the TEE of the $SU(2)_3$ Chern-Simons theory is $S_{\mathrm{topo}}[T^1]=-\log (\sqrt{5}/(2\sin(\pi/5)))$\cite{bondersonnon}, and $\exp(-2S_{\mathrm{topo}}[T^{1}])$ is not an integer. Hence Eq.~\eqref{conjecture} can not hold because the GSD should be an integer. However, we note that for some nonabelian theories, the conjecture still holds. For example, for the bosonic Moore-Read quantum Hall state in $(2+1)$D, $\mathrm{GSD}[T^2]=4$ (which consists of 3 states from the even parity sector and 1 state from the odd parity sector), and $S_{\mathrm{topo}}[T^1]=-\log 2$, hence Eq.~\eqref{conjecture} holds in this case.

\subsubsection{EE for Arbitrary Genus}
\label{EEGWW3}

\begin{table*}[t]
	\begin{center}
		\begin{tabular}{ c | c | c | c | c  }
			\hline
			& & $S^2$ & $T^2$ & $[(0,n_0), \cdots, (g^*, n_{g^*})]$\\ \hline\hline
			\multirow{2}*{$\frac{n}{2\pi}BF$} & $S^{\mathrm{TQFT}}_{\mathrm{c}}$  & $-\log n$ & $-\log n$ & $-b_0\log n$\\\cline{2-5}
			& $S_{\mathrm{topo}}$	& $-\log n$ & $-\log n$ & $-b_0\log n$ \\\hline
			\multirow{2}*{$\frac{n}{2\pi}BF+\frac{np}{4\pi}BB$} &$S^{\mathrm{TQFT}}_{\mathrm{c}}$& $-\log n$ & $-\log \gcd(n,p)$& $(-b_0+\frac{\chi}{2})\log \gcd(n,p)-\frac{\chi}{2}\log n$ \\\cline{2-5}
			&$S_{\mathrm{topo}}$& $-\log \gcd(n,p)$ & $-\log \gcd(n,p)$ & $-b_0\log \gcd(n,p)$\\
			\hline
		\end{tabular}
		\caption{Constant part and topological part of the entanglement entropy for generalized Walker-Wang models. $S^{\mathrm{TQFT}}_{\mathrm{c}}$ is the constant part of the EE for the TQFT, while $S_{\mathrm{topo}}$ is the TEE for a general theory which belongs to the same phase of the TQFT. $b_0$ is the zeroth Betti number of entanglement surface $b_0=\sum_{g=0}^{g^*}n_g$. $\chi=\sum_{g=0}^{g^*}(2-2g)n_g$ is the Euler characteristic of the entanglement surface. In particular, we have $S_{\mathrm{topo}}(S^2)=S_{\mathrm{topo}}(T^2)$.}
		\label{Summary}
	\end{center}
\end{table*}

Following the same procedure used for the torus, we calculate the EE across a general entanglement surface with genus $g$. (The results are summarized in Table \ref{Summary}.) For each hole $i$ ($i=1,\cdots,g$) of the entanglement surface, we introduce a pair of additional indices $\alpha_i$ and $\beta_i$  that count the number of loops (modulo $n$) winding around the non-contractible cycles around the hole in region A and region $\mathrm{A}^\mathrm{c}$, respectively. Then the wavefunction is 
\begin{equation}\label{55}
\begin{split}
|\psi\rangle=&\mathfrak{C}\sum_{\mathcal{C}_{\mathrm{E}}}\sum_{a=1}^{\frac{N_{\mathrm{A}}(\mathcal{C}_{\mathrm{E}})}{n^g}}\sum_{b=1}^{\frac{N_{\mathrm{A}^{\mathrm{c}}}(\mathcal{C}_{\mathrm{E}})}{n^g}}\sum_{\alpha_1\cdots \alpha_g=0}^{n-1}\sum_{\beta_1\cdots \beta_g=0}^{n-1}\\
&\prod_{i=1}^g e^{\frac{i2\pi p \alpha_i \beta_i}{n}}|\mathrm{A}_a^{\mathcal{C}_{\mathrm{E}}}\rangle_{\boldsymbol{\alpha}}|\mathrm{A}_{b}^{\mathrm{c}\mathcal{C}_{\mathrm{E}}}\rangle_{\boldsymbol{\beta}}.\\~\\~\\
\end{split}
\end{equation}
We collect the set of indices  $\alpha_1,\cdots, \alpha_g$ into a index vector $\boldsymbol{\alpha}$. We first consider the configurations in region A. Since each hole is associated with an index $\alpha_i$, which can take $n$ different values, the complete set of indices $\boldsymbol{\alpha}$ can take $n^g$ different values. Hence, the $N_{\mathrm{A}}(\mathcal{C}_{\mathrm{E}})$ configurations are partitioned into $n^g$ classes, where each class contains $N_{\mathrm{A}}(\mathcal{C}_{\mathrm{E}})/n^g$ configurations. For this reason the summation in Eq.~\eqref{55} reaches only up to $N_{\mathrm{A}}(\mathcal{C}_{\mathrm{E}})/n^g$. For region $\mathrm{A}^\mathrm{c}$, similar arguments hold. Then the reduced density matrix on a genus $g$ surface takes the form
\begin{widetext}
\begin{equation}\label{Reduceddensitymatrixnp1-1}
\begin{split}
\rho_\mathrm{A}=&|\mathfrak{C}|^2 \sum_{\mathcal{C}_\mathrm{E}}\frac{N_{\mathrm{A}^{\mathrm{c}}}(\mathcal{C}_\mathrm{E})}{n^g}
\sum_{\alpha_1,\cdots, \alpha_g=0}^{n-1}
\sum_{\tilde{\alpha}_1,\cdots, \tilde{\alpha}_g=0}^{n-1}
\sum_{\gamma_1,\cdots, \gamma_g=0}^{n-1}
\sum_{a,\tilde{a}=1}^{N_{\mathrm{A}}(\mathcal{C}_\mathrm{E})/n^g}
\prod_{i=1}^g
e^{\frac{i2\pi p (\alpha_i-\tilde{\alpha}_i)\gamma_i}{n}}
|\mathrm{A}_a^{\mathcal{C}_\mathrm{E}}\rangle_{\boldsymbol{\alpha}}\langle \mathrm{A}_{\tilde{a}}^{\mathcal{C}_\mathrm{E}}|_{\tilde{\boldsymbol{\alpha}}}\\
=& |\mathfrak{C}|^2 \sum_{\mathcal{C}_\mathrm{E}}N_{\mathrm{A}^{\mathrm{c}}}(\mathcal{C}_\mathrm{E})\sum_{\alpha_1,\cdots, \alpha_g=0}^{n-1}
\sum_{\tilde{\alpha}_1,\cdots, \tilde{\alpha}_g=0}^{n-1}\sum_{a,\tilde{a}=1}^{N_{\mathrm{A}}(\mathcal{C}_\mathrm{E})/n^g}\prod_{i=1}^g \delta\bigg(\alpha_i-\tilde{\alpha}_i=0\mod \frac{n}{\gcd(n,p)} \bigg)|\mathrm{A}_a^{\mathcal{C}_\mathrm{E}}\rangle_{\boldsymbol{\alpha}}\langle \mathrm{A}_{\tilde{a}}^{\mathcal{C}_\mathrm{E}}|_{\tilde{\boldsymbol{\alpha}}}
\\
=& \sum_{\mathcal{C}_\mathrm{E}}\sum_{\boldsymbol{\alpha}, \tilde{\boldsymbol{\alpha}}}\sum_{a,\tilde{a}=1}^{N_{\mathrm{A}}(\mathcal{C}_\mathrm{E})/n^g}\bigg[\rho^{\mathcal{C}_\mathrm{E}}_{\mathrm{A}}\bigg]_{a\boldsymbol{\alpha},\tilde{a}\tilde{\boldsymbol{\alpha}}}|\mathrm{A}_a^{\mathcal{C}_\mathrm{E}}\rangle_{\boldsymbol{\alpha}}\langle \mathrm{A}_{\tilde{a}}^{\mathcal{C}_\mathrm{E}}|_{\tilde{\boldsymbol{\alpha}}},
\end{split}
\end{equation}
where
\begin{equation}\label{rhomatrix}
\begin{split}
\bigg[\rho^{\mathcal{C}_\mathrm{E}}_{\mathrm{A}}\bigg]_{a\boldsymbol{\alpha},\tilde{a}\tilde{\boldsymbol{\alpha}}}=& |\mathfrak{C}|^2N_{\mathrm{A}^{\mathrm{c}}}(\mathcal{C}_\mathrm{E})\bigotimes_{i=1}^{g}\bigg[\openone _{\frac{n}{\gcd(n,p)}}\otimes J_{\gcd(n,p)}\bigg]_{\alpha_i\tilde{\alpha}_i}\bigotimes\bigg[J_{\frac{N_{\mathrm{A}}(\mathcal{C}_{\mathrm{E}})}{n^g}}\bigg]_{a\tilde{a}}\\
=& |\mathfrak{C}|^2N_{\mathrm{A}^{\mathrm{c}}}(\mathcal{C}_\mathrm{E})\gcd(n,p)^g\frac{N_{\mathrm{A}}(\mathcal{C}_{\mathrm{E}})}{n^g}\bigg[\openone _{\frac{n^g}{\gcd(n,p)^g}}\oplus \mathbf{0}_{N_{\mathrm{A}}(\mathcal{C}_{\mathrm{E}})-\frac{n^g}{\gcd(n,p)^g}}\bigg]_{a\boldsymbol{\alpha},\tilde{a}\boldsymbol{\tilde{\alpha}}}\\
=& \frac{\gcd(n,p)^g}{n^{|\Sigma|+g-1}}\bigg[\openone _{\frac{n^g}{\gcd(n,p)^g}}\oplus \mathbf{0}_{N_{\mathrm{A}}(\mathcal{C}_{\mathrm{E}})-\frac{n^g}{\gcd(n,p)^g}}\bigg]_{a\boldsymbol{\alpha},\tilde{a}\boldsymbol{\tilde{\alpha}}}.
\end{split}
\end{equation}
\end{widetext}
In the second line of Eq.~\eqref{Reduceddensitymatrixnp1-1}, we summed over $\gamma_1, \cdots, \gamma_g$ using Eq.~\eqref{sumovergamma}. In the last line of Eq.~\eqref{Reduceddensitymatrixnp1-1} and the first line of Eq.~\eqref{rhomatrix}, we reorganized the coefficients $|\mathrm{A}_a^{\mathcal{C}_\mathrm{E}}\rangle_{\boldsymbol{\alpha}}\langle \mathrm{A}_{\tilde{a}}^{\mathcal{C}_\mathrm{E}}|_{\tilde{\boldsymbol{\alpha}}}$ into a matrix form, where  $\openone _{\frac{n}{\gcd(n,p)}}$ is the identity matrix due to the delta function, and $J_{\gcd(n,p)}$ is because all elements of  $\alpha=\frac{n}{\gcd(n,p)}k, \tilde{\alpha}=\frac{n}{\gcd(n,p)}\tilde{k}$ with $k, \tilde{k}=0, 1, \cdots, \gcd(n,p)-1$ are enumerated, and similar for  $J_{\frac{N_{\mathrm{A}}(\mathcal{C}_{\mathrm{E}})}{n^g}}$. In the second line of Eq.~\eqref{rhomatrix}, we expand the tensor product. In the last line, we use the normalization condition $|\mathfrak{C}|^2N_{\mathrm{A}^{\mathrm{c}}}(\mathcal{C}_\mathrm{E})N_{\mathrm{A}}(\mathcal{C}_{\mathrm{E}})=\frac{1}{n^{|\Sigma|-1}}$. 
We see that all of the non-zero eigenvalues of the entanglement spectrum are given by $1/N_{n,p,g;|\Sigma |}$, where
\begin{equation}
N_{n,p,g;|\Sigma |}\equiv
\frac{n^{|\Sigma |-\chi/2}}{\mathrm{gcd}(n,p)^g}, \qquad \chi=2-2g.
\end{equation}
$\chi$ is the Euler characteristic of $\Sigma$. Thus, the EE across a general surface of genus $g$ is:
\begin{equation}
\begin{split}
S[(0,0),&(1,0),\ldots,(g-1,0),(g,1)]	\\
=&|\Sigma|\log n -g\log \gcd(n,p)-(1-g)\log n\\
=&|\Sigma|\log n-\frac{\chi}{2}\log \frac{n}{\gcd(n,p)}-\log\gcd(n,p).\label{Finial}
\end{split}
\end{equation}
Equation \eqref{Finial} is consistent with Eq.~\eqref{TEEcentralresult}. We summarize $S_{\mathrm{topo}}(\mathrm{A})$ and $S^{\mathrm{TQFT}}_{\mathrm{c}}(\mathrm{A})$ for various systems and various entanglement surfaces in Table~\ref{Summary}.

We note that although Eq.~\eqref{Finial} is the EE for a low energy TQFT, there is still an area law term. Since the TQFT is independent of the metric of the entanglement surface, one may naively expect that the area law term should vanish. The reason that the area law term appears in Eq.~\eqref{Finial} is that we formulated our theory on a lattice, which explicitly broke the scaling symmetry (i.e., changing the area of the cut changes the number of links passing through $\Sigma$). However symmetry under area-preserving diffeomorphisms was unaffected by the lattice regularization (changing the shape of the cut does not change the number of links passing through $\Sigma$). Because of this, we get terms that scale like the area of the cut (area law term), but no further shape-dependent terms. Therefore, we expect, and indeed find, that the mean curvature term vanishes for the TQFT ($F_2'\to 0$).

\section{Summary and Future Directions}
\label{DiscussionSummary}

In this paper, we have analyzed the general structure of the EE for gapped phases of matter whose low energy physics is described by TQFTs in (3+1)D. The EE for gapped phases generally obeys the area law. The area law part of the EE is not universal, while the constant part of EE contains topological information. Hence we have focused on the constant part of the EE.

For TQFTs, our analysis relied on the SSA inequalities. We found that they are strong enough to constrain the possible expressions for the EE. One of our main results is Eq.~\eqref{TEEcentralresult}: the EE across a general surface can be reduced to a linear combination of $S^{\mathrm{TQFT}}_{\mathrm{c}}[S^2]$ and $S^{\mathrm{TQFT}}_{\mathrm{c}}[T^2]$. We have identified the topological and universal contribution to the entanglement entropy, i.e., the topological entanglement entropy (TEE). We also analyzed the behavior of various terms in the EE when the theory was deformed away from the fixed point, and argued that a generalization of the KPLW prescription allows to extract the TEE.  

We then provided independent calculations of the entanglement entropy for the GWW class of TQFTs. We determined the ground state wave functions of the GWW models, which allowed us to calculate the EE. The results confirm our more general analysis of the EE. We showed that twisting terms in the Lagrangian can in general change the topological entanglement entropy. We then  conjectured a relationship between the topological entanglement entropy and the ground state degeneracy of Abelian theories in $(d+1)$ dimensions. 

Since we have only considered gapped systems without global symmetry, one natural question for future work is whether one can use the entanglement of the ground state wavefunction to probe topological phases with global symmetries, such as SPT order and symmetry enriched topological order in higher dimensions. In particular, for systems with SPT order, there is no intrinsic topological order and the ground state wavefunction is only short range entangled, hence the TEE is trivial. However, it has been realized that the entanglement spectrum serves as a useful tool to probe SPT order. In Refs.~\onlinecite{2010PhRvB..81f4439P} and \onlinecite{turner2011topological}, the entanglement spectra of one dimensional spin and fermion systems were studied, where the nontrivial degeneracy of the spectra revealed nontrivial SPT order. In Refs.~\onlinecite{2011PhRvB..83x5132H} and \onlinecite{2013PhRvB..87c5119F}, the existence of in-gap states in the single body entanglement spectrum was proven to reveal the nontrivial topology of a topological band insulator. Furthermore, there are extensive theoretical and numerical studies on the entanglement spectrum of quantum Hall systems\cite{2008PhRvL.101a0504L,2009PhRvL.103a6801R,2011PhRvB..84t5136C,PhysRevB.85.115321,PhysRevB.86.245310,2012PhRvB..85l5308S,2014arXiv1406.6262E} and fractional Chern insulators\cite{2012PhRvB..85g5116W}. It would be beneficial to complement this with a more systematic investigation of the entanglement spectrum as a probe of SPT order in higher dimensions in the future. 

\section*{Acknowledgments}

We thank F. Burnell, M. Mezei, S. Pufu and S. Sondhi for useful comments. B. A. Bernevig wishes to thank Ecole Normale Superieure, UPMC Paris, and the Donostia International Physics Center for their generous sabbatical hosting during some of the stages of this work. 
BAB acknowledges support for the analytic work from NSF EAGER grant DMR – 1643312, ONR - N00014-14-1-0330,  NSF-MRSEC DMR-1420541.  The computational part of the Princeton work was performed under department of Energy de-sc0016239, Simons Investigator Award, the Packard Foundation, and the Schmidt Fund for Innovative Research.

\appendix
\counterwithin{figure}{section}

\section{Review of Entanglement Entropy and Spectrum}
\label{DefinitionOfEntanglementEntropy}

In this appendix, we review the definition of the entanglement entropy, and review the notation that we use in this work.

To define the entanglement entropy, we first partition the space into two parts, A, and its complement, B, via an entanglement surface $\Sigma$.\footnote{Because we are interested in (3+1)D systems, the entanglement surface $\Sigma$ is a two dimensional surface.} For a given pure quantum state $|\psi\rangle$, the wave function can be decomposed as
\begin{eqnarray}
	|\psi\rangle=\sum_{ab}W_{ab}|\mathrm{A}_a\rangle|\mathrm{A}^{\mathrm{c}}_b\rangle,
\end{eqnarray}
where $a$ labels normalized basis states of the Hilbert space $\ch_\mathrm{A}$ localized in region A and $b$ labels normalized basis states of the Hilbert space $\ch_{\mathrm{A}^{\mathrm{c}}}$ localized in region $\mathrm{A}^\mathrm{c}$. We perform a singular value decomposition (SVD) of the matrix $W$ as $W_{ab}= U_{ac}D_{cd}V^\dagger_{db}$ and define new bases $|\mathrm{A}'_c\rangle= U_{ac}|\mathrm{A}_a\rangle$ and $|\mathrm{A}^{\mathrm{c}'}_d\rangle= V_{db}^\dagger|\mathrm{A}^{\mathrm{c}}_b\rangle$. $D_{cd}$ is a diagonal matrix with positive entries, but not all the diagonal elements need be nonzero. The number of nonzero elements is the rank of $W$, and the nonzero ``singular values" are denoted as $e^{-\xi_{\lambda}/2}$.  $\xi_{\lambda}$ are termed the entanglement energies, and the whole set of entanglement energies is the entanglement spectrum $\{\xi_{\lambda}\}_{\lambda=1,\cdots,\mathrm{Rank}(W)}$. Zero singular values correspond to infinite entanglement energies. Thus,
\begin{eqnarray}\label{A2}
	|\psi\rangle=\sum_{\lambda=1}^{\mathrm{Rank}(W)}e^{-\xi_{\lambda}/2}|\mathrm{A}'_{\lambda}\rangle|\mathrm{A}^{\mathrm{c}'}_{\lambda}\rangle.
\end{eqnarray} 
To compute the entanglement entropy, we trace over the states in region $\mathrm{A}^{\mathrm{c}}$ to obtain a reduced density matrix of region A,
\begin{eqnarray}
	\rho_\mathrm{A}=\Tr_{\ch_{\mathrm{A}^{\mathrm{c}}}}|\psi\rangle\langle\psi|=\sum_{\lambda=1}^{\mathrm{Rank}(W)}e^{-\xi_{\lambda}}|\mathrm{A}'_{\lambda}\rangle\langle \mathrm{A}'_{\lambda}|.
\end{eqnarray}
The entanglement entropy is defined as the von Neumann entropy of the reduced density matrix $\rho_\mathrm{A}$ (see Refs.~\onlinecite{zeng2015quantum} and \onlinecite{nielsen2010quantum} for a review),
\begin{equation}
S(\mathrm{A})=-\Tr_{\ch_\mathrm{A}}\rho_\mathrm{A}\log \rho_\mathrm{A}=-\sum_{\lambda=1}^{\mathrm{Rank}(W)}e^{-\xi_{\lambda}}\log e^{-\xi_{\lambda}}.
\end{equation}
Heuristically, the entanglement entropy measures how much the degrees of freedom in the two regions A and $\mathrm{B}$ are correlated. 

In this paper, we denote the entanglement entropy of subregion A (whose boundary is $\Sigma$) as either $S(\mathrm{A})$ or $S[\Sigma]$, using either parentheses or square brackets to highlight the sub region or the entanglement surface, respectively. 

\section{Local Contributions to the Entanglement Entropy}
\label{Curvature}

In this appendix, we review the general properties of the entanglement entropy. Following the discussions in Ref.~\onlinecite{grover2011entanglement}, we provide some detailed and quantitative analyses on how the non-universal and shape dependent terms can enter into the constant part of the EE.

The simplest property of the EE is $S(\mathrm{A})=S(\mathrm{A}^{\mathrm{c}})$, which says the entropy computed for region A is equal to the entropy computed for its complement $\mathrm{A}^{\mathrm{c}}$. This is also true for the full entanglement spectrum, and follows directly from Eq.~\eqref{A2}.

We assume that in a gapped system with finite correlation length, the EE can be decomposed into a local part and a topological part,
\begin{equation}\label{SeperationofEEToLocalTopo}
S(\mathrm{A})=S_{\mathrm{local}}(\mathrm{A})+S_{\mathrm{topo}}(\mathrm{A}).
\end{equation}
The local part $S_{\mathrm{local}}(\mathrm{A})$ only depends on the local degrees of freedom near the entanglement surface, and therefore can be written in the form of an integral over local variables. Since the only local functions on $\Sigma$ are the metric $h_{\mu\nu}$, the extrinsic curvature (second fundamental form) $K_{\mu\nu}$, and the covariant derivatives of $K_{\mu\nu}$ (covariant derivatives of $h_{\mu\nu}$ are zero by definition), Refs.~\onlinecite{grover2011entanglement,Liu2013,Liu2014} argued that $S_{\mathrm{local}}$ should be expressible in
terms of local geometric quantities of the entanglement surface $\Sigma$, i.e.,
\begin{eqnarray}
S_{\mathrm{local}}(\mathrm{A})=\int_{\Sigma} d^2x\sqrt{h}F(K_{\mu\nu}, \nabla_{\rho}K_{\mu\nu},..., h_{\mu\nu}),
\end{eqnarray}
where $F$ is a local function of $K_{\mu\nu}$ and $h_{\mu\nu}$ and their covariant derivatives. \footnote{Suppose the submanifold is given by the embedding $\phi: \Sigma\to M$, concretely, $\phi: y^i\to x^{\mu}=(z^*, y^i)$ where $z^*$ is a fixed number specifying the position of hypersurface in the perpendicular direction of the embedded space. Let the metric in $M$ be $g_{\mu\nu}$, the induced metric therefore is $h_{ij}\equiv (\phi^* g)_{ij}=\frac{\partial x^{\mu}}{\partial y^i}\frac{\partial x^{\nu}}{\partial y^j} g_{\mu\nu}$. Let $n^{\mu}$ be the normal unit vector of the surface $\Sigma$, then the extrinsic curvature $K_{\mu\nu}$ of $\Sigma$ is $K_{\mu\nu}=\nabla_{\mu}n_{\nu}-n_{\mu}n_{\rho}\nabla^{\rho}n_{\nu}$.  See Appendix D of Ref.~\onlinecite{carroll2004spacetime} for more details.}

In contrast, the topological part of the EE, $S_{\mathrm{topo}}(\mathrm{A})$, is precisely the contribution that cannot be written as an integral of local variables near the entanglement surface. (In particular, the Euler characteristic term does not contribute to $S_{\mathrm{topo}}(\mathrm{A})$.)  $S_{\mathrm{topo}}(\mathrm{A})$ should be invariant under smooth deformations of the entanglement surface, and should also be invariant under smooth deformations of the Hamiltonian of the system (provided the gap does not close). Therefore, reminiscent of two-dimensional systems, $S_{\mathrm{topo}}(\mathrm{A})$ is expected to be the constant part of the EE. However, in three spatial dimensions, there are subtleties as we will explore below.

Before moving on, it is important for us to first specify for which systems the EE separates into a local and a topological part. Systems such as the toric code and its generalizations (e.g. Dijkgraaf Witten models), as well as the Walker-Wang models \cite{Walker2012} and their generalizations (e.g., the generalized Walker-Wang models which we study in Sec.~\ref{EntanglementEntropyOfTQFT}) satisfy this decomposition. There are some systems for which this decomposition is obviously not valid. For instance, the systems constructed by layer stacking of two-dimensional systems do not satisfy Eq.~\eqref{SeperationofEEToLocalTopo}. The constant part of entropy depends on the thickness $L_z$ of the layered direction, i.e., $-\gamma_{2\mathrm{D}}L_z$, where $\gamma_{2\mathrm{D}}$ is the topological entropy of a two-dimensional layer. Another class of systems beyond our discussion are fracton models\cite{2011PhRvA..83d2330H}, whose entanglement entropy does not satisfy Eq.~\eqref{SeperationofEEToLocalTopo}. Apart from the area law term and the constant term, the entanglement entropies of these model generically contain a term linearly proportional  to the size of the subregion \cite{2018PhRvB..97l5101M, 2018PhRvB..97l5102H}. Since the decomposition Eq.~\eqref{SeperationofEEToLocalTopo} does not lead to a linear subleading term, its presence in the layered models and the fracton models suggest the decomposition Eq.~\eqref{SeperationofEEToLocalTopo} does not hold.

Since the definition of the EE dictates that $S(\mathrm{A})=S(\mathrm{A}^{\mathrm{c}})$, this should also be true of the local part of the EE. To compute $S(\mathrm{A})$, one can expand $F(K_{\mu\nu}, \nabla_{\rho} K_{\mu\nu}, ..., h_{\mu\nu})$ as
\begin{eqnarray}\label{B3}
\begin{split}
&F(K_{\mu\nu}, \nabla_{\rho} K_{\mu\nu}, ...,h_{\mu\nu})\\
&=F_0+F_1K^{\mu}_{\mu} +F_2 [K_{\mu\nu}K^{\mu\nu}-(K^{\mu}_{\mu})^2]\\&~~~~+ F'_2(K^{\mu}_{\mu})^2+ F_{3}\nabla_{\mu}\nabla_{\nu}K^{\mu\nu}+ ...,
\end{split}
\end{eqnarray}
where $\nabla_{\mu}$ is the covariant derivative induced from $h_{\mu\nu}$, and the indices are raised and lowered via $h_{\mu\nu}$ and its inverse $h^{\mu\nu}$. All indices are contracted so that the formula Eq.~\eqref{B3} is independent of the choice of the coordinates. Demanding that $S(\mathrm{A})=S(\mathrm{A}^{\mathrm{c}})$ constrains the form of the function $F$. To see this, we may simply transform $x_1\to -x_1$ and $x_2\to x_2$, under which $ K_{\mu\nu}\to -K_{\mu\nu}$ and $h_{\mu\nu}\to h_{\mu\nu}$. \footnote{$x_1\to -x_1$ and $x_2\to x_2$ changes the orientation of the entanglement surface $\Sigma$. Since the principle curvature is an odd function of the orientation of the surface and the eigenvalues of the extrinsic curvature are two principle curvatures, we conclude that the extrinsic curvature is odd under $x_1\to -x_1$ and $x_2\to x_2$.} Then $S(\mathrm{A})=S(\mathrm{A}^{\mathrm{c}})$ implies
\begin{equation}
F(K_{\mu\nu}, \nabla_{\rho} K_{\mu\nu}, ..., h_{\mu\nu})=F(-K_{\mu\nu}, \nabla_{\rho}K_{\mu\nu}, ..., h_{\mu\nu}).
\end{equation}
 After integration, keeping only those terms even under reflection, we find that the local part of the EE has the form
\begin{equation}
S_{\mathrm{local}}(A)=F_0 |\Sigma|-F_2  4\pi \chi +4F'_2\int_{\Sigma} d^2x \sqrt{h}H^2+...,
\end{equation}
where $|\Sigma|$ is the area of the entanglement surface. The part proportional to $F_2$ gives the Euler characteristic $\chi(\Sigma)$ of the surface $\Sigma$, defined by $\int_{\Sigma}d^2x \sqrt{h}[K_{\mu\nu}K^{\mu\nu}-(K^{\mu}_{\mu})^2]=-4\pi \chi(\Sigma)$. This term is invariant under any smooth deformation of the entanglement surface because the Euler characteristic is a topological invariant of $\Sigma$. The part proportional to $F'_2$ gives the integral of the square of the mean curvature $H=(k_1+k_2)/2$ (since $2H=K_{\mu}^{\mu}$), where $k_1,k_2$ are the two principal curvatures of $\Sigma$, i.e., the eigenvalues of $K_{\mu\nu}$.  This term, though independent of the size of $\Sigma$, depends on its shape. This shows that the local part of the EE has constant terms, which contrasts with the familiar case in (2+1)D. Therefore, computing the EE and extracting the constant part is not a promising way to extract topological information about the underlying theory.\footnote{In $(2+1)$D, by applying the same analysis, one can show that there is no constant term in the EE which can be written as an integral of local curvature when the space dimension $d$ is even. This is because the term of dimension $1/L^{d-1}$ acquires a minus sign when the coordinates of entanglment surface are reversed. In particular, in (2+1)D, the constant term in entanglement entropy is topological.}

The above analysis shows that for a generic gapped system (which is not at an RG fixed point), the structure of the entanglement entropy is
\begin{eqnarray}
S(\mathrm{A})&=&F_0 |\Sigma| +S_{\mathrm{topo}}(A)-4\pi F_2\chi(\Sigma)\nonumber\\&&+4F'_2\int_{\Sigma} d^2x \sqrt{h}H^2+\mathcal{O}(1/|\Sigma|).
\end{eqnarray}
In the main text, we denote the constant part of the EE as $S_c(\mathrm{A})=S_{\mathrm{topo}}(\mathrm{A})-4\pi F_2\chi(\Sigma)+4F'_2\int_{\Sigma} d^2x \sqrt{h}H^2$. 

The above analysis gives all the possible terms that \emph{can} exist, but does not require that they are non-vanishing for a given theory. In Ref.~\onlinecite{Lewkowycz2013}, the authors computed the entanglement entropy for massive bosons and massive fermions in (3+1)D across $S^2$. Their results show a constant term in the entanglement entropy. For a massive scalar with mass $m$ and curvature coupling term $\frac{1}{2}\xi R \phi^2$, $S_c(\mathrm{A})=(\xi-\frac{1}{6})\log(m\delta)$, where $\delta$ is the cut off. For a massive Dirac fermion with mass $m$, $S_c(\mathrm{A})=\frac{1}{18}\log(m\delta)$. Obviously, these entropies are not topological (they depend on the cutoff and on mass parameters), which shows that non-universal contributions to the local term in fact do exist.

\section{Derivation of the Reduction Formula}\label{app:recurrencederiv}

	\begin{figure}[H]
		\centering
		\subfigure[]{
			\includegraphics[width=0.7\columnwidth]{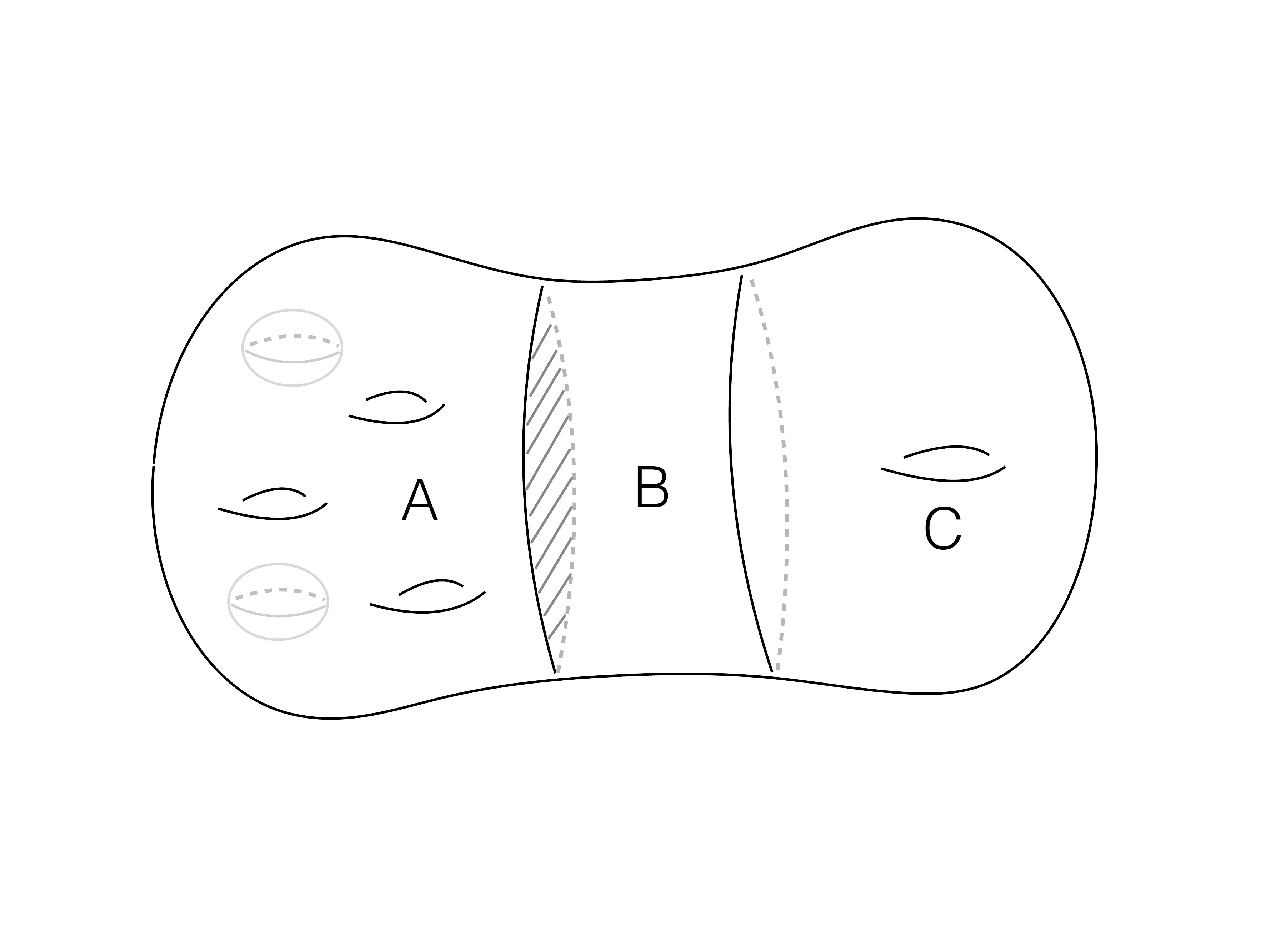}
			\label{Recurrence_Genus_1}}
		\subfigure[]{
			\includegraphics[width=0.7\columnwidth]{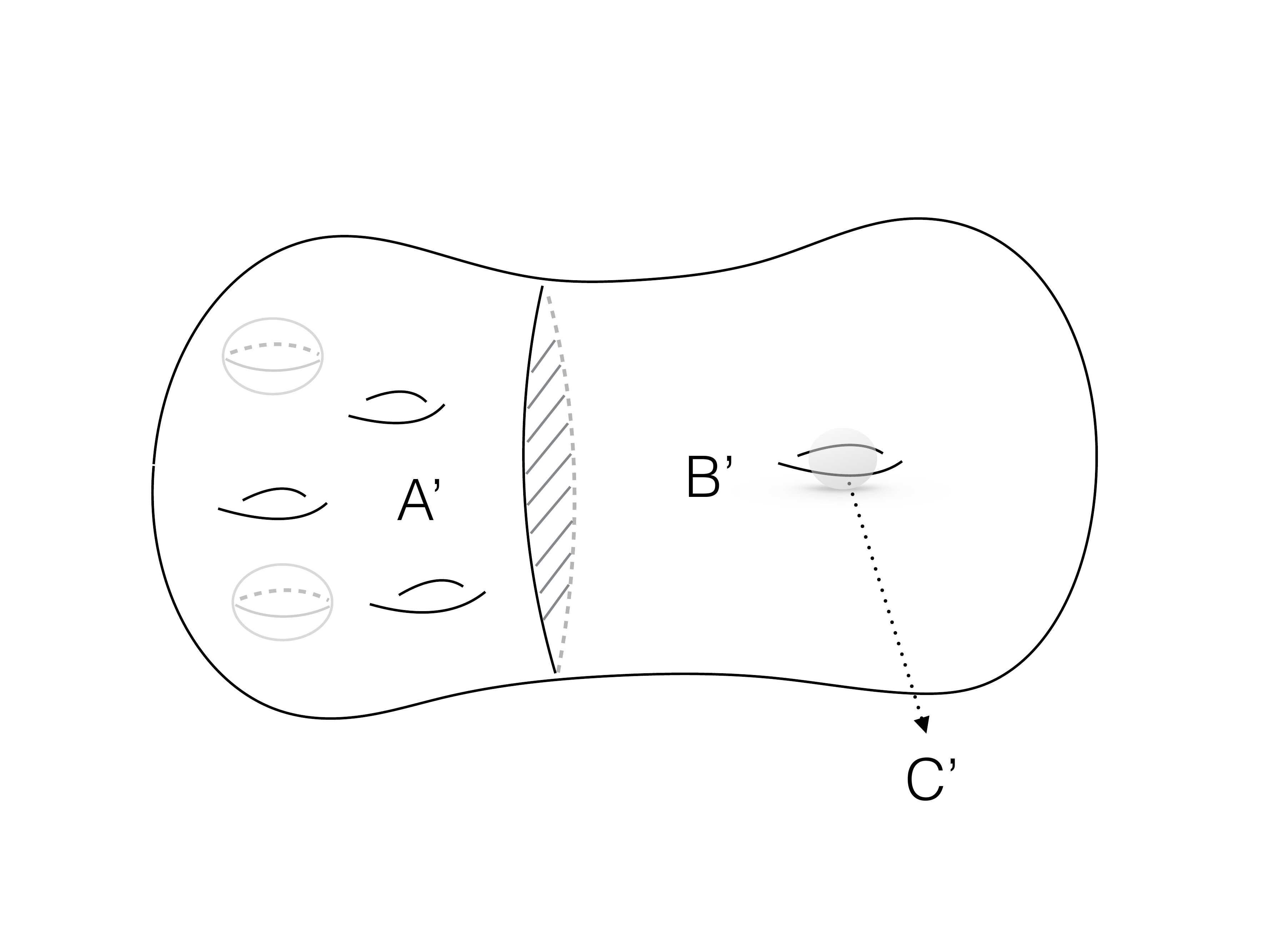}
			\label{Recurrence_Genus_2}}
		\caption{Entanglement surfaces used in the application of strong sub-additivity to derive the recurrence relation Eq.~\eqref{genusfinial}. In (a), A is a general 3-manifold (as an example, we draw A with 1 genus 3 surface and 2 genus 0 surfaces), B is 3-ball and C is a solid torus. In (b), $\mathrm{A}^\prime$ is a general 3-manifold (as an example, we draw $\mathrm{A}^\prime$ with 1 genus 3 surface and 2 genus 0 surfaces), $\mathrm{B}^\prime$ is a solid torus, and $\mathrm{C}^\prime$ is a 3-ball, which is located exactly at the hole of $\mathrm{B}^\prime$.}
	\end{figure}

In this appendix we present the complete derivation of the entropy reduction formula Eq.~\eqref{TEEcentralresult}. We will use the SSA inequality in two steps. First, in Subsection~\ref{Recurrence1} we derive and solve a recurrence relation for the dependence of $S^{\mathrm{TQFT}}_c$ on the genus of the entanglement cut. Second, in Subsection~\ref{Recurrence2} we derive an additional recurrence relation for the dependence of $S^{\mathrm{TQFT}}_c$ on the number of disconnected components of the entanglement surface. We solve this recurrence relation to obtain our main result Eq.~\eqref{TEEcentralresult}. Our derivation expands upon the discussion in Ref.~\onlinecite{grover2011entanglement} in that we obtain explicit formulas for the entropy of arbitrary multiply-connected entanglement surfaces.

\subsection{Recurrence for Genus}
\label{Recurrence1}

In order to find the dependence of the TEE on the data $\{n_g\}$, we need to consider the configuration of entanglement surfaces as shown in Fig.~C.\ref{Recurrence_Genus_1}: We start with a general connected 3-manifold with boundary specified by $[(0,n_0),\ldots,(g^*,n_{g^*})]$. The 3-manifold is cut into three regions A, B and C.  B is a 3-ball, C is a solid torus and A occupies the remainder of the manifold. A is connected to B and disconnected from C. Suppose A connects with B via a disk (shown as a shaded region) which belongs to a genus $(g^*-1)$\footnote{Since $\mathrm{C}\cup\mathrm{B}$ has a genus 1 surface boundary.} boundary of A and also belongs to the genus 0 boundary of B. Then the boundary of region A is specified by $[(0,n_0),\ldots,(g^*-1,n_{g^*-1}+1),(g^*,n_{g^*}-1)]$, where we adopt the labeling scheme defined in Sec.~\ref{Structure}.

\begin{figure*}[t]
	\subfigure[]{\includegraphics[width=0.7\columnwidth]{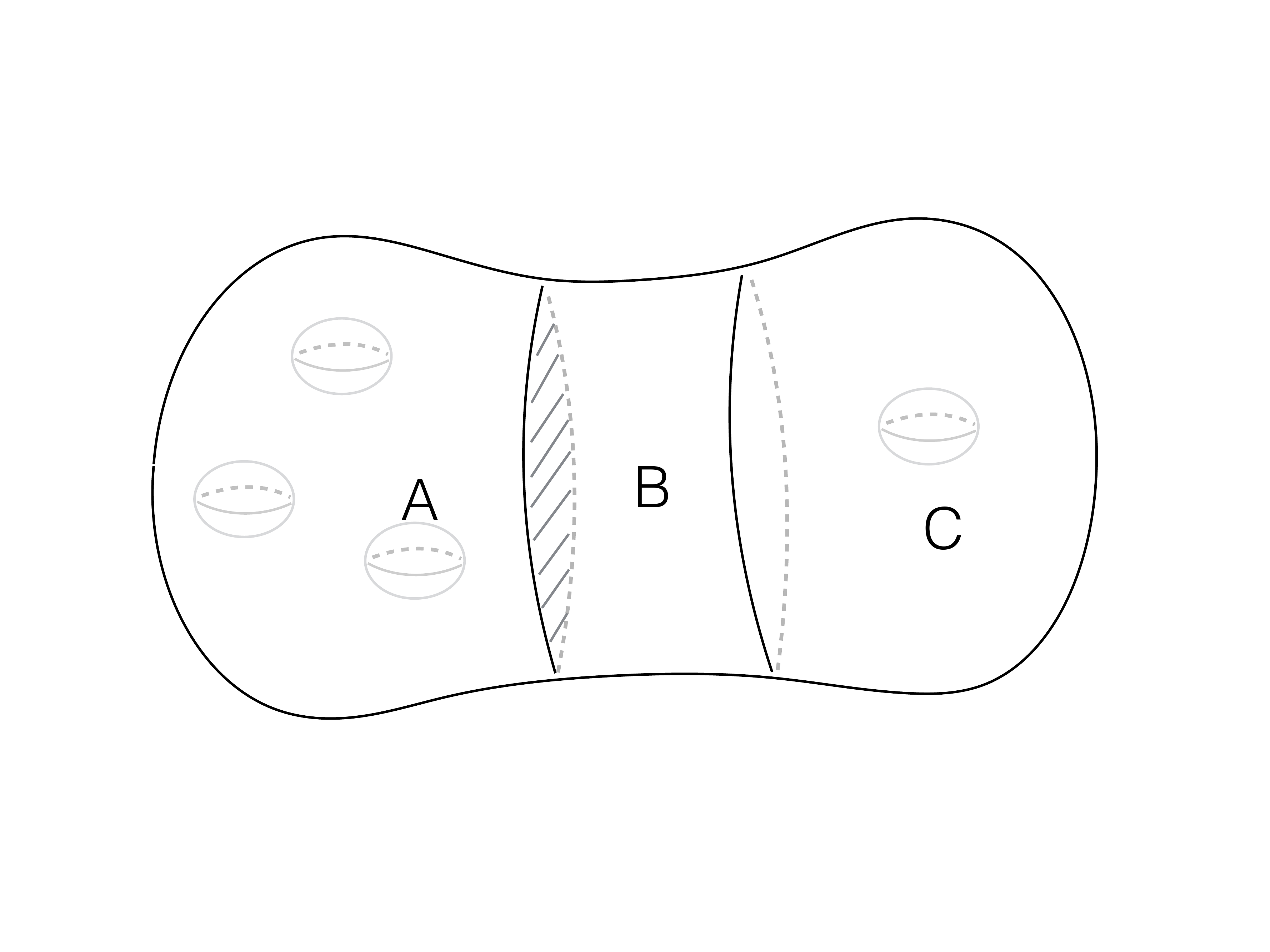}
		\label{Recurrence_b0_1}}
	\subfigure[]{
		\includegraphics[width=0.7\columnwidth]{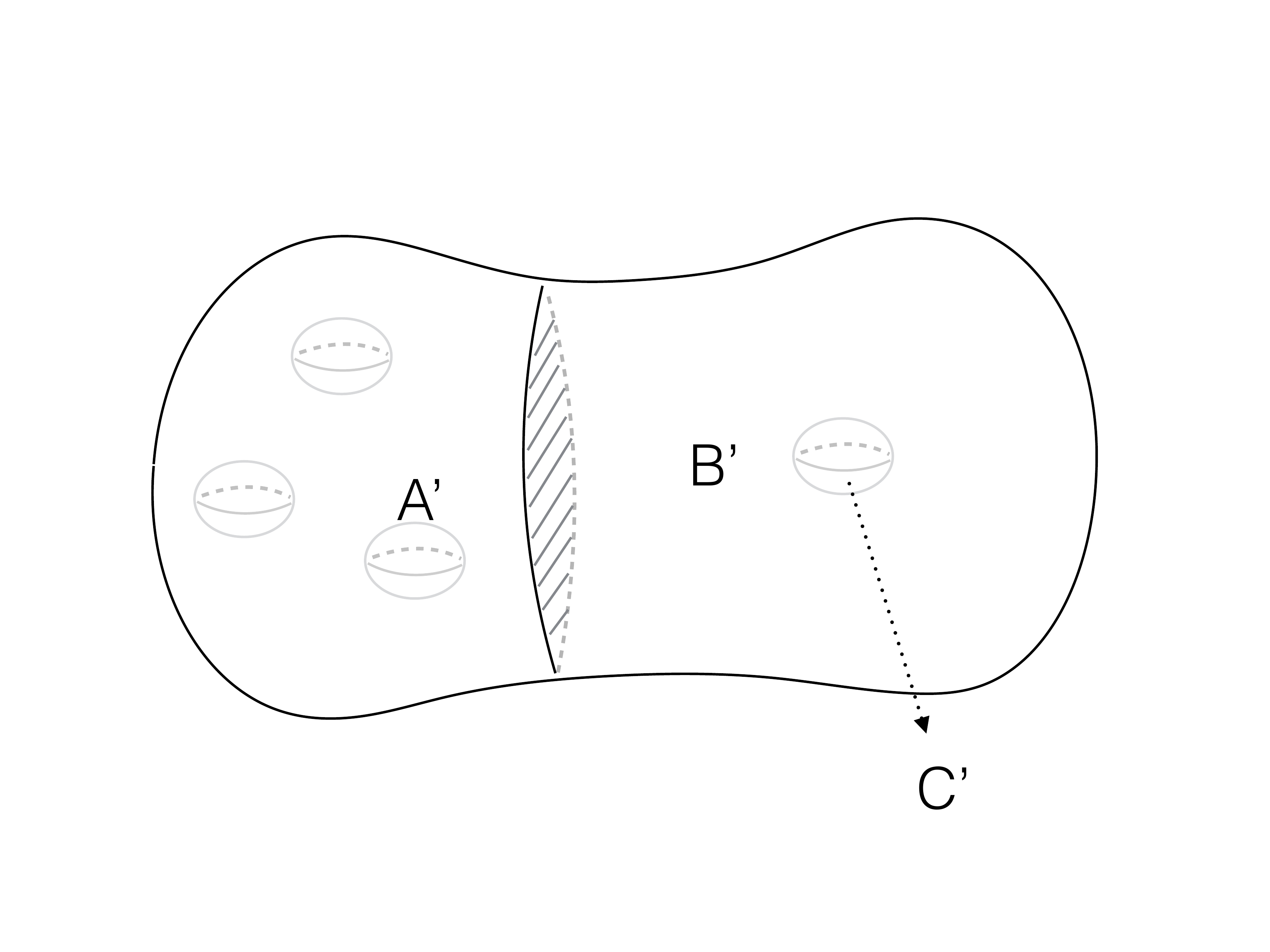}
		\label{Recurrence_b0_2}}
	\caption{Entanglement surfaces used in the application of strong sub-additivity to derive Eq.~\eqref{b0smaller2}. In (a), A is a 3-manifold with multiple genus zero surfaces, B is a 3-ball, C is a 3-ball with small 3-ball removed. In (b), $\mathrm{A}^\prime$ is an open 3-manifold with multiple genus zero surfaces, $\mathrm{B}^\prime$ is a 3-ball with a small 3-ball removed and $\mathrm{C}^\prime$ is a 3-ball located exactly in the empty 3-ball inside $\mathrm{B}^\prime$.}
\end{figure*}

\begin{widetext}
	We list the constant part of the EE of all regions by their topologies as follows:
	\begin{equation}
	\begin{split}
	S^{\mathrm{TQFT}}_{\mathrm{c}}(\mathrm{A})&=S^{\mathrm{TQFT}}_{\mathrm{c}}[(0,n_0),\ldots,(g^*-1,n_{g^*-1}+1),(g^*,n_{g^*}-1)],	\\
	S^{\mathrm{TQFT}}_{\mathrm{c}}(\mathrm{B})&=S^{\mathrm{TQFT}}_{\mathrm{c}}[(0,1)],	\\
	S^{\mathrm{TQFT}}_{\mathrm{c}}(\mathrm{C})&=S^{\mathrm{TQFT}}_{\mathrm{c}}[(0,0),(1,1)],	\\
	S^{\mathrm{TQFT}}_{\mathrm{c}}(\mathrm{AB})&=S^{\mathrm{TQFT}}_{\mathrm{c}}[(0,n_0),\ldots,(g^*-1,n_{g^*-1}+1),(g^*,n_{g^*}-1)],	\\
	S^{\mathrm{TQFT}}_{\mathrm{c}}(\mathrm{BC})&=S^{\mathrm{TQFT}}_{\mathrm{c}}[(0,0),(1,1)],	\\
	S^{\mathrm{TQFT}}_{\mathrm{c}}(\mathrm{ABC})&=S^{\mathrm{TQFT}}_{\mathrm{c}}[(0,n_0),\ldots,(g^*-1,n_{g^*-1}),(g^*,n_{g^*})]	.
	\end{split}
	\end{equation}
	Then the SSA inequality for regions A, B, and C in Eq.~\eqref{SSA} reads
	\begin{equation}\label{genuslarge}
	\begin{split}
	S^{\mathrm{TQFT}}_{\mathrm{c}}&[(0,n_0),\ldots,(g^*-1,n_{g^*-1}+1),(g^*,n_{g^*}-1)]	\\&\ge 
	S^{\mathrm{TQFT}}_{\mathrm{c}}[(0,n_0),\ldots,(g^*,n_{g^*})]+S^{\mathrm{TQFT}}_{\mathrm{c}}[(0,1)]-S^{\mathrm{TQFT}}_{\mathrm{c}}[(0,0),(1,1)].
	\end{split}
	\end{equation}
	We could have taken A and B to be connected via a disk which belongs to a genus $i$ $(i\le g^*-1)$ boundary of A and also belongs to the genus 0 boundary of B. Following an identical procedure, we conclude:
	\begin{equation}
	\begin{split}
	&S^{\mathrm{TQFT}}_{\mathrm{c}}[(0,n_0),\ldots,(i,n_i+1),(i+1,n_{i+1}-1),\ldots,(g^*,n_{g^*})]+S^{\mathrm{TQFT}}_{\mathrm{c}}[(0,0),(1,1)]	\\
	&\qquad\ge S^{\mathrm{TQFT}}_{\mathrm{c}}[(0,n_0),\ldots,(i,n_i),(i+1, n_{i+1}),\ldots,(g^*, n_{g^*})]+S^{\mathrm{TQFT}}_{\mathrm{c}}[(0,1)].
	\end{split}
	\end{equation}
	For simplicity, we will only need to adopt the choice where $i=g^*-1$.

	We proceed to consider another configuration illustrated in Fig.~C.\ref{Recurrence_Genus_2}: We start with a general 3-manifold with boundary specified by $[(0,n_0),\ldots,(g^*,n_{g^*})]$. The 3-manifold is cut into two regions, $\mathrm{A}^\prime$ and $\mathrm{B}^\prime$. $\mathrm{B}^\prime$ is a solid torus, and $\mathrm{A}^\prime$ is the rest of the manifold. We assume $\mathrm{A}^\prime$ connects with $\mathrm{B}^\prime$ via a disk (shown as a shaded region) in the genus $(g^*-1)$ boundary of $\mathrm{A}^\prime$ and the genus 1 boundary of $\mathrm{B}^\prime$. Hence the boundary of $\mathrm{A}^\prime$ is labeled by $[(0,n_0),\ldots,(g^*-1,n_{g^*-1}+1),(g^*,n_{g^*}-1)]$. In addition, we denote the 3-ball located in the ``hole" of $\mathrm{B}^\prime$ as  $\mathrm{C}^\prime$. 
	
	We list the constant part of the EE of all regions as follows:
	\begin{equation}
	\begin{split}
	S^{\mathrm{TQFT}}_{\mathrm{c}}(\mathrm{A}^\prime)=&S^{\mathrm{TQFT}}_{\mathrm{c}}[(0,n_0),\ldots,(g^*-1,n_{g^*-1}+1),(g^*,n_{g^*}-1)],\\
	S^{\mathrm{TQFT}}_{\mathrm{c}}(\mathrm{B}^\prime)=&S^{\mathrm{TQFT}}_{\mathrm{c}}[(0,0),(1,1)],\\
	S^{\mathrm{TQFT}}_{\mathrm{c}}(\mathrm{C}^\prime)=&S^{\mathrm{TQFT}}_{\mathrm{c}}[(0,1)],\\
	S^{\mathrm{TQFT}}_{\mathrm{c}}(\mathrm{A}^\prime \mathrm{B}^\prime)=&S^{\mathrm{TQFT}}_{\mathrm{c}}[(0,n_0),\ldots,(g^*,n_{g^*})],\\
	S^{\mathrm{TQFT}}_{\mathrm{c}}(\mathrm{B}^\prime \mathrm{C}^\prime)=&S^{\mathrm{TQFT}}_{\mathrm{c}}[(0,1)],\\
	S^{\mathrm{TQFT}}_{\mathrm{c}}(\mathrm{A}^\prime \mathrm{B}^\prime \mathrm{C}^\prime)=&S^{\mathrm{TQFT}}_{\mathrm{c}}[(0,n_0),\ldots,(g^*-1,n_{g^*-1}+1),(g^*,n_{g^*}-1)].
	\end{split}
	\end{equation}
	The SSA for $\mathrm{A}^\prime$, $\mathrm{B}^\prime$ and $\mathrm{C}^\prime$ in Fig.~C.\ref{Recurrence_Genus_2} reads in this case:
	\begin{equation}\label{genussmall}
	\begin{split}
	S^{\mathrm{TQFT}}_{\mathrm{c}}[(0,n_0)&,\ldots,(g^*-1,n_{g^*-1}+1),(g^*,n_{g^*}-1)] \\&\le S^{\mathrm{TQFT}}_{\mathrm{c}}[(0,n_0),\ldots,(g^*,n_{g^*})]+S^{\mathrm{TQFT}}_{\mathrm{c}}[(0,1)]-S^{\mathrm{TQFT}}_{\mathrm{c}}[(0,0),(1,1)] .
	\end{split}
	\end{equation}
	
	Combining inequalities Eq.~\eqref{genuslarge} and Eq.~\eqref{genussmall}, we find the following equality
	\begin{equation}\label{genusequal2}
	\begin{split}
	&S^{\mathrm{TQFT}}_{\mathrm{c}}[(0,n_0),\ldots,(g^*-1,n_{g^*-1}+1),(g^*,n_{g^*}-1)]
	\\&=S^{\mathrm{TQFT}}_{\mathrm{c}}[(0,n_0),\ldots,(g^*,n_{g^*})]+S^{\mathrm{TQFT}}_{\mathrm{c}}[(0,1)] - S^{\mathrm{TQFT}}_{\mathrm{c}}[(0,0),(1,1)].
	\end{split}
	\end{equation}
	This relates the constant part of the EE of a given subsystem to that of a system whose boundary has lower genus. Applying Eq.~\eqref{genusequal2} repeatedly, we find
	\begin{equation}\label{genusfinial}
	\begin{split}
	&S^{\mathrm{TQFT}}_{\mathrm{c}}[(0,n_0),(1,n_1),...,(g^*,n_{g^*})]
	=S^{\mathrm{TQFT}}_{\mathrm{c}}[(0,\sum_{i=0}^{g^*}n_i)]+\sum_{i=1}^{g^*}in_i\Big(S^{\mathrm{TQFT}}_{\mathrm{c}}[(0,0),(1,1)]-S^{\mathrm{TQFT}}_{\mathrm{c}}[(0,1)]\Big).
	\end{split}
	\end{equation}
	
	In summary, we can reduce the constant part of the EE of an arbitrary surface $S^{\mathrm{TQFT}}_{\mathrm{c}}[(0,n_0),(1,n_1),...,(g^*,n_{g^*})]$ to a linear combination of $S^{\mathrm{TQFT}}_{\mathrm{c}}[(0,n)]$ and $S^{\mathrm{TQFT}}_{\mathrm{c}}[(0,0),(1,1)]$.

\subsection{Recurrence for $b_0$}
\label{Recurrence2}

We can further simplify $S^{\mathrm{TQFT}}_{\mathrm{c}}[(0,\sum_{i=0}^{g^*}n_i)]$ in Eq.~\eqref{genusfinial}, by using $S^{\mathrm{TQFT}}_{\mathrm{c}}[(0,n)] = n S^{\mathrm{TQFT}}_{\mathrm{c}}[(0,1)]$. Here we derive this relation by making use of the SSA in a manner similar to that of the derivation above.

We consider the configuration shown in Fig.~C.\ref{Recurrence_b0_1}, where $\mathrm{A}$ is a 3-manifold with $(n-1)$ genus zero surfaces, $\mathrm{B}$ is a 3-ball and $\mathrm{C}$ is a 3-ball with a small 3-ball inside it removed. The constant parts of the EE for these three manifolds are
\begin{equation}
\begin{split}
S^{\mathrm{TQFT}}_{\mathrm{c}}(A)=&S^{\mathrm{TQFT}}_{\mathrm{c}}[(0, n-1)],	\\
S^{\mathrm{TQFT}}_{\mathrm{c}}(B)=&S^{\mathrm{TQFT}}_{\mathrm{c}}[(0, 1)],	\\
S^{\mathrm{TQFT}}_{\mathrm{c}}(C)=&S^{\mathrm{TQFT}}_{\mathrm{c}}[(0, 2)],	\\
S^{\mathrm{TQFT}}_{\mathrm{c}}(AB)=&S^{\mathrm{TQFT}}_{\mathrm{c}}[(0, n-1)],	\\
S^{\mathrm{TQFT}}_{\mathrm{c}}(BC)=&S^{\mathrm{TQFT}}_{\mathrm{c}}[(0, 2)],	\\
S^{\mathrm{TQFT}}_{\mathrm{c}}(ABC)=&S^{\mathrm{TQFT}}_{\mathrm{c}}[(0, n)].
\end{split}
\end{equation}
The SSA inequality reads
\begin{equation}\label{b0larger}
\begin{split}
&S^{\mathrm{TQFT}}_{\mathrm{c}}[(0, n-1)]+S^{\mathrm{TQFT}}_{\mathrm{c}}[(0, 2)]\ge S^{\mathrm{TQFT}}_{\mathrm{c}}[(0, n)]+S^{\mathrm{TQFT}}_{\mathrm{c}}[(0, 1)].
\end{split}
\end{equation}

We can furthermore consider another configuration shown in Fig.~C.\ref{Recurrence_b0_2}, where $A'$ is a 3-manifold with $(n-1)$ genus-0 surfaces, $\mathrm{B}^\prime$ is a 3-ball with small 3-ball removed, and $\mathrm{C}^\prime$ is a 3-ball locating exactly in the empty 3-ball inside $\mathrm{B}^\prime$. The constant parts of the EE for these three manifolds are
\begin{equation}
\begin{split}
S^{\mathrm{TQFT}}_{\mathrm{c}}(\mathrm{A}^\prime)&=S^{\mathrm{TQFT}}_{\mathrm{c}}[(0, n-1)],	\\
S^{\mathrm{TQFT}}_{\mathrm{c}}(\mathrm{B}^\prime)&=S^{\mathrm{TQFT}}_{\mathrm{c}}[(0, 2)],	\\
S^{\mathrm{TQFT}}_{\mathrm{c}}(\mathrm{C}^\prime)&=S^{\mathrm{TQFT}}_{\mathrm{c}}[(0, 1)],	\\
S^{\mathrm{TQFT}}_{\mathrm{c}}(\mathrm{A}^\prime \mathrm{B}^\prime)&=S^{\mathrm{TQFT}}_{\mathrm{c}}[(0, n)],	\\
S^{\mathrm{TQFT}}_{\mathrm{c}}(\mathrm{B}^\prime \mathrm{C}^\prime)&=S^{\mathrm{TQFT}}_{\mathrm{c}}[(0, 1)],	\\
S^{\mathrm{TQFT}}_{\mathrm{c}}(\mathrm{A}^\prime \mathrm{B}^\prime \mathrm{C}^\prime)&=S^{\mathrm{TQFT}}_{\mathrm{c}}[(0, n-1)].
\end{split}
\end{equation}
Then SSA inequality reads
\begin{equation}\label{b0smaller1}
\begin{split}
&S^{\mathrm{TQFT}}_{\mathrm{c}}[(0, n)]+S^{\mathrm{TQFT}}_{\mathrm{c}}[(0, 2)]\le S^{\mathrm{TQFT}}_{\mathrm{c}}[(0, n+1)]+S^{\mathrm{TQFT}}_{\mathrm{c}}[(0, 1)].
\end{split}
\end{equation}
Combining Eq.~\eqref{b0larger} and Eq.~\eqref{b0smaller1}, one obtains
\begin{equation}\label{b0smaller2}
\begin{split}
&S^{\mathrm{TQFT}}_{\mathrm{c}}[(0, n)]+S^{\mathrm{TQFT}}_{\mathrm{c}}[(0, 2)] = S^{\mathrm{TQFT}}_{\mathrm{c}}[(0, n+1)]+S^{\mathrm{TQFT}}_{\mathrm{c}}[(0, 1)].
\end{split}
\end{equation}
Since $S^{\mathrm{TQFT}}_{\mathrm{c}}[(0,0)]=0$, we have
\begin{eqnarray}
S^{\mathrm{TQFT}}_{\mathrm{c}}[(0, n)]=nS^{\mathrm{TQFT}}_{\mathrm{c}}[(0,1)].
\end{eqnarray}
Combining this result with Eq.~\eqref{genusfinial}, we have\footnote{As remarked in Sec.~\ref{DefinitionOfEntanglementEntropy}, we use $S(\mathrm{A})$ to denote the EE of region A, and $S[\Sigma]$ to denote the EE of region with boundary $\Sigma$, such as $S[S^2]$ when entanglement surface is $\Sigma=S^2$. Both notations refer to the same thing.}
\begin{equation}\label{TEEcentralresultapp}
\begin{split}
S^{\mathrm{TQFT}}_{\mathrm{c}}[(0,n_0),\ldots,(g^*,n_{g^*})]	&=\sum_{i=0}^{g^*}n_iS^{\mathrm{TQFT}}_{\mathrm{c}}[(0,1)]+\sum_{i=1}^{g^*}in_i\Big(S^{\mathrm{TQFT}}_{\mathrm{c}}[(0,0),(1,1)]-S^{\mathrm{TQFT}}_{\mathrm{c}}[(0,1)]\Big)\\
&=\sum_{i=0}^{g^*}(1-i)n_iS^{\mathrm{TQFT}}_{\mathrm{c}}[(0,1)]+\sum_{i=1}^{g^*}in_iS^{\mathrm{TQFT}}_{\mathrm{c}}[(0,0),(1,1)]\\\
&=b_0 S^{\mathrm{TQFT}}_{\mathrm{c}}[(0,0),(1,1)]+\frac{\chi}{2}\Big(S^{\mathrm{TQFT}}_{\mathrm{c}}[(0,1)]-S^{\mathrm{TQFT}}_{\mathrm{c}}[(0,0),(1,1)]\Big)\\
&=b_0 S^{\mathrm{TQFT}}_{\mathrm{c}}[T^2]+\frac{\chi}{2}\Big(S^{\mathrm{TQFT}}_{\mathrm{c}}[S^2]-S^{\mathrm{TQFT}}_{\mathrm{c}}[T^2]\Big),
\end{split}
\end{equation}

where $\chi=\sum_{i=0}^{g^*}(2-2i)n_i$ is the Euler characteristic of the entanglement surface, which in the previous examples of this appendix is $\partial (\mathrm{ABC})$. This is precisely Eq.~\eqref{TEEcentralresult} in the main text. In the last line, we have changed the notation for clarity: $S^2$ is a 2-sphere and $T^2$ is a 2-torus. We emphasize that Eq.~\eqref{TEEcentralresult} gives the constant part of the EE for a TQFT. In particular, Eq.~\eqref{TEEcentralresult} shows that the constant part of the EE across an arbitrary entanglement surface is reduced to that across the sphere $S^2$ and that across the torus $T^2$. \footnote{Notice that $S^{\mathrm{TQFT}}_{\mathrm{c}}(A)$ is an additive variable, i.e., $S^{\mathrm{TQFT}}_{\mathrm{c}}(A\cup A')=S^{\mathrm{TQFT}}_{\mathrm{c}}(A)+S^{\mathrm{TQFT}}_{\mathrm{c}}(A')$ if $A\cap A'=\emptyset$. This fact also follows from the vanishing of mutual information, i.e., $I(A\cup A')=S(A)+S(A')-S(A\cup A')=0$ if $A\cap A'=\emptyset$. This is because the area part cancels out in $I(A\cup A')$, and $I(A\cup A')=0$ yields exactly the additivity of the constant part of the entanglement entropy for a TQFT $S^{\mathrm{TQFT}}_{\mathrm{c}}(A)$. }\\~\\~\\~\\

\end{widetext}

\section{Vanishing of the Mean Curvature Contribution in KPLW Prescription}
\label{APPKPLWH2}

In this appendix, we explain why the mean curvature terms cancel in the KPLW combination Eq.~\eqref{KPLWT2}, therefore justifying Eq.~\eqref{KPLWH22} in the main text.

\begin{figure}[H]
	\includegraphics[width=0.5\textwidth]{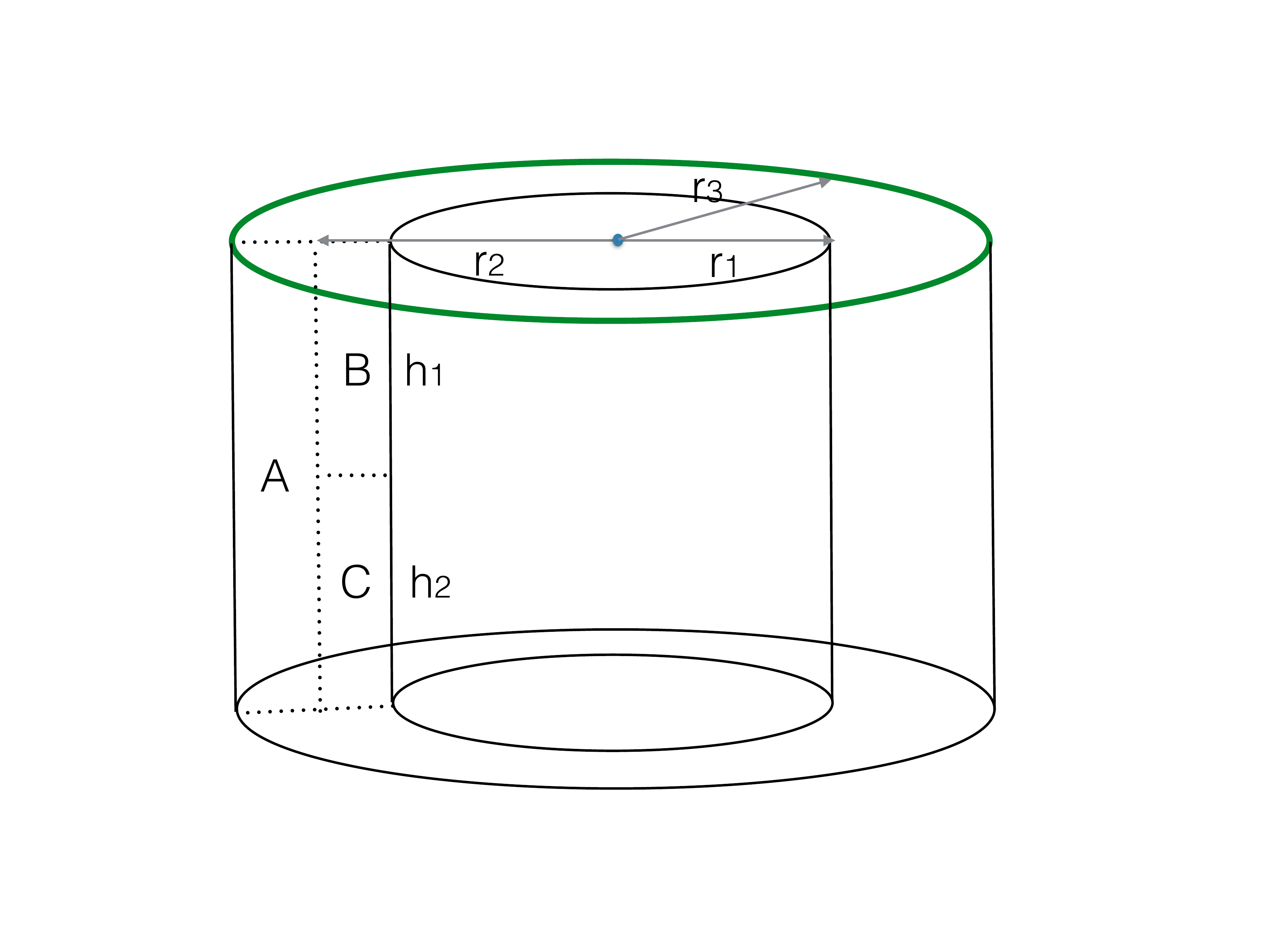}
	\centering
	\caption{KPLW prescription of regularized entanglement surface $T^2$.}
	\label{KPT2regularized}
\end{figure}

In the main text, we argued that the KPLW combination of the area law term and the Euler characteristic term vanish separately, hence we only need to consider the topological term and the mean curvature term, i.e.,
\begin{equation}
	S_{\mathrm{KPLW}}[T^2]=S_{\mathrm{topo}}[T^2]+4F'_2\int_{\substack{\partial \mathrm{A}+\partial \mathrm{B}+\partial \mathrm{C}\\-\partial \mathrm{AB}-\partial \mathrm{AC}\\-\partial \mathrm{BC}+\partial \mathrm{ABC}}}d^2x \sqrt{h}H^2.\label{AppKPLWH22}
\end{equation}
Eq.~\eqref{AppKPLWH22} suggests that the mean curvature term in the KPLW combination is invariant under deformations of the entanglement surface since, as argued in the main text, both $S_{\mathrm{KPLW}}[T^2]$ and $S_{\mathrm{topo}}[T^2]$ in Eq.~\eqref{AppKPLWH22} are topological invariants. Therefore, we only need to show that Eq.~\eqref{KPLWH22} vanishes for one particular entanglement surface that is topologically equivalent to that in Fig.~\ref{KPT2} in the main text, such as Fig.~\ref{KPT2regularized}.  Then by topological invariance, Eq.~\eqref{KPLWH22} vanishes for general configurations.

For the configuration in Fig.~\ref{KPT2regularized}, we can compute the mean curvature straightforwardly. The mean curvature is $H=(k_1+k_2)/2$, where $k_1$ and $k_2$ are the two principal curvatures at each point of the entanglement surface. We distinguish three types of points on the cylinder in Fig.~\ref{KPT2regularized}. 

			 \textit{Points on the top/bottom of a cylinder}: the surface is locally flat, $k_1=k_2=0$. Hence, $H=(k_1+k_2)/2=0$.
			 
			 \textit{Points on the side of a cylinder}: $k_1=\pm 1/r, k_2=0$, where $r$ is the radius of the cylinder, and the $\pm$ sign depends on whether it is inner or outer side surface.  Hence, $H=(k_1+k_2)/2=\pm 1/2r$. In the following, we will pick the $+$ sign. 
			 
\begin{figure}[H]
	\includegraphics[width=0.5\textwidth]{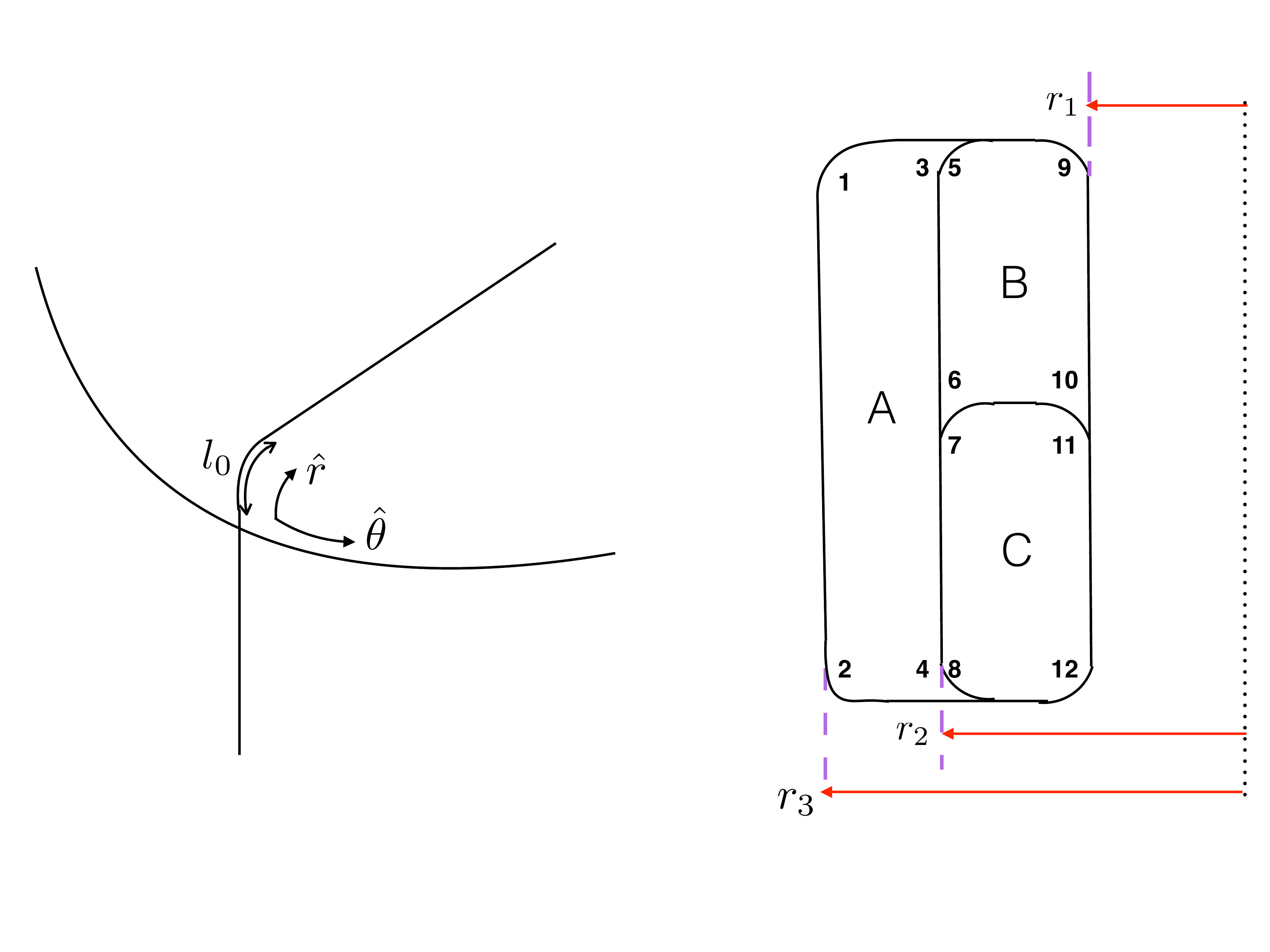}
	\centering
	\caption{Left: Regularization of a rectangular hinge with small arcs. Right: One choice of regularization of each hinge in Fig.~\ref{KPT2regularized}. The numbers label various hinges.}
	\label{regularization}
\end{figure}

			 \textit{Points on the hinge of a cylinder}: One of the hinges of the regular cylinders in Fig.~\ref{KPT2regularized} is shown as the thick green loop. On every point of the hinge, the Gauss curvature is the same. To find it, we apply the Gauss-Bonnet theorem to a cylinder. Because the Gauss curvature on the side and top/bottom of the cylinder vanishes, integration over the entire surface of the cylinder is reduced to the integration over the hinge. Hence the Gauss-Bonnet theorem dictates
			\begin{eqnarray}\label{GBCylinder}
			2\int_{\mathrm{hinge}} \frac{1}{r_3}k d\sigma=2\pi \chi[C]=4\pi,
			\end{eqnarray}
			where $C$ is the full cylinder, $r_3$ is the radius of the cylinder. $1/r_3$ is the principle curvature along the hinge and $k$ is the principal curvature along the direction perpendicular to the hinge.  In order to perform the two-dimensional surface integral, we need to regularize the one-dimensional hinge by smoothing it into an arc of infinitesimal radius, as shown in Fig.~\ref{regularization}. Assuming the length of the arc is $l_0$, Eq.~\eqref{GBCylinder} implies $\int_0^{l_0} k dl=1$, which reduces to $k=1/l_0$. The principal curvature for an ideal hinge (which corresponds to $l_0\to 0$) is infinite, and we regularize it with the small parameter $l_0$ to handle the computation.

To compute the integral of the mean curvature squared over various surfaces in Fig.~\ref{KPT2regularized}, we first introduce some notation. Let $r_1$ be the inner radius of region B/C, $r_2$ be the outer radius of region B/C, $r_3$ be the outer radius of region A, $h_1$ be the height of region B, and $h_2$ be the height of region C. We adopt the same finite regularization for every hinge, although this is not essential. For region A, the integration $\int_{\partial\mathrm{A}}H^2$ splits into three parts: the top/bottom, the side and the hinges. Since the top/bottom surface are flat, they do not contribute to the mean curvature integral. The mean curvature of the outer side surface is $1/2r_3$, and that of the inner side surface is $-1/2r_2$. The integration of the mean curvature over the outer and inner side of $\partial \mathrm{A}$ is 
\begin{widetext}
\begin{equation}
2\pi r_3 (h_1+h_2)\bigg(\frac{1}{2r_3}\bigg)^2+2\pi r_2 (h_1+h_2)\bigg(\frac{-1}{2r_2}\bigg)^2=\frac{\pi(h_1+h_2)}{2r_3}+\frac{\pi (h_1+h_2)}{2r_2}.
\end{equation} 
The mean curvature of the outer hinge is $(1/r_3+1/l_0)/2$, while according to our choice of regularization in Fig.~\ref{regularization}, the mean curvature of the inner hinge is $(1/l_0-1/r_2)/2$ because the principle curvature along the $\hat{\theta}$ direction (the meaning of $\hat{\theta}$ and $\hat{r}$ are specified in Fig.~\ref{regularization})  is $-1/r_2$ and the principle curvature along the $\hat{r}$ direction is $1/l_0$ (because we evaluate the curvature from the inside). The integration of the mean curvature over the hinges is
\begin{eqnarray}
2\times 2\pi r_3 l_0 \bigg(\frac{1}{2r_3}+\frac{1}{2l_0}\bigg)^2+ 2\times 2\pi r_2 l_0 \bigg(\frac{-1}{2r_2}+\frac{1}{2l_0}\bigg)^2,
\end{eqnarray} 
where the factor of $2$ in the front comes from equal contribution of the hinges from the top and bottom respectively. Collecting the above results, we have
\begin{eqnarray}
\begin{split}
\int_{\partial\mathrm{A}}H^2=\frac{\pi(h_1+h_2)}{2r_3}+\frac{\pi(h_1+h_2)}{2r_2}+\frac{\pi (r_3+l_0)^2}{r_3l_0}+\frac{\pi (r_2-l_0)^2}{r_2l_0}.
\end{split}
\end{eqnarray}
For convenience, we list the mean curvature of each hinge in the following table.
\begin{center}
	\begin{tabular}{ c | c  }
			\hline
			Hinge & Mean curvature \\ \hline
			1 & $1/2r_3+1/2l_0$ \\ \hline
			2 & $1/2r_3+1/2l_0$ \\ \hline
			3 & $-1/2r_2+1/2l_0$ \\ \hline
			4 & $-1/2r_2+1/2l_0$ \\ \hline
			5 & $1/2r_2+1/2l_0$ \\ \hline
			6 & $1/2r_2+1/2l_0$ \\ \hline
			7 & $1/2r_2+1/2l_0$ \\ \hline
			8 & $1/2r_2+1/2l_0$ \\ \hline
			9 & $-1/2r_1+1/2l_0$ \\ \hline
			10 & $-1/2r_1+1/2l_0$ \\ \hline
			11 & $-1/2r_1+1/2l_0$ \\ \hline
			12 & $-1/2r_1+1/2l_0$ \\ \hline
   \end{tabular}
\end{center}
where the labels of hinges are shown in Fig.~\ref{regularization}. 
For region B, the side surface contribution is 
\begin{eqnarray}
2\pi r_2 h_1 \bigg(\frac{1}{2 r_2}\bigg)^2+2\pi r_1 h_1\bigg(\frac{-1}{2r_1}\bigg)^2=\frac{\pi h_1}{2r_2}+\frac{\pi h_1}{2r_1}
\end{eqnarray}
The hinge contribution is 
\begin{eqnarray}
\begin{split}
&2\times 2\pi r_2 l_0 \bigg(\frac{1}{2r_2}+\frac{1}{2l_0}\bigg)^2+2\times 2\pi r_1l_0\bigg(\frac{-1}{2r_1}+\frac{1}{2l_0}\bigg)^2
=\frac{\pi(r_2+l_0)^2}{r_2l_0}+\frac{\pi(r_1-l_0)^2}{r_1l_0}
\end{split}
\end{eqnarray}
Hence the total contribution from region B is 
\begin{eqnarray}
\int_{\partial\mathrm{B}}H^2=\frac{\pi h_1}{2r_2}+\frac{\pi h_1}{2r_1}+\frac{\pi(r_2+l_0)^2}{r_2l_0}+\frac{\pi(r_1-l_0)^2}{r_1l_0}
\end{eqnarray}
For region C, the side surface contribution is 
\begin{eqnarray}
2\pi r_2 h_2 \bigg(\frac{1}{2 r_2}\bigg)^2+2\pi r_1 h_2\bigg(\frac{-1}{2r_1}\bigg)^2=\frac{\pi h_2}{2r_2}+\frac{\pi h_2}{2r_1}
\end{eqnarray}
The hinge contribution is
\begin{eqnarray}
\begin{split}
&2\times 2\pi r_2 l_0 \bigg(\frac{1}{2r_2}+\frac{1}{2l_0}\bigg)^2+2\times 2\pi r_1l_0\bigg(\frac{-1}{2r_1}+\frac{1}{2l_0}\bigg)^2=\frac{\pi(r_2+l_0)^2}{r_2l_0}+\frac{\pi(r_1-l_0)^2}{r_1l_0}
\end{split}
\end{eqnarray}
Hence the total contribution from region C is 
\begin{eqnarray}
\int_{\partial\mathrm{C}}H^2=\frac{\pi h_2}{2r_2}+\frac{\pi h_2}{2r_1}+\frac{\pi (r_2+l_0)^2}{r_2l_0}+\frac{\pi (r_1-l_0)^2}{r_1l_0}
\end{eqnarray}
For region AB, the side surface contribution is 
\begin{eqnarray}
2\pi r_3(h_1+h_2)\bigg(\frac{1}{2r_3}\bigg)^2+2\pi r_1 h_1 \bigg(\frac{-1}{2r_1}\bigg)^2+2\pi r_2 h_2 \bigg(\frac{-1}{2r_2}\bigg)^2=\frac{\pi(h_1+h_2)}{2r_3}+\frac{\pi h_1}{2r_1}+\frac{\pi h_2}{2r_2}
\end{eqnarray}
The hinge contribution is 
\begin{eqnarray}
2\times 2\pi r_3 l_0\bigg(\frac{1}{2r_3}+\frac{1}{2l_0}\bigg)^2+2\times 2\pi r_1l_0 \bigg(-\frac{1}{2r_1}+\frac{1}{2l_0}\bigg)^2+2\pi r_2 l_0\bigg(-\frac{1}{2r_2}-\frac{1}{2l_0}\bigg)^2+2\pi r_2 l_0 \bigg(-\frac{1}{2r_2}+\frac{1}{2l_0}\bigg)^2
\end{eqnarray}
Notice that the third term corresponds to  the opposite of hinge $7$ (which is not hinge $6$). Hence the total contribution from region AB is 
\begin{eqnarray}
	\int_{\partial\mathrm{AB}}H^2=\frac{\pi(h_1+h_2)}{2r_3}+\frac{\pi h_1}{2r_1}+\frac{\pi h_2}{2r_2}+\frac{\pi (r_3+l_0)^2}{r_3l_0}+\frac{\pi (r_1-l_0)^2}{r_1l_0}+\frac{\pi(r_2+l_0)^2}{2r_2l_0}+\frac{\pi(r_2-l_0)^2}{2r_2l_0}
\end{eqnarray}
For region AC, the side surface contribution is 
\begin{eqnarray}
2\pi r_3(h_1+h_2)\bigg(\frac{1}{2r_3}\bigg)^2+2\pi r_2 h_1 \bigg(\frac{-1}{2r_2}\bigg)^2+2\pi r_1 h_2 \bigg(\frac{-1}{2r_1}\bigg)^2=\frac{\pi(h_1+h_2)}{2r_3}+\frac{\pi h_1}{2r_2}+\frac{\pi h_2}{2r_1}
\end{eqnarray}
The hinge contribution is 
\begin{eqnarray}
2\times 2\pi r_3 l_0\bigg(\frac{1}{2r_3}+\frac{1}{2l_0}\bigg)^2+2\times 2\pi r_1l_0 \bigg(-\frac{1}{2r_1}+\frac{1}{2l_0}\bigg)^2+2\pi r_2 l_0\bigg(-\frac{1}{2r_2}-\frac{1}{2l_0}\bigg)^2+2\pi r_2 l_0 \bigg(-\frac{1}{2r_2}+\frac{1}{2l_0}\bigg)^2
\end{eqnarray}
Hence the total contribution from region AC is
\begin{eqnarray}
\int_{\partial\mathrm{AC}}H^2=\frac{\pi(h_1+h_2)}{2r_3}+\frac{\pi h_2}{2r_1}+\frac{\pi h_1}{2r_2}+\frac{\pi (r_3+l_0)^2}{r_3l_0}+\frac{\pi (r_1-l_0)^2}{r_1l_0}+\frac{\pi(r_2+l_0)^2}{2r_2l_0}+\frac{\pi(r_2-l_0)^2}{2r_2l_0}
\end{eqnarray}
For region BC, the side surface contribution is 
\begin{eqnarray}
2\pi r_2 (h_1+h_2) \bigg(\frac{1}{2 r_2}\bigg)^2+2\pi r_1 (h_1+h_2)\bigg(\frac{-1}{2r_1}\bigg)^2=\frac{\pi (h_1+h_2) }{2r_2}+\frac{\pi (h_1+h_2) }{2r_1}
\end{eqnarray}
The hinge contribution is 
\begin{eqnarray}
2\times 2\pi r_2 l_0 \bigg(\frac{1}{2r_2}+\frac{1}{2l_0}\bigg)^2+2\times 2\pi r_1l_0\bigg(\frac{-1}{2r_1}+\frac{1}{2l_0}\bigg)^2=\frac{\pi(r_2+l_0)^2}{r_2l_0}+\frac{\pi(r_1-l_0)^2}{r_1l_0}
\end{eqnarray}
Hence the total contribution from region BC is
\begin{eqnarray}
\int_{\partial\mathrm{BC}}H^2=\frac{\pi(h_1+h_2)}{2r_2}+\frac{\pi(h_1+h_2)}{2r_1}+\frac{\pi (r_2+l_0)^2}{r_2l_0}+\frac{\pi (r_1-l_0)^2}{r_1l_0}
\end{eqnarray}
Finally, for region ABC, the side surface contribution is 
\begin{eqnarray}
2\pi r_3 (h_1+h_2) \bigg(\frac{1}{2 r_3}\bigg)^2+2\pi r_1 (h_1+h_2)\bigg(\frac{-1}{2r_1}\bigg)^2=\frac{\pi (h_1+h_2) }{2r_3}+\frac{\pi (h_1+h_2) }{2r_1}
\end{eqnarray}
The hinge contribution is 
\begin{eqnarray}
2\times 2\pi r_3 l_0 \bigg(\frac{1}{2r_3}+\frac{1}{2l_0}\bigg)^2+2\times 2\pi r_1l_0\bigg(\frac{-1}{2r_1}+\frac{1}{2l_0}\bigg)^2=\frac{\pi(r_3+l_0)^2}{r_3l_0}+\frac{\pi(r_1-l_0)^2}{r_1l_0}
\end{eqnarray}
Hence the total contribution from region ABC is
\begin{eqnarray}
\int_{\partial\mathrm{ABC}}H^2=\frac{\pi(h_1+h_2)}{2r_3}+\frac{\pi(h_1+h_2)}{2r_1}+\frac{\pi (r_3+l_0)^2}{r_3l_0}+\frac{\pi (r_1-l_0)^2}{r_1l_0}
\end{eqnarray}

In summary, we obtain the contribution of mean curvature squared of seven regions as follows.
\begin{eqnarray}
\begin{split}
	\int_{\partial\mathrm{A}}H^2&=\frac{\pi(h_1+h_2)}{2r_3}+\frac{\pi(h_1+h_2)}{2r_2}+\frac{\pi (r_3+l_0)^2}{r_3l_0}+\frac{\pi (r_2-l_0)^2}{r_2l_0}.,\\
\int_{\partial\mathrm{B}}H^2&=\frac{\pi h_1}{2r_2}+\frac{\pi h_1}{2r_1}+\frac{\pi(r_2+l_0)^2}{r_2l_0}+\frac{\pi(r_1-l_0)^2}{r_1l_0},\\
	\int_{\partial\mathrm{C}}H^2&=\frac{\pi h_2}{2r_2}+\frac{\pi h_2}{2r_1}+\frac{\pi (r_2+l_0)^2}{r_2l_0}+\frac{\pi (r_1-l_0)^2}{r_1l_0},\\
\int_{\partial\mathrm{AB}}H^2&=\frac{\pi(h_1+h_2)}{2r_3}+\frac{\pi h_1}{2r_1}+\frac{\pi h_2}{2r_2}+\frac{\pi (r_3+l_0)^2}{r_3l_0}+\frac{\pi (r_1-l_0)^2}{r_1l_0}+\frac{\pi(r_2+l_0)^2}{2r_2l_0}+\frac{\pi(r_2-l_0)^2}{2r_2l_0},\\
	\int_{\partial\mathrm{AC}}H^2&=\frac{\pi(h_1+h_2)}{2r_3}+\frac{\pi h_2}{2r_1}+\frac{\pi h_1}{2r_2}+\frac{\pi (r_3+l_0)^2}{r_3l_0}+\frac{\pi (r_1-l_0)^2}{r_1l_0}+\frac{\pi(r_2+l_0)^2}{2r_2l_0}+\frac{\pi(r_2-l_0)^2}{2r_2l_0},\\
\int_{\partial\mathrm{BC}}H^2&=\frac{\pi(h_1+h_2)}{2r_2}+\frac{\pi(h_1+h_2)}{2r_1}+\frac{\pi (r_2+l_0)^2}{r_2l_0}+\frac{\pi (r_1-l_0)^2}{r_1l_0},\\
	\int_{\partial\mathrm{ABC}}H^2&=\frac{\pi(h_1+h_2)}{2r_3}+\frac{\pi(h_1+h_2)}{2r_1}+\frac{\pi (r_3+l_0)^2}{r_3l_0}+\frac{\pi (r_1-l_0)^2}{r_1l_0}.
\end{split}
\end{eqnarray}
It is straightforward to check that the combination Eq.~\eqref{KPLWH22} vanishes. Hence the relation Eq.~\eqref{KPLWTEE} in the main text holds.

\end{widetext}

\section{Review of Lattice TQFT}
\label{TriangulationofTQFT}

In this section, we briefly review the lattice formulation of  TQFTs. We begin with a triangulation of spacetime. The letters $i$, $j$, $k$ etc. label the vertices of a spacetime lattice. Combinations of vertices denote the simplicies of the lattice. For instance, $(ij)$ is the 1-simplex (bond) whose ends are vertices $i$ and $j$. $(ijk)$ is a 2-simplex (triangle) whose vertices are $i$, $j$ and $k$. 
Gauge fields live on these simplicies. In our paper, 1-form gauge fields $A$ live on 1-simplicies; 2-form gauge fields $B$ live on 2-simplicies; etc. In the language of discrete theories, $A(ij)$, $B(ijk)$ are the 1-cochain and 2-cochain associated with the indicated 1-simplex and 2-simplex, respectively. Exterior derivatives are defined by:
\begin{equation}
\begin{split}
dA(ijk) =& A(jk) - A(ik) + A(ij),	\\
dB(ijkl) =& B(jkl) - B(ikl) + B(ijl) - B(ijk).
\end{split}
\end{equation}
Note that the vertices are ordered such that $i<j<k<l$.

We further illustrate the values that the cochains $A(ij)$ and $B(ijk)$ can take using canonical quantization. Let us first consider the GWW model described by Eq.~\eqref{GWW} on a continuous spacetime with $U(1)$ gauge group. It is known that there are $n$ surface operators $\exp(is\oint_{\Sigma} B), s=0, 1, \cdots, n-1$\cite{Gaiotto2015,seiberg2014coupling}, and $\exp(in\oint_{\Sigma} B)=1$ is a trivial operator for an arbitrary closed surface $\Sigma$.  Hence $\oint_{\Sigma}B=\frac{2\pi q}{n}$, where $q\in \mathbb{Z}_n$ and $\Sigma$ is any closed surface. The fact that $\exp(in\oint_{\Sigma} B)$ is a trivial operator can be verified via canonical quantization.  To perform canonical quantization, we first use the gauge transformation Eq.~\eqref{gaugetransformation} to fix the gauge $A_t=0, B_{tx}=0, B_{ty}=0, B_{tz}=0$. The commutation relations from canonical quantization are
\begin{equation}\label{commutationrelation}
\begin{split}
[A_x(t, x, y, z),&B_{yz}(t, x',y',z')]\\&=-i\frac{2\pi}{n}\delta(x-x')\delta(y-y')\delta(z-z').
\end{split}
\end{equation}
and similarly for other components. Using Eq.~\eqref{commutationrelation}, we find that $\exp(in\oint_{\Sigma} B)$ commutes with all other gauge invariant operators. Specifically, we compute the commutation relation between the surface operator $\exp(in\oint_{\Sigma} B)$ and the line operator $\exp(il \oint_{\gamma}A+ilp\int_{\Sigma_2}B)$. Here $\Sigma$ is a closed surface in a spatial slice, and $\Sigma_2$ is an open surface with boundary $\gamma$. Both $\Sigma_2$ and $\gamma$ are living in the spatial slice. We find 
\begin{equation}
\begin{split}
e^{in\oint_{\Sigma}B}&e^{il\oint_{\gamma}A+ilp\int_{\Sigma_2}B}\\&=e^{i\frac{2\pi }{n}n l N_{\Sigma, \gamma}} e^{il\oint_{\gamma}A+ilp\int_{\Sigma_2}B}e^{in\oint_{\Sigma}B}\\
&=e^{il\oint_{\gamma}A+ilp\int_{\Sigma_2}B}e^{in\oint_{\Sigma}B},
\end{split}
\end{equation}
where $N_{\Sigma, \gamma}$ is the intersection number of the surface $\Sigma$ and the loop $\gamma$. Since the phase factor coming from the commutation relation is always 1, $\exp(in\oint_{\Sigma}B)$ commutes with all line operators. Since it also commutes with $\exp(il\oint_{\Sigma'}B)$ for any $l$ and $\Sigma'$, we conclude that $\exp(in\oint_{\Sigma}B)$ commutes with all the gauge invariant operators. Therefore, it must be a constant operator, $e^{in\oint_{\Sigma}B}=e^{i \theta}$ where $\theta$ is a constant number. We further show that $e^{in\oint_{\Sigma}B}=1$. To show this, we act $e^{in\oint_{\Sigma}B}$ on a state $|0\rangle$ where $B=0$ everywhere (more concretely, if the spacetime is discrete, $B=0$ on every 2-simplex). Since $e^{in\oint_{\Sigma}B}$ measures the value of $B$-field of the state, and $B$-field is zero everywhere,
\begin{eqnarray}\label{E4}
e^{i\theta}|0\rangle=e^{in\oint_{\Sigma}B}|0\rangle=|0\rangle
\end{eqnarray}
Hence the constant number $e^{i\theta}=1$ everywhere. This proves that $e^{in\oint_{\Sigma}B}=1$. 

Similarly, $\exp(in \oint_{\gamma}A+inp\int_{\Sigma_2}B)$ commutes with all other operators as well. 
\begin{equation}
\begin{split}
&e^{in \oint_{\gamma}A+inp\int_{\Sigma_2}B}e^{il\oint_{\Sigma}B}\\&=e^{-i\frac{2\pi }{n}n l N_{\Sigma, \gamma}} e^{il\oint_{\Sigma}B}e^{in \oint_{\gamma}A+inp\int_{\Sigma_2}B}\\
&=e^{il\oint_{\Sigma}B}e^{in \oint_{\gamma}A+inp\int_{\Sigma_2}B}.
\end{split}
\end{equation}
and
\begin{equation}
\begin{split}
&e^{in \oint_{\gamma}A+inp\int_{\Sigma_2}B}e^{il \oint_{\gamma'}A+ilp\int_{\Sigma'_2}B}\\&=e^{-i\frac{2\pi }{n}n l p (N_{\gamma, \Sigma'_2}-N_{\gamma', \Sigma_2})} e^{il \oint_{\gamma'}A+ilp\int_{\Sigma'_2}B}e^{in \oint_{\gamma}A+inp\int_{\Sigma_2}B}\\
&=e^{il \oint_{\gamma'}A+ilp\int_{\Sigma'_2}B}e^{in \oint_{\gamma}A+inp\int_{\Sigma_2}B}.
\end{split}
\end{equation}
Therefore $e^{in \oint_{\gamma}A+inp\int_{\Sigma_2}B}$ commutes with all gauge invariant operators as well, which implies $e^{in \oint_{\gamma}A+inp\int_{\Sigma_2}B}=e^{i\eta}$ where $e^{i\eta}$ is a constant. Using the same analysis for the operator $e^{in\oint_{\Sigma}B}$, we find $e^{in \oint_{\gamma}A+inp\int_{\Sigma_2}B}=1$. 
 
On a triangulated lattice, since $\Sigma$ is any two dimensional surface, $\exp(in\oint_{\Sigma}B)=1$ implies that  $\exp(in\oint_{(ijkl)} B)=1$ for any 3-simplex $(ijkl)$. Using the Stokes formula, $\oint_{(ijkl)} B=\int_{(ijkl)}dB=(dB)(ijkl)=B(ijk)-B(ijl)+B(ikl)-B(jkl)$ where we used the fact that integrating $dB$ over the volume of 3-simplex $(ijkl)$ is just evaluating the $dB$ on $(ijkl)$ itself. Hence $\exp(in\oint_{(ijkl)} B)=1$ implies that  $B(ijk)-B(ijl)+B(ikl)-B(jkl)\in\frac{2\pi}{n}\mathbb{Z}_n$ for any 3-simplex $(ijkl)$. Since the choice of $(ijkl)$ is arbitrary, we conclude that on each 2-simplex $(ijk)$, $B(ijk)$ takes values in $\frac{2\pi}{n}\mathbb{Z}_n$. Similarly, on each 1-simplex $(ij)$, $A(ij)$ takes values in $\frac{2\pi}{n}\mathbb{Z}_n$ for any $i,j$. 

Next, we comment on the delta functions obtained from integrating out the $A$ fields as in Eq.~\eqref{BFWF}. For simplicity, we work with a level $n=2$ BF/GWW theory. On each 4-simplex with vertices labeled by $(i,j,k,l,s)$, the action is
\begin{eqnarray}
\frac{2}{2\pi}(AdB)(ijkls)=\frac{2}{2\pi}A(ij)dB(jkls).
\end{eqnarray} 
Integrating over $A$ means summing over all configurations of $A(ij)=0, \pi$. Hence the path integral is
\begin{eqnarray}
&&\frac{1}{2}\sum_{A(ij)=0,\pi}\exp\bigg[i\frac{2}{2\pi}A(ij)dB(jkls)\bigg]\nonumber\\
&&=\frac{1}{2}\bigg\{1+\exp\big[idB(jkls)\big]\bigg\}\equiv \delta\big[dB(jkls)\big].
\end{eqnarray}
This explains the meaning of the delta function in the discrete theory, and we refer to the $B$ field as flat if the above delta function constraint is satisfied, i.e. if $dB(jkls)=0\mod{2\pi}$.

Although we write TQFT actions as integrals in the continuum in the main text, they can actually be translated into lattice actions using the conventions we have introduced in this appendix. The wave functions defined via the path integral in Eqs.~\eqref{WavefunctionBF} and \eqref{Wavefunction} are then wave functions on the lattice.

\section{Surfaces in the dual lattice}
\label{Appclosedsurface}

In this appendix, we argue that the simplices on which $\tilde{B}=\pi$ in the dual lattice form continuous surfaces. Continuous means that connected simplices in the dual lattice join via edges, rather than via vertices. Specifically,
\begin{enumerate}
	\item In three-dimensional space, if a real space 2-cochain $B(ijk)$ satisfies the flatness condition
	$
	dB(ijkl)= B(jkl)-B(ikl)+B(ijl)-B(ijk)
	= 0 \mod 2\pi
	$
	then its dual $\widetilde{B}=\pi$ on a closed loop in the dual lattice. 
	\item In $(3+1)$-dimensional spacetime, if a real space 2-cochain $B(ijk)$ satisfies the flatness condition $
	dB(ijkl)= B(jkl)-B(ikl)+B(ijl)-B(ijk)
	= 0 \mod 2\pi
	$
	then its dual $\widetilde{B}=\pi$ on a continuous and closed surface in the dual lattice. 
\end{enumerate}
The first statement is proven in the main text. In the following, we will present a more algebraic proof of the first statement, which is easier to generalize to $(3+1)$-dimensions, allowing for a proof of the second statement.

\begin{figure}[H]
	\centering
	\includegraphics[width=0.5\textwidth]{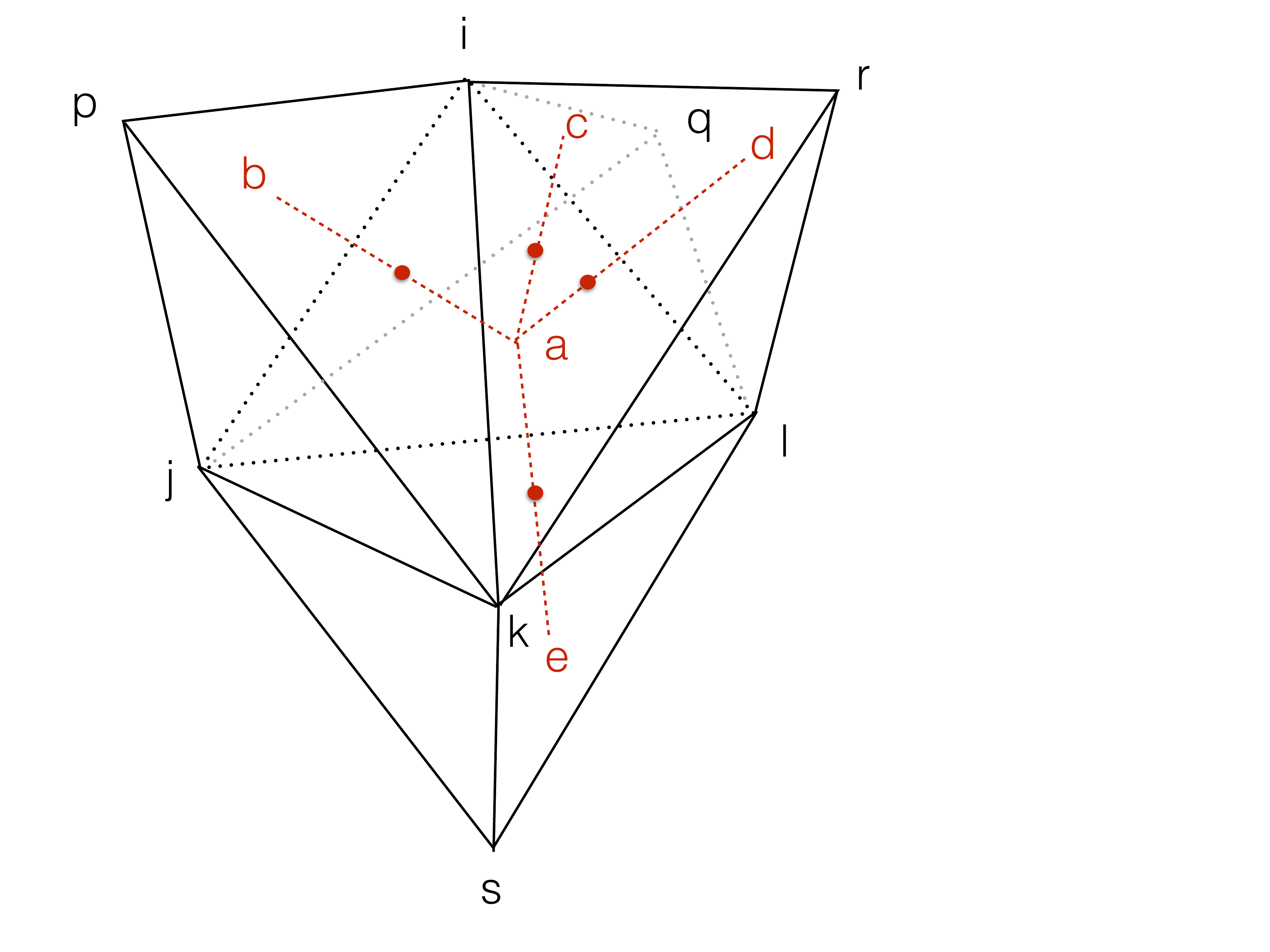}
	\caption{Dual lattice of a tetrahedron $(ijkl)$. $(ijkp), (ijlq), (iklr), (jkls)$ are four adjacent tetrahedra to $(ijkl)$, which are dual to $(b), (c), (d), (e), (a)$ respectively. The red dots are the intersection between 2-simplices in the real lattice and the 1-simplices in the dual lattice. For example, the red dot on $(ab)$ is the intersection point of $(ab)$ and $(ijk)$.}
	\label{dualsurface}
\end{figure}

We first redraw the simplex in Fig.~\ref{DualLattice} with some additional details, as shown in Fig.~\ref{dualsurface}. To construct the duals of simplices in three-dimensional \emph{space}, we begin by considering the tetrahedron $(ijkl)$, in addition to its neighbors $(ijkp), (ijlq), (iklr),$ and $(jkls)$. 
 3-simplices in the real lattice are dual to points in the dual lattice: for example $(ijkl)$ is dual to the point $(a)$, and similarly $(ijkp)$ is dual to $(b)$, $(ijlq)$ is dual to $(c)$, $(iklr)$ is dual to $(d)$, and $(jkls)$ is dual to $(e)$. 2-simplices in the real lattice are dual to 1-simplices (bonds). For example, $(ijk)$ is the intersection of $(ijkl)$ and $(ijkp)$, i.e., $(ijk)=(ijkl)\cap (ijkp)$. Therefore,
the dual of $(ijk)$ is the bond $(ab)$, joining the dual of $(ijkl)$ and $(ijkp)$. Similarly, we are able to identify the duals of all other simplices. We list the result in the following table:
\vspace{3mm}

\begin{tabular}{cc}
	\begin{minipage}{.5\linewidth}
	\begin{tabular}{ c | c  }
		\hline
		Real & Dual \\ \hline
		$(ijkl)$ & $(a)$ \\ \hline
		$(ijkp)$ & $(b)$ \\ \hline
		$(ijlq)$ & $(c)$ \\ \hline
		$(iklr)$ & $(d)$ \\ \hline
		$(jkls)$ & $(e)$ \\ \hline
	\end{tabular}
	\end{minipage} &\qquad
	
	\begin{minipage}{.5\linewidth}
		\begin{tabular}{c|c}
			\hline
			Real & Dual \\\hline
			$(ijk)$ & $(ab)$ \\ \hline
			$(ijl)$ & $(ac)$\\\hline
			$(ikl)$ & $(ad)$\\\hline
			$(jkl)$ & $(ae)$\\\hline
		\end{tabular}
	\end{minipage} 
\end{tabular}
\vspace{3mm}

\noindent The flatness condition implies that there are even number of 2-simplices among the four faces of the tetrahedron $(ijkl)$ on which $B=\pi$. It follows that there are an even number $\widetilde{B}=\pi$ bonds among the four dual lattice bonds $(ab), (ac), (ad), (ae)$. Thus these form closed loops in the dual lattice. This proves the first statement.

We proceed to prove the second statement. In $(3+1)$ dimensions, spacetime is triangulated into 4-simplices. Let us consider a 4-simplex labeled by the five vertices $(ijklm)$ where $m$ is in the extra dimension compared with $3$D case shown in Fig.~\ref{dualsurface}. To find the dual of 2-simplices, we will begin -- as above -- by considering the 4-simplices adjacent to $(ijklm)$ which share one 3-simplex with $(ijklm)$. Introducing the additional vertices $p$, $q$, $r$, $s$, and $t$\footnote{Notice that $t$ is in the additional dimension as well.}, these 4-simplices are: $(ijkmp)$, $(ijlmq)$, $(iklmr)$, $(jklms)$, and $(ijklt)$. Dual simplices in $(3+1)$ dimensional \emph{spacetime} are determined as follows: 4-simplices in the real lattice are dual to points in the dual lattice; $(ijklm)$ is dual to a point $(a)$, $(ijkmp)$  is dual to $(b)$, $(ijlmq)$  is dual to $(c)$, $(iklmr)$  is dual to $(d)$, $(jklms)$  is dual to $(e)$, and $(ijklt)$ is dual to $(f)$\footnote{Notice that $(f)$ is in the additional dimension of the dual lattice.}. 3-simplices in the real lattice are dual to bonds in the dual lattice. For instance, since $(ijkm)$ is the intersection of $(ijklm)$ and $(ijkmp)$, i.e., $(ijkm)=(ijklm)\cap (ijkmp)$, the dual of $(ijkl)$ is the bond $(ab)$, joining the dual of $(ijklm)$ and $(ijkmp)$. Similarly, $(ijlm)$ is dual to $(ac)$, $(iklm)$ is dual to $(ad)$, $(jklm)$ is dual to $(ae)$, and $(ijkl)$ is dual to $(af)$. We further proceed to consider the dual of 2-simplices, applying the same method. For instance, since the 2-simplex $(ijk)$ is the common simplex of $(ijkm)$ and $(ijkl)$, i.e., $(ijk)=(ijkl)\cap (ijkm)$, the dual of $(ijk)$ is the surface $(abf)$ joining the dual of $(ijkl)$ and $(ijkm)$. Similarly, we can identify the duals of the remaining 2-simplices. We list all the results in the following table:

\vspace{3mm}
\begin{tabular}{ccc}
	\begin{minipage}{.33\linewidth}
		\begin{tabular}{c|c}
			\hline
			Real & Dual \\ \hline
			$(ijklm)$ & $(a)$ \\ \hline
			$(ijkmp)$ & $(b)$ \\ \hline
			$(ijlmq)$ & $(c)$ \\ \hline
			$(iklmr)$ & $(d)$ \\ \hline
			$(jklms)$ & $(e)$ \\ \hline
			$ (ijklt)$ & $(f)$ \\ \hline
		\end{tabular}
	\end{minipage} &
	
	\begin{minipage}{.33\linewidth}
		\begin{tabular}{c|c}
			\hline
			Real & Dual \\ \hline
			$(ijkm)$ & $(ab)$\\\hline
			$(ijlm)$ & $(ac)$\\\hline
			$(iklm)$ & $(ad)$\\\hline
			$(jklm)$& $(ae)$\\\hline
			$(ijkl)$&$(af)$\\\hline
		\end{tabular}
	\end{minipage}&
	
	 \begin{minipage}{.33\linewidth}
     \begin{tabular}{c|c}
     	\hline
     	Real & Dual \\ \hline
     	$(ijk)$& $(abf)$\\\hline
     	$(ijl)$&$(acf)$\\\hline
     	$(ijm)$&$(abc)$\\\hline
     	$(ikl)$&$(adf)$\\\hline
     	$(ikm)$&$(abd)$\\\hline
     	$(ilm)$&$(acd)$\\\hline
     	$(jkl)$&$(aef)$\\\hline
     	$(jkm)$&$(abe)$\\\hline
     	$(jlm)$&$(ace)$\\\hline
     	$(klm)$&$(ade)$\\\hline
     \end{tabular}
	 \end{minipage}
\end{tabular}
\vspace{3mm}

\noindent The four surfaces $(abf), (acf), (adf), (aef)$ are dual to the four faces $(ijk), (ijl), (ikl), (jkl)$ of the tetrahedron $(ijkl)$. All of these dual surfaces share a common link $(af)$. The flatness condition $dB(ijkl)=B(jkl)-B(ikl)+B(ijl)-B(ijk)=0\mod 2\pi$ implies that an even number of faces of the tetrahedron $(ijkl)$ are occupied. Thus, there are an even number of surfaces among $(abf), (acf), (adf), (aef)$ occupied in the dual lattice. Since all these occupied surfaces in the dual lattice share a common edge $(af)$, it follows from our definition of continuity (at the beginning of this appendix) that surfaces in the dual lattice are continuous. Furthermore, the continuous surfaces formed by the occupied simplices in the dual lattice are closed, because for any bond in the dual lattice, for example $(af)$, there exist even (among four) number of occupied dual-lattice 2-simplices adjacent to it. While for an open dual-lattice surface, there exist at least one dual-lattice bond such that there are only odd number of the adjacent dual-lattice 2-simplices occupied, which violate the flatness condition for the $B$-cochain. Hence the dual-lattice surface is closed. This proves the second statement.  

For completeness, we comment on how two loops can intersect in the dual space lattice, and how two surfaces can intersect in the dual spacetime lattice. We first prove by construction that two loops in the dual spatial lattice can intersect at a vertex: suppose one dual lattice loop includes the occupied bonds $(ab), (ac)$, and the other dual lattice loop includes the occupied bonds $(ad), (ae)$. Hence these two loops intersect at the vertex $(a)$. We now argue that if two surfaces in the dual spacetime lattice contain the same point, then they must share a bond. Let us assume two surfaces intersect (at least) at $(a)$. Since all the 2-simplices in the dual lattice including the vertex $(a)$ are $(abc), (abd), (acd), (abe), (ace), (ade), (abf), (acf), (adf)$ and $(aef)$, by enumerating all possibilities, we find the two surfaces must share at least one bond. Without loss of generality, suppose one surface includes the 2-simplices $(abc)$ and $(abd)$ (notice that  $(abc)$ and $(abd)$ join via the bond $(ab)$ and therefore form a continuous surface in the dual lattice). The surface thus includes the three bonds $(ab), (ac)$, and $(ad)$ emanating from $(a)$. Any other surface that contains $(a)$, would include, just like this surface, three of bonds emanating from $(a)$. Thus, as $(a)$ is the only shared part of five bonds $(ab), (ac), (ad), (ae)$, and $(af)$, two surfaces that include $(a)$ have to share at least one of these bonds, as they occupy three bonds each. In summary, two loops can intersect at vertices in the dual space lattice, and two surfaces can intersect at bonds (but not vertices) in the dual spacetime lattice.

\section{Mutual and Self-Linking Numbers}
\label{AppMutualandSelfLinkingNumbers}

\begin{figure}[H]
	\includegraphics[width=0.5\textwidth]{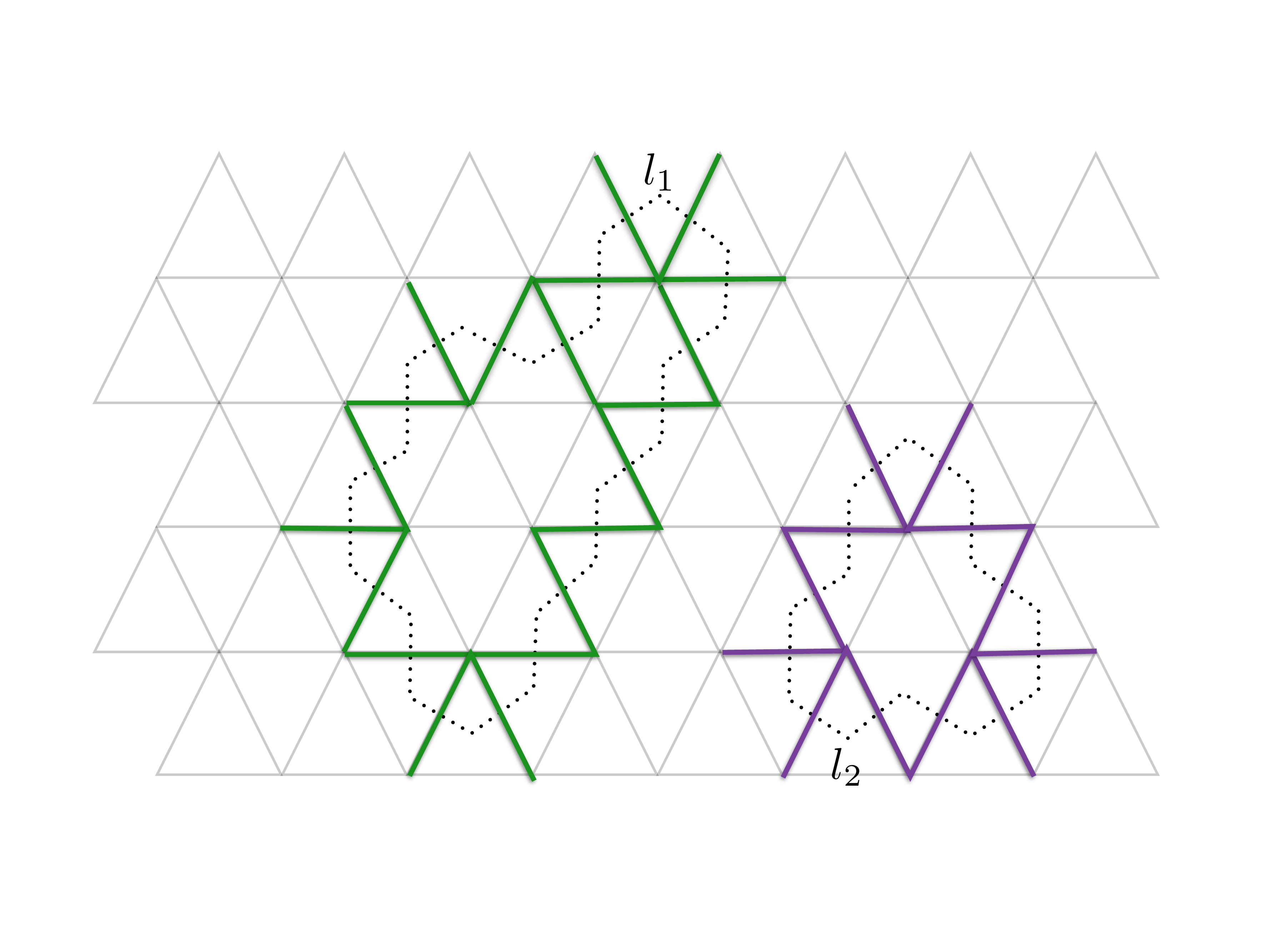}
	\centering
	\caption{We illustrate the geometric meaning of the Hodge dual in a two-dimensional space example. Suppose $A$ is a 1-cochain, which equals $\pi$ on 1-simplices in the dual lattice and 0 elsewhere. $A=\pi *_2\Sigma(l_1)+\pi*_2 \Sigma(l_2)$, where $l_1$ and $l_2$ are loops in the dual lattice drawn in dashed lines. $\Sigma(l_1)$ and $\Sigma(l_2)$ are 1-cochains living on the 1-simplices in the dual lattice.  $*_2$ is a lattice version of Hodge star, which transforms the 1-cochain living on the dual lattice (dashed lines) to a 1-cochain living on the lattice (green and purple bold lines). Correspondingly, $A=\pi *_2\Sigma(l_1)+\pi*_2 \Sigma(l_2)$ is a 1-cochain living on the green and purple bold lines. We use the dual lattice configuration $S_i, l_i$ to label the $B, A$-cochains because the dual lattice configurations are easier to visualize. 
		The interpretation of the 2-cochain $B$ can be straightforwardly generalized to three spatial dimensions.}
	\label{Hodgedualillustration}
\end{figure}

In this section, we provide all details needed to evaluate the integral Eq.~\eqref{linking}. As a simple case, we assume a configuration where $B=\pi$ only at two surfaces $S_1, S_2$ in the dual lattice of $\mathcal{M}_4$, with their boundaries given by the loops $l_1=\partial S_1, l_2=\partial S_2$ on the dual lattice of $\partial\mathcal{M}_4$. We can write this succinctly as
\begin{equation}\label{BS1S2}
B=\pi*_4\Sigma(S_1)+\pi*_4\Sigma(S_2),
\end{equation} 
where $*_4$ is the discretized version of Hodge star in four spacetime dimensions; its meaning is explained pictorially in Fig.~\ref{Hodgedualillustration}. Let us comment on Eq.~\eqref{BS1S2} in detail. On $\partial\mathcal{M}_4$, $B$ is a 2-cochain, which can be 0 or $\pi$; while on the dual lattice of $\partial \mathcal{M}_4$, the $\pi$-valued 1-cochains $\Sigma(l_i)$ (which are the dual of real-\emph{space} 2-cochains) form loops $l_i, i=1,2$. Moreover, on the spacetime $\mathcal{M}_4$, $B$ is still a 2-cochain valued in $0$ or $\pi$; while on the dual lattice of $\mathcal{M}_4$, the $\pi$-valued 2-cochains $\Sigma(S_i)$ (which are the dual of the real \emph{spacetime} 2-cochains) form surfaces $S_i, i=1,2$ whose boundaries are $l_i, i=1,2$. Notice that the closed dual-lattice surfaces which do not intersect with the spatial slice do not contribute to the wavefunction.  Further $*_4\Sigma(S_i)$ is a 2-cochain on the original lattice (dual to $S_i$), which is 1 on the dual of $S_i$, and 0 elsewhere. Hence, the role of the Hodge star is to transform the cochain defined on the dual lattice to the cochain defined on the real lattice. In Fig.~\ref{Hodgedualillustration} we illustrate the geometric meaning of these notions with an example in lower dimensions. Returning to the integral in the wavefunction Eq.~\eqref{linking}, we thus have
\begin{widetext}
	\begin{equation}
	\begin{split}
	\int_{\mathcal{M}_4} B\wedge B
	=&\pi^2 \int_{\mathcal{M}_4} \Big(*_4\Sigma(S_1)+*_4\Sigma(S_2)\Big)\wedge\Big(*_4\Sigma(S_1)+*_4\Sigma(S_2)\Big)\\
	=& 2\pi^2\int_{\mathcal{M}_4}*_4\Sigma(S_1)\wedge *_4\Sigma(S_2)+\pi^2\int_{\mathcal{M}_4}*_4\Sigma(S_1)\wedge *_4\Sigma(S_1)+\pi^2\int_{\mathcal{M}_4}*_4\Sigma(S_2)\wedge *_4\Sigma(S_2)\\
	=& 2\pi^2\mathrm{link}(l_1,l_2)+\pi^2 \mathrm{link}(l_1,l_1)+\pi^2 \mathrm{link}(l_2,l_2),
	\end{split}
	\label{AppBwedgeB}
	\end{equation}
\end{widetext}
where ${\mathrm{link}(l_1,l_2)}$ is the linking number between two loops $l_1$ and $l_2$.  This leads to Eq.~\eqref{BwedgeB} in the main text.

We will derive the last equality of Eq.~\eqref{AppBwedgeB} in Appendix~\ref{AppIntersectionLinking}, and provide a detailed discussion of the self-linking numbers of one single loop in Appendix~\ref{selflinkingnumber}.

\subsection{Intersection and Linking}
\label{AppIntersectionLinking}

We prove a statement relating the intersection form in the bulk and the linking number on the boundary, which in turn explains the last equality in Eq.~\eqref{AppBwedgeB}.

As explained below Eq.~\eqref{BS1S2}, $*_4\Sigma(S_i)$ is a 2-cochain in the real \emph{spacetime}, which equals 1 if it is evaluated on any triangulation of $S_i$ (in the dual spacetime lattice) and 0 if evaluated elsewhere.  Similarly, $*_3\Sigma(l_i)$ is still a 2-cochain in the real \emph{space}, which equals 1 if it is evaluated on the $l_i$ (in the dual space lattice) and 0 if evaluated elsewhere. Furthermore, if $l_i$ is on the boundary of $S_i$ (notice that both $l_i$ and $S_i$ are in the dual lattice), we have a relation between these two 2-simplices,\footnote{We can understand this formula by constructing examples using the method in appendix \ref{Appclosedsurface}. Let $(abf), (acf)\in S$ be two dual-lattice 2-simplices in the dual-lattice open surface $S$ in $4$D, which join via $(af)$. The boundary is along $(ab)$ and $(ac)$ direction, joined via $(a)$.  $(ab), (ac)\in l$ form a loop in 3D, which is the boundary of $S$. We need to compare the real space configuration of $S$ and $l$ by taking their duals. From the correspondence of real simplices and dual simplices listed in appendix \ref{Appclosedsurface}, in 3D, $(ab), (ac)$ are dual to $(ijk), (ijl)$ respectively, and in 4D, $(abf), (acf)$ are dual to $(ijk), (ijl)$ respectively. We find that their real lattice configurations are the same, hence $*_4\Sigma(S)=*_3\Sigma(l)$.}
\begin{eqnarray}
*_4\Sigma(S_i)=*_3\Sigma(\partial S_i)=*_3\Sigma(l_i).
\end{eqnarray}
We also notice that $B$ is flat, i.e., $d*_4\Sigma(S_i)=d*_3\Sigma(l_i)=0, ~i=1,2$ which come from the Gauss law for $B$-cochain Eq.~\eqref{constraintBijk}. This means the duals of the $B=\pi$ 2-simplices form two-dimensional surfaces in the spacetime, and form one-dimensional loops (which are the boundary of two-dimensional dual lattice surfaces) in the space, as shown in Fig.~\ref{M4S3}. We want to prove,
\begin{equation}
\int_{\mathcal{M}_4} *_4\Sigma(S_1)\wedge *_4\Sigma(S_2)=\int_{l_1\cap \partial^{-1}l_2}1\equiv\mathrm{link}(l_1,l_2),\label{NeedtoProve}
\end{equation}
where $\partial^{-1}l_2$ denotes a surface in the dual lattice of $\partial \mathcal{M}_4$ whose boundary is $l_2$. In the last equality, we used the definition of the linking number between two loops.

The relation \eqref{NeedtoProve} can be shown as follows. Keeping in mind that $*_3\Sigma(l)$ is a delta function that is nonzero on $l$ only, we find 
\begin{eqnarray}
\int_{l_1\cap \partial^{-1}l_2}1=\int_{\mathcal{M}_3}*_3\Sigma(l_1)\wedge d^{-1}*_3\Sigma(l_2).
\end{eqnarray}
Noticing that $\mathcal{M}_3=\partial \mathcal{M}_4$,
\begin{eqnarray}\label{F6}
&&\int_{\mathcal{M}_3}*_3\Sigma(l_1)\wedge d^{-1}*_3\Sigma(l_2)\nonumber\\&&=\int_{\partial \mathcal{M}_4}*_3\Sigma(l_1)\wedge d^{-1}*_3\Sigma(l_2)\nonumber\\
&& =\int_{ \mathcal{M}_4}d\Big(*_4\Sigma(S_1)\wedge d^{-1}*_4\Sigma(S_2)\Big)\nonumber\\&&=\int_{ \mathcal{M}_4}*_4\Sigma(S_1)\wedge *_4\Sigma(S_2).
\end{eqnarray}
In the second equality, we used $*_4\Sigma(S_i)=*_3\Sigma(l_i), i=1,2$. To get the last equality, we used the flatness condition $d*_4\Sigma(S_i)=d*_3\Sigma(l_i)=0, i=1,2$. Hence
\begin{eqnarray}\label{F7}
\int_{\mathcal{M}_4} *_4\Sigma(S_1)\wedge *_4\Sigma(S_2)=\int_{l_1\cap \partial^{-1}l_2}1.
\end{eqnarray}
Combining Eqs.~\eqref{NeedtoProve}, \eqref{F6} and \eqref{F7}, we find
\begin{eqnarray}
\int_{\mathcal{M}_4}B\wedge B&=&2\pi^2 \mathrm{link}(l_1, l_2)+\pi^2 \mathrm{link}(l_1, l_1)\nonumber\\&&+\pi^2 \mathrm{link}(l_2, l_2).
\end{eqnarray}

\subsection{Self-linking Number}
\label{selflinkingnumber}

\begin{figure}
	\includegraphics[width=0.5\textwidth]{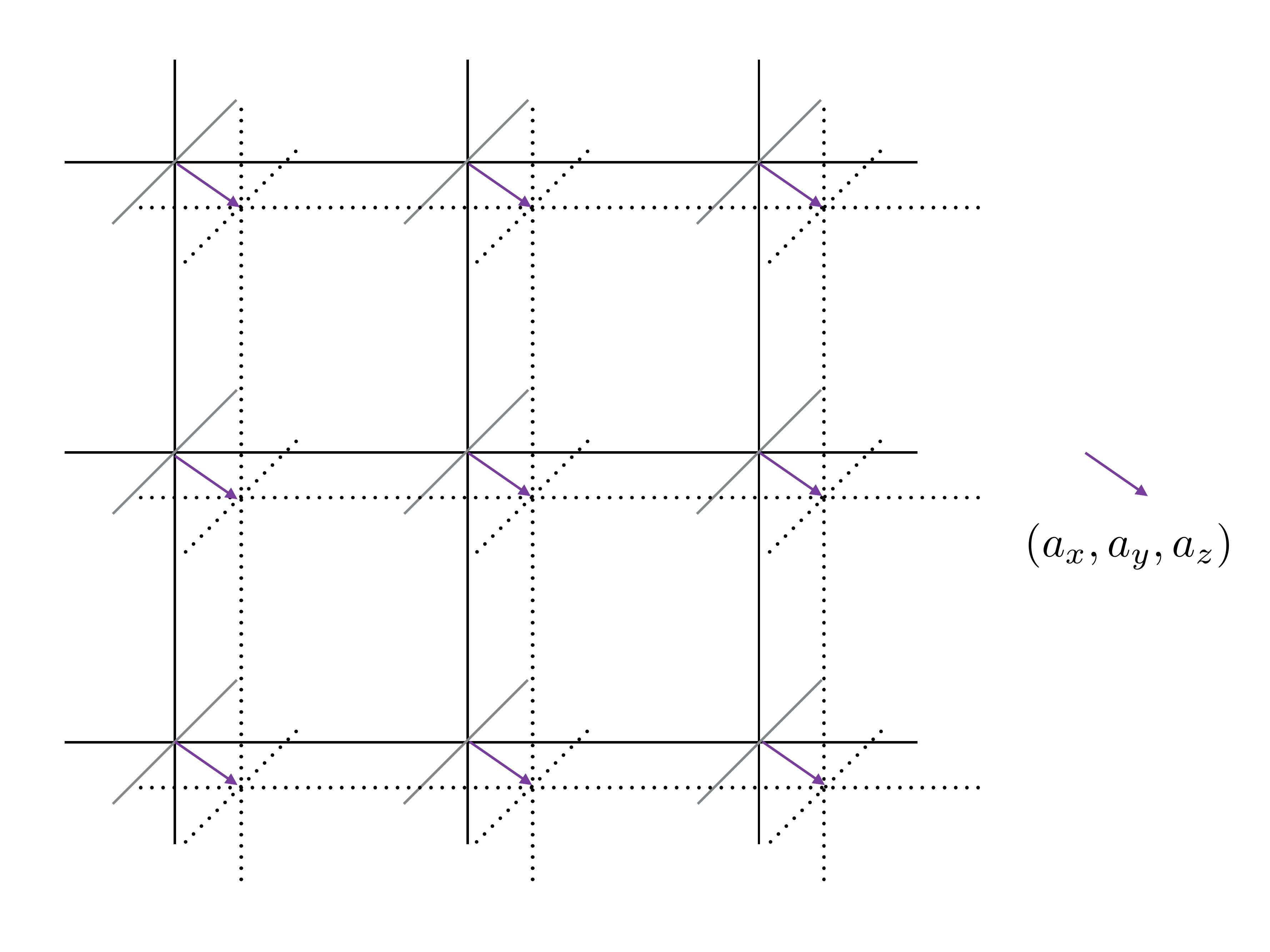}
	\centering
	\caption{Regularization of a spatial lattice. The blue arrow represents the constant vector $(a_x, a_y, a_z)$. The dashed lattice is obtained from the solid lattice by the translation $(x,y,z)\to (x+a_x, y+a_y, z+a_z)$.}
	\label{axayaz}
\end{figure}

\begin{figure}
	\includegraphics[width=0.5\textwidth]{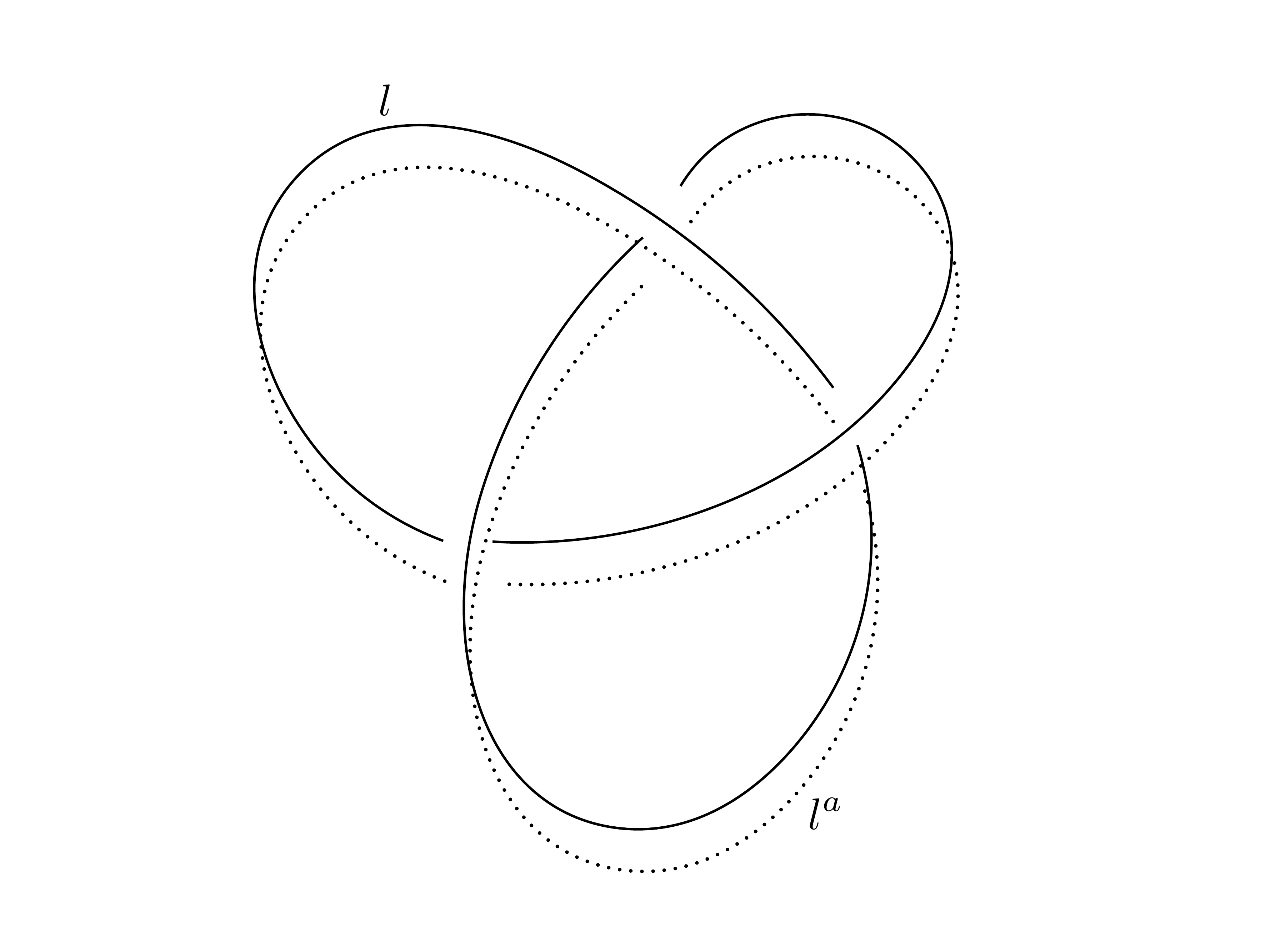}
	\centering
	\caption{An example of lattice regularization of a trefoil knot. $l$ is a knot (drawn in the dual lattice), while $l^a$ is the knot obtained by lattice regularization. The underlying lattice is omitted for clarity.}
	\label{loopreg}
\end{figure}

In this subsection, we define the self-linking number of a loop $l$, i.e., the $\mathrm{link}(l, l)$. To define the self-linking number, we need to regularize the loop into two nearby loops. This can be achieved by point splitting regularization\footnote{The point splitting method is widely used in studying lattice systems, such as in Ref.~\onlinecite{PhysRevX.5.041013,chen2014symmetry}.}. We separate each point of the spatial lattice into two points, for example
\begin{eqnarray}\label{regu}
(x,y,z)\to 
\begin{cases}
(x,y,z)\\
(x+a_x, y+a_y, z+a_z)
\end{cases},
\end{eqnarray}
where $(a_x, a_y, a_z)$ is a constant vector in space chosen to be the same for all loops. The original loop $l$ splits into two loops $l$ and $l^a$. See Fig.~\ref{axayaz} for an illustration of lattice regularization and Fig.~\ref{loopreg} for an illustration of the regularization of a loop. The mutual-linking number between two loops is well defined, and it is natural to identify the self-linking number of $l$ to be the mutual-linking number between $l$ and $l^a$, i.e.,
\begin{eqnarray}\label{selflinkingnumber1}
\mathrm{link}(l, l)\equiv\mathrm{link}(l, l^a).
\end{eqnarray}
We notice that the definition Eq.~\eqref{selflinkingnumber1} depends on the regularization Eq.~\eqref{regu}. But as long as we use the same regularization for all the loops $l$ [i.e., $(a_x, a_y, a_z)$ is a position-independent constant vector], Eq.~\eqref{selflinkingnumber1} is consistent [i.e., translating $l$ (without change its shape) does not change the self-linking number $\mathrm{link}(l, l)$ of $l$].

The definition of the self-linking number of a loop (knot) depends on the point splitting regularization [i.e., changing the constant vector  $(a_x, a_y, a_z)$ changes the regularization, and hence changes the self linking number], and so does the wavefunction. However, the entanglement entropy is independent of the self-linking number, hence it is independent of the point splitting regularization.

\section{$N_{\mathrm{A}}(\mathcal{C}_{\mathrm{E}})N_{\mathrm{A}^{\mathrm{c}}}(\mathcal{C}_{\mathrm{E}})$ is Independent of $\mathcal{C}_{\mathrm{E}}$}
\label{Appmoreon41}

In this appendix, we give a more detailed derivation of Eq.~\eqref{1over2Sigma}. We first show that $N_{\mathrm{A}}(\mathcal{C}_{\mathrm{E}})N_{\mathrm{A}^{\mathrm{c}}}(\mathcal{C}_{\mathrm{E}})$ is independent of $\mathcal{C}_{\mathrm{E}}$. We further explain the fact that the number of configurations on the entanglement surface $\Sigma$ is $2^{|\Sigma|-1}$. 

\begin{figure}[H]
	\centering
	\subfigure[]{
		\includegraphics[width=0.45\columnwidth]{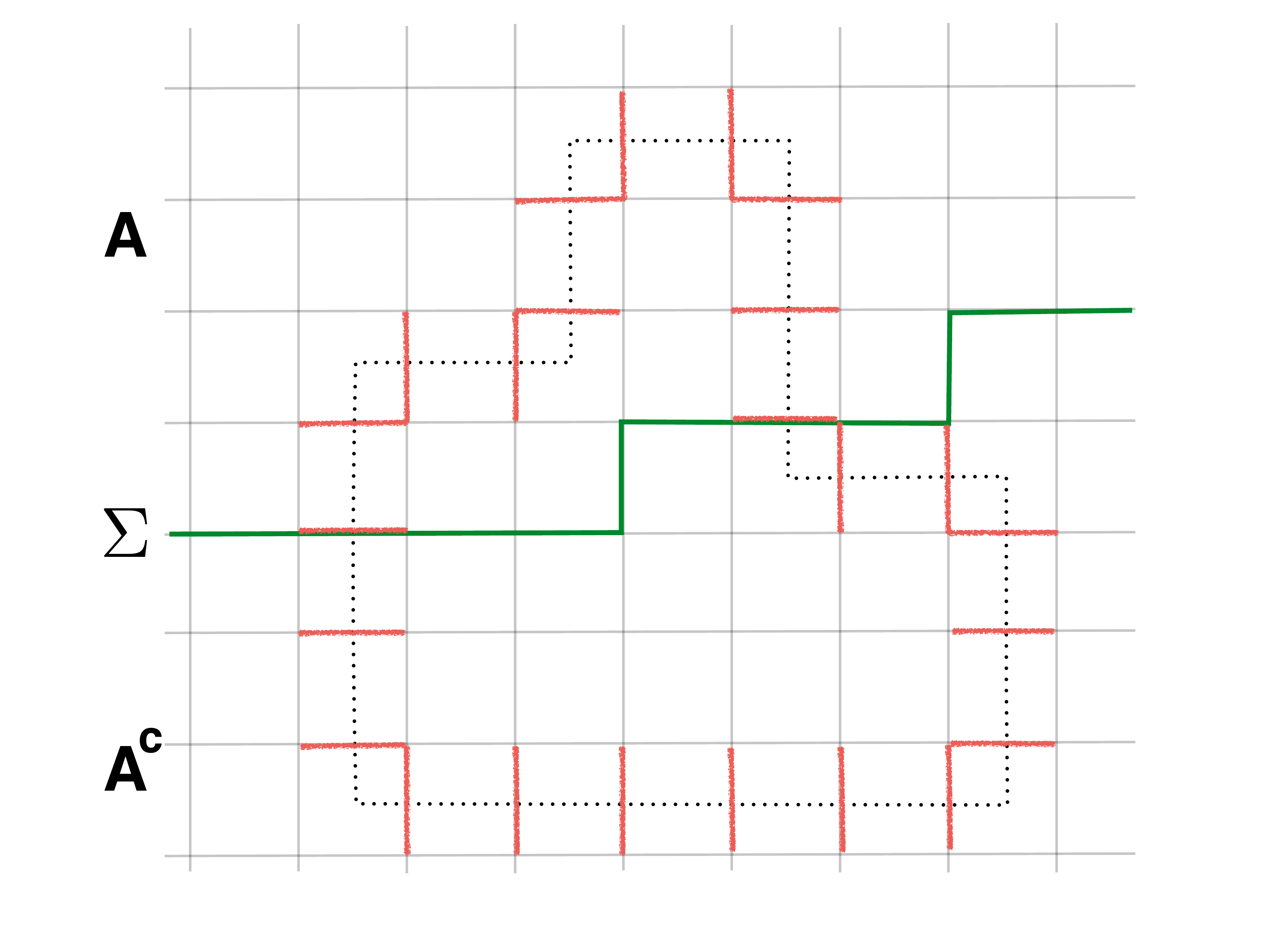}
		\label{41explaination1}}
	\subfigure[]{
		\includegraphics[width=0.45\columnwidth]{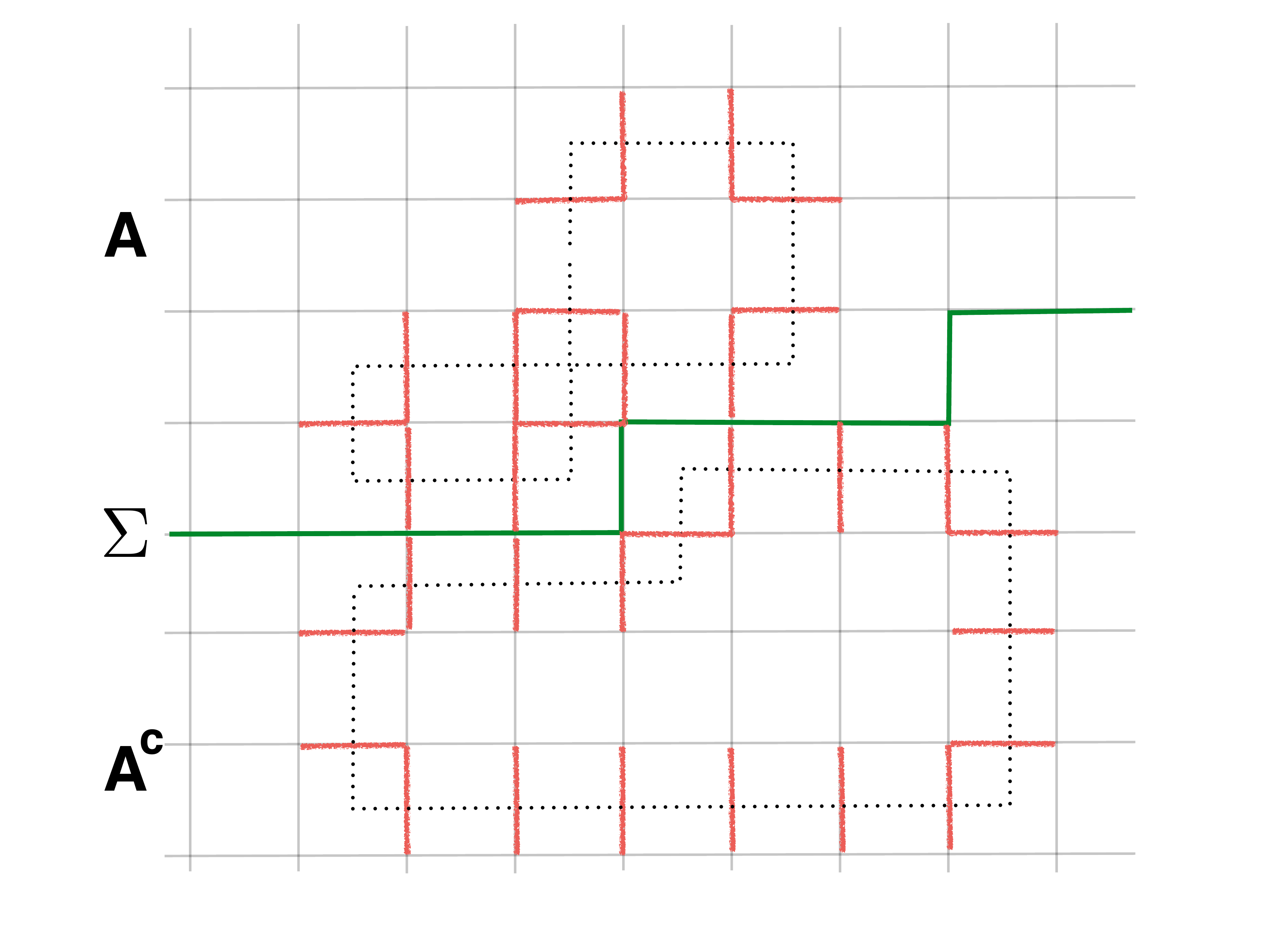}
		\label{41explaination2}}
	\caption{A configuration associated with nontrivial $\mathcal{C}_{\mathrm{E}}$ (on panel (a)) can be reduced to a configuration associated with trivial $\mathcal{C}_{\mathrm{E}}$ (on panel (b)). }
	\label{41explaination}
\end{figure}

We start by establishing a one-to-one correspondence between a configuration $\mathcal{C}_{\mathrm{E}}$ and a configuration with no dual lattice loops across the entanglement surface. We find that it is more illuminating to demonstrate this using a two-dimensional square lattice (but similar arguments work for triangular lattice as well), as shown in Fig.~\ref{41explaination}, which is a spatial slice of the $(2+1)$D spacetime. For simplicity, we consider the $n=2$ case only, where each bond\footnote{In this section, we will use bonds instead of 1-simplices because simplices are not defined on the square lattice.} is either occupied ($B=\pi\mod 2\pi$) or unoccupied ($B=0\mod 2\pi$). In panel (a), we present a general configuration with one occupied loop\footnote{The loop configuration is given by the flatness condition $dB=0\mod 2\pi$. On a $2$D spatial lattice, $B$ is a 1-form and the flatness condition is $(dB)(i,i+x, i+y, i+x+y)=B(i,i+x)+B(i+x, i+x+y)-B(i+y, i+x+y)-B(i, i+y)=0\mod 2\pi$. On a $3$D spatial lattice, $B$ is a 2-form and the flatness condition is $(dB)(i, i+x, i+y, i+z, i+x+y, i+x+z, i+y+z, i+x+y+z)=B(i, i+x, i+x+y, i+y)-B(i+z, i+z+x, i+z+x+y, i+z+y)+B(i, i+z, i+x+z, i+x)-B(i+y, i+y+z, i+y+x+z, i+y+x)+B(i, i+y, i+y+z, i+z)-B(i+x, i+x+y, i+x+y+z, i+x+z)=0\mod 2\pi$. } in the dual lattice (the dotted line). The corresponding configuration in the real lattice is given by the red bonds. The entanglement cut $\Sigma$ consists of the green bonds, where two are occupied (bonds which are both green and red). In panel (b), we present a related configuration with no bonds occupied on $\Sigma$. We denote the boundary configuration on the entanglement surface $\Sigma$ with no bonds occupied as $\mathcal{C}_0$. The configuration in (b) is obtained from the configuration in (a) by cutting the loop at $\Sigma$ in the dual lattice and completing the loops along $\Sigma$ within the two regions A and $\mathrm{A}^{\mathrm{c}}$ separately. Therefore, we have shown that every bulk configuration with non-trivial boundary $\mathcal{C}_{\mathrm{E}}$ can be reduced to a bulk configuration with trivial boundary configuration $\mathcal{C}_0$. However, we note that there can be multiple ways of cutting and completing the loops (which is more obvious in three spatial dimensions), and the reduction may not be unique. Hence we have shown that 
\begin{eqnarray}\label{H1}
N_{\mathrm{A}^{\mathrm{c}}}(\mathcal{C}_\mathrm{E})N_{\mathrm{A}}(\mathcal{C}_\mathrm{E})\leq N_{\mathrm{A}^{\mathrm{c}}}(\mathcal{C}_0)N_{\mathrm{A}}(\mathcal{C}_0).
\end{eqnarray} 
\begin{figure}[H]
	\centering
	\subfigure[]{
		\includegraphics[width=0.45\columnwidth]{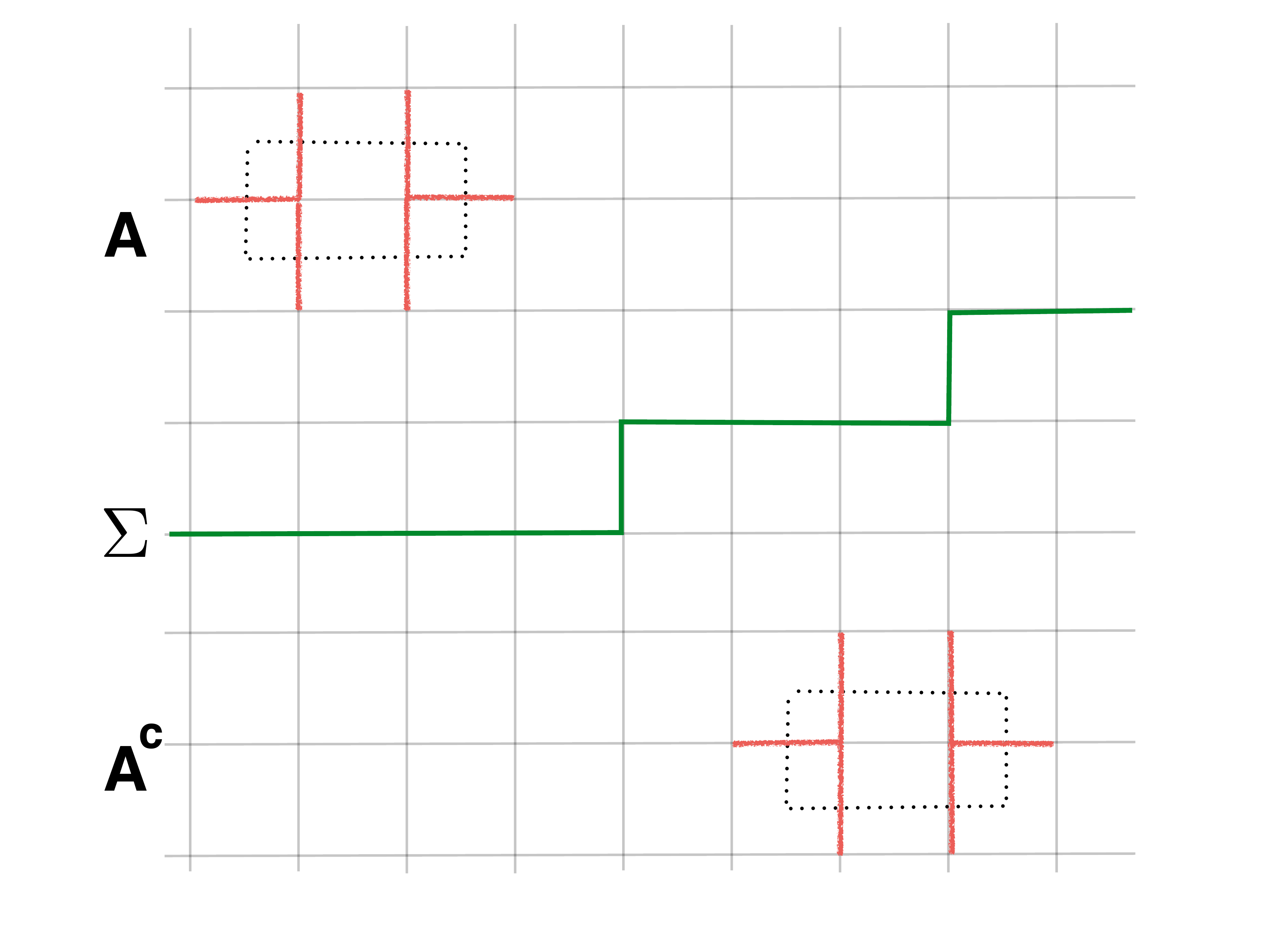}
		\label{41explaination3}}
	\subfigure[]{
		\includegraphics[width=0.45\columnwidth]{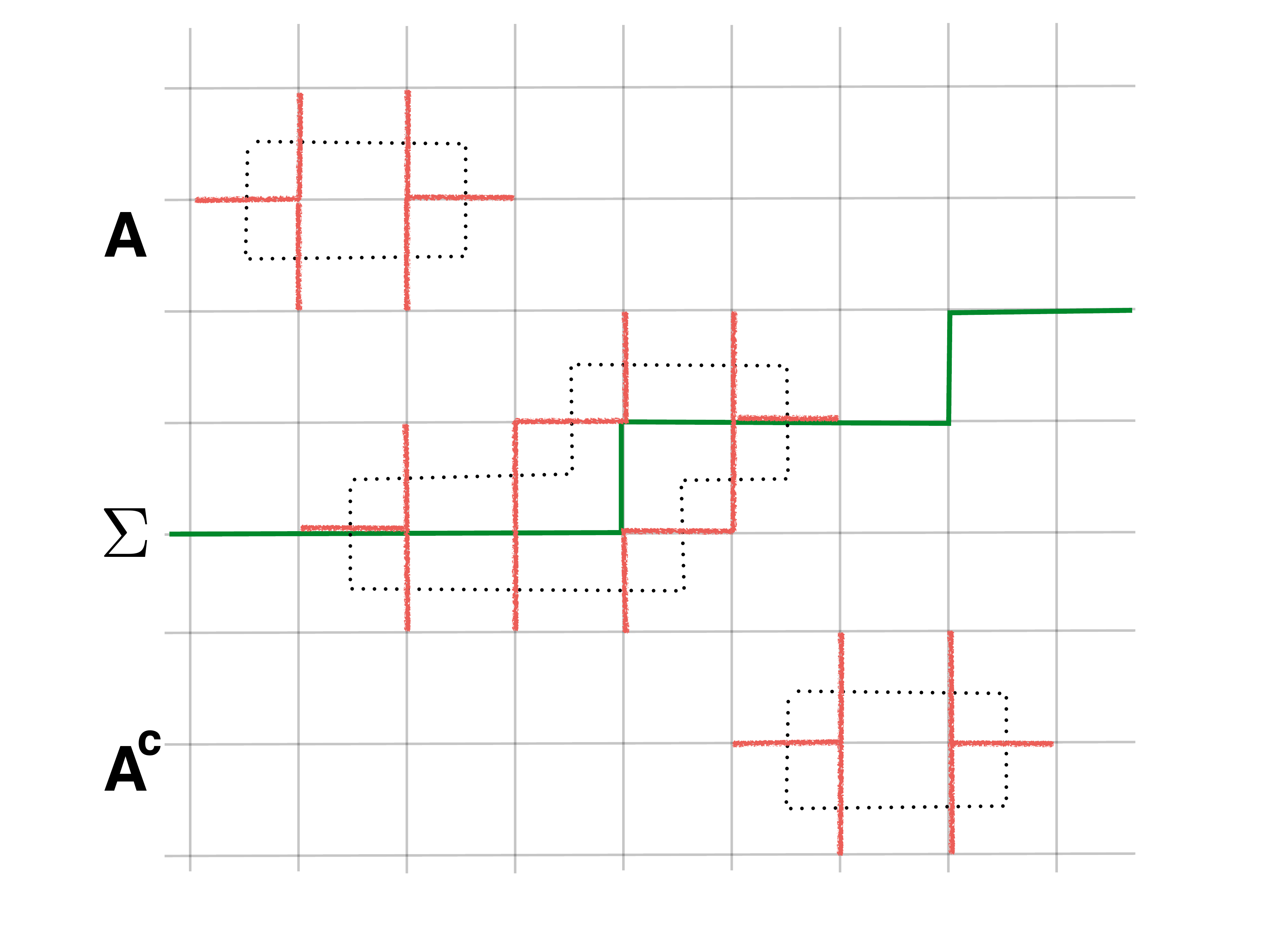}
		\label{41explaination4}}
	\caption{A configuration associated with trivial $\mathcal{C}_{\mathrm{E}}$ (on panel (a)) can be reduced to a configuration associated with a nontrivial $\mathcal{C}_{\mathrm{E}}$ (on panel (b)). }
	\label{41explainationreverse}
\end{figure}

To complete the one-to-one correspondence, we have to consider the opposite deformation: every bulk configuration with trivial boundary configuration $\mathcal{C}_0$ can be changed to a bulk configuration with a specified non-trivial boundary configuration $\mathcal{C}_{\mathrm{E}}$. We use Fig.~\ref{41explainationreverse} to illustrate this process. In panel (a), we present a configuration with no bonds occupied on $\Sigma$, corresponding to the trivial boundary configuration $\mathcal{C}_0$. In panel (b), we draw a specific configuration in which two bonds are occupied. The two occupied bonds on $\Sigma$ are connected via a ``thin" loop along the two sides of $\Sigma$. Therefore, a bulk configuration with nontrivial boundary configuration $\mathcal{C}_{\mathrm{E}}$ can be obtained from a bulk configuration with trivial boundary configuration $\mathcal{C}_{0}$ by adding a ``thin" loop along the two sides of the entanglement cut. However, we note that starting from a configuration with $\mathcal{C}_0$, there can be multiple ways to add the thin loops to obtain a corresponding configuration with a nontrivial $\mathcal{C}_{\mathrm{E}}$. Hence, we have shown that
\begin{eqnarray}\label{H2}
N_{\mathrm{A}^{\mathrm{c}}}(\mathcal{C}_0)N_{\mathrm{A}}(\mathcal{C}_0)\leq N_{\mathrm{A}^{\mathrm{c}}}(\mathcal{C}_\mathrm{E})N_{\mathrm{A}}(\mathcal{C}_\mathrm{E}).
\end{eqnarray}
Combining the inequalities \eqref{H1} and \eqref{H2}, we obtain 
\begin{eqnarray}\label{H3}
N_{\mathrm{A}^{\mathrm{c}}}(\mathcal{C}_\mathrm{E})N_{\mathrm{A}}(\mathcal{C}_\mathrm{E})= N_{\mathrm{A}^{\mathrm{c}}}(\mathcal{C}_0)N_{\mathrm{A}}(\mathcal{C}_0).
\end{eqnarray}
Equation~\eqref{H3} shows that $N_{\mathrm{A}^{\mathrm{c}}}(\mathcal{C}_\mathrm{E})N_{\mathrm{A}}(\mathcal{C}_\mathrm{E})$ is independent of the configuration $\mathcal{C}_\mathrm{E}$, as expected.

\begin{figure}[H]
	\centering
	\includegraphics[width=1\columnwidth]{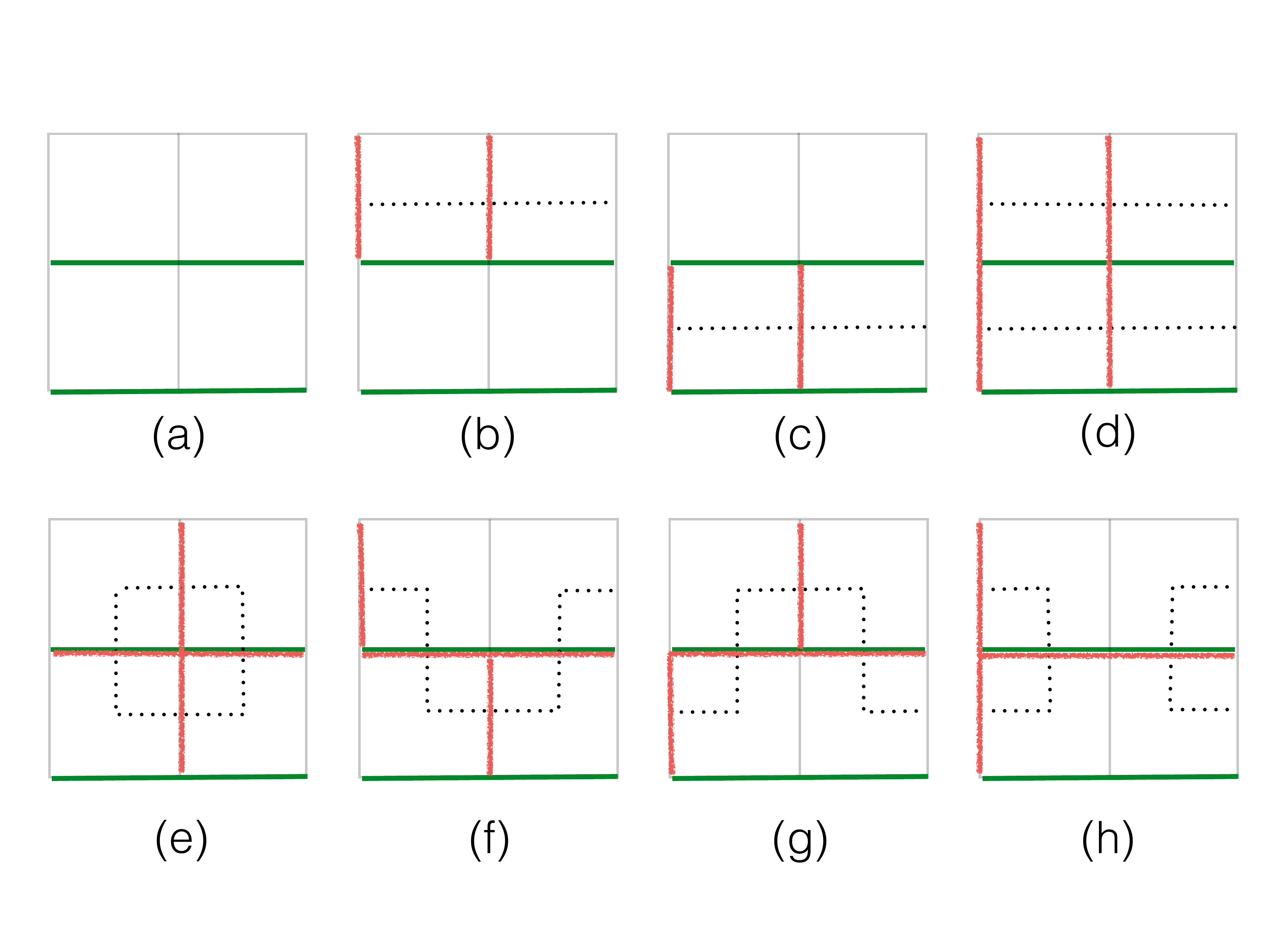}
	\caption{Configurations on a $2\times 2$ lattice with periodic boundary conditions. There are two entanglement cuts, denoted by two green lines. The occupied bonds in the real lattice are shown in red, and occupied bonds in the dual lattice are shown as dotted lines. (a), (b), (c), (d) are configurations with no bonds occupied on the entanglement cut. (e), (f), (g), (h) are configurations with two bonds occupied on the entanglement cut. }
	\label{2by2}
\end{figure}

In addition to the general arguments, it is beneficial to consider an example. In Fig.~\ref{2by2}, we present all the configurations on a $2\times 2$ lattice associated with $\mathcal{C}_0$ (no bonds occupied on the entanglement surface) and with $\mathcal{C}_{\mathrm{E}}$ (two bonds in the middle occupied on the entanglement surface). The configuration such as 
\includegraphics[scale=0.015]{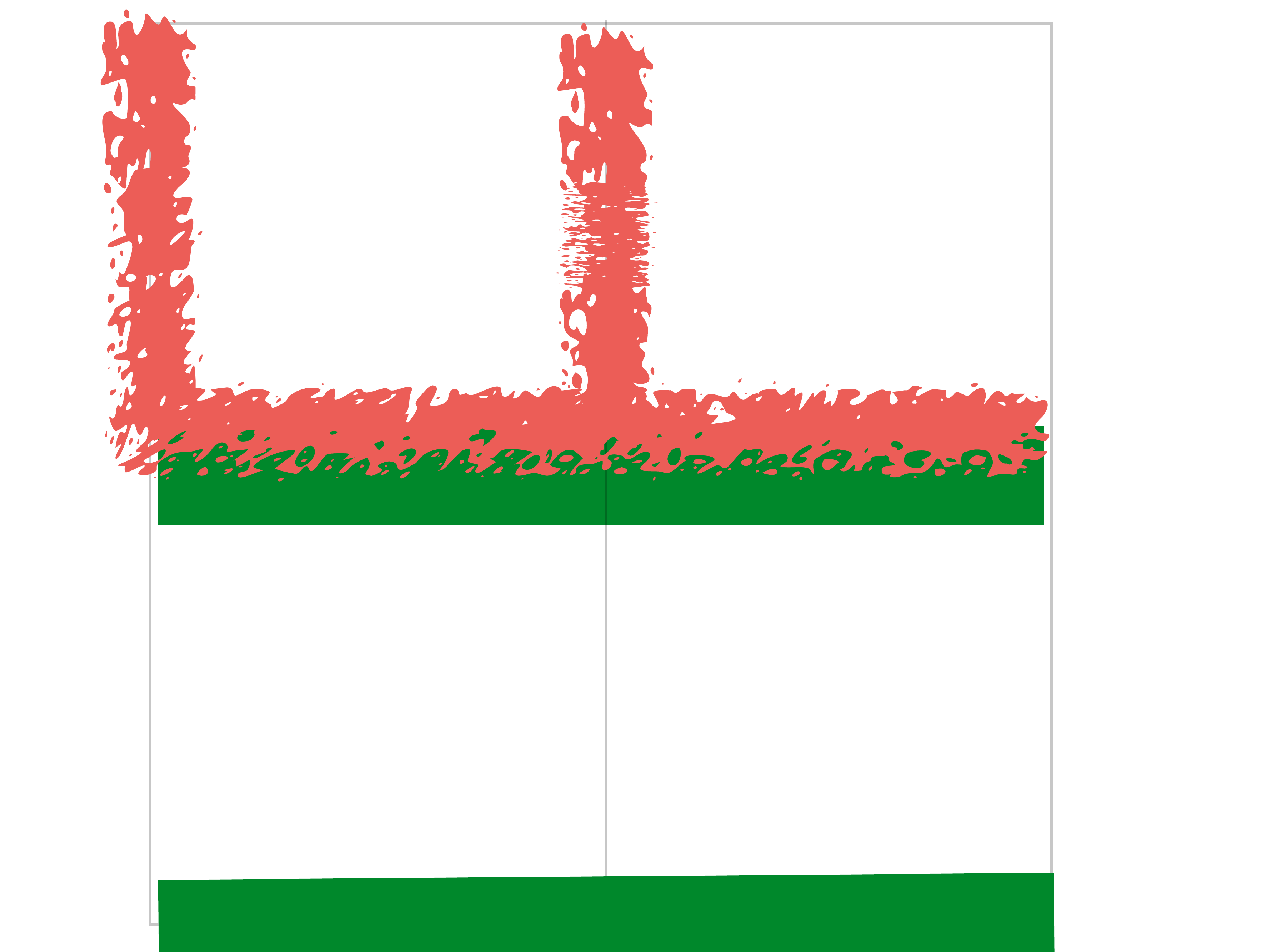}
does not exist because the configuration in the dual lattice is not a loop. In each case, there are 4 configurations, which agrees with our general analysis $N_{\mathrm{A}^{\mathrm{c}}}(\mathcal{C}_\mathrm{E})N_{\mathrm{A}}(\mathcal{C}_\mathrm{E})= N_{\mathrm{A}^{\mathrm{c}}}(\mathcal{C}_0)N_{\mathrm{A}}(\mathcal{C}_0)$.

We further show that the total number of configurations on $\mathcal{C}_{\mathrm{E}}$ is $2^{|\Sigma|-1}$ for the $n=2$ theory, where $|\Sigma|$ is the number of simplices (bonds) on $\Sigma$. (The discussion in this paragraph works for both triangular and square lattices, and we will use the notations simplices and cochains here.) Notice that since each $B$-cochain can take 2 values, i.e., $0\mod 2\pi$ or $\pi\mod 2\pi$, the naive counting of configurations of $\mathcal{C}_{\mathrm{E}}$ is $2^{|\Sigma|}$. However, since the simplices where $B=\pi\mod 2\pi$ form loops in the dual lattice, there must be an even number of simplices occupied on $\Sigma$. This reduces the total number of $\mathcal{C}_{\mathrm{E}}$ configurations by half. Therefore, there are  $2^{|\Sigma|-1}$ possible configurations on the entanglement surface. Applying the normalization condition Eq.~\eqref{normalization}, we complete the demonstration of Eq.~\eqref{1over2Sigma}.

\section{A Case Study of the Conjecture Between GSD and TEE}
\label{AppConjectureCaseStudy}

In this appendix, we examine the conjecture Eq.~\eqref{conjecture} for the BF theory with level $n$ in $(d+1)$D by explicitly computing both the GSD on $d$-dimensional torus $T^d$ and the constant part of the EE across $T^{d-1}$ (which we believe is the topological part for the BF theory). 

The action of the BF theory with level $n$ on the spacetime $T^d\times S^1$ is
\begin{eqnarray}
\mathcal{S}_{\mathrm{BF}}=\int_{T^d\times S^1} \frac{n}{2\pi}B\wedge dA,
\end{eqnarray}
where $A$ is a 1-form gauge field and $B$ is a $(d-1)$-form gauge field. The gauge transformations are $A\to A+dg, ~B\to B+d\lambda$ where $\lambda$ is a $u(1)$ valued $(d-1)$-form gauge field, and $g$ is a compact scalar (i.e., $g\simeq g+2\pi$). The gauge invariant operators, which wrap around the non-contractible cycles of the spatial torus $T^d$, are
\begin{widetext}
\begin{eqnarray}\label{Operatorsgeneral}
V^k_{T_{i_1\cdots i_{d-1}}}&=&\exp\Big(ik\oint_{T_{i_1\cdots i_{d-1}}}B\Big), ~k\in\{0, 1, \cdots , n-1\},\nonumber\\
W^l_{T_i}&=&\exp\Big(il\oint_{T_i}A\Big), ~l\in\{0, 1, \cdots, n-1\},
\end{eqnarray}
\end{widetext}
and their combinations. In the first equation $T_{i_1\cdots i_{d-1}}$ is a $(d-1)$-dimensional torus extending along the $i_1\cdots i_{d-1}$ directions and in the second equation $T_i$ is a 1-dimensional circle extending along the $i$-th direction. (The fact that $V^n_{T_{i_1\cdots i_{d-1}}}$ and $W^n_{T_i}$ are trivial operators will be explained in the following.) We will use canonical quantization to determine the commutation relation between these operators, from which we can determine the ground state degeneracy $\mathrm{GSD}[T^d]$. 

\begin{widetext}
To perform the canonical quantization, we first fix the gauge as $A_0=0, ~B_{0i_1\cdots i_{d-2}}=0$ for any $i_1\cdots i_{d-2}$ using the gauge transformations $A\to A+dg, ~B\to B+d\lambda$. Moreover, the Gauss constraints are $\varepsilon^{0i_1\cdots i_{d-1}i_d}\partial_{i_{d-1}}A_{i_d}=0$ for any $i_1\cdots i_{d-2}$, and $\varepsilon^{0i_1\cdots i_{d-1}i_d}\partial_{i_1}B_{i_2\cdots i_d}=0$ where summation over repeated indices is implied. We have used the definition of totally anti-symmetric tensor
\begin{equation}
\varepsilon^{i_1\cdots i_{d-1}}=
\begin{cases}
+1, & \text{if } i_1\cdots i_{d-1}\mathrm{~ is ~an~even~permutation~of~}0\cdots d-2\\
-1, & \text{if } i_1\cdots i_{d-1}\mathrm{~ is ~an~odd~permutation~of~}0\cdots d-2\\
0&\text{otherwise}.
\end{cases}
\end{equation} 
The Lagrangian, after gauge fixing, is 
\begin{eqnarray}
\mathcal{L}_{\mathrm{BF}}=\frac{n}{2\pi}\frac{(-1)^{d-1}}{(d-1)!}\varepsilon^{i_1\cdots i_d}B_{i_1\cdots i_{d-1}}\partial_0 A_{i_d},
\end{eqnarray}
where $B_{i_1\cdots i_{d-1}}$ and $A_{i_d}$ obey the Gauss constraints. The canonical quantization conditions on the gauge fields are
\begin{eqnarray}
\begin{split}
\Bigg[\frac{(-1)^{d-1}}{(d-1)!}\varepsilon^{i_1\cdots i_d}B_{i_1\cdots i_{d-1}}(t, \vec{x}), A_{j_d}(t, \vec{y})\Bigg]=\frac{2\pi i}{n}\delta_{i_dj_d}\delta(\vec{x}-\vec{y}).
\end{split}
\end{eqnarray}
From this canonical relation, one can determine the commutation relation of the line and higher volume operators by applying the Baker-Campbell-Hausdorff formula. We find
\begin{eqnarray}\label{Appcommutationrelation}
V^k_{T_{i_1\cdots i_{d-1}}}W^l_{T_{i_d}}=e^{(-1)^di 2\pi k l /n}W^l_{T_{i_d}}V^k_{T_{i_1\cdots i_{d-1}}}.
\end{eqnarray}
From Eq.\eqref{Appcommutationrelation}, we can see that $\exp(in\oint_{T_{i_1\cdots i_{d-1}}}B)$ commutes with any line operator $\exp(ik\oint_{T_i}A)$, and also trivially commutes with any surface operator $\exp(ik\oint_{T_{j_1\cdots j_{d-1}}}B)$. Therefore, $\exp(in\oint_{T_{i_1\cdots i_{d-1}}}B)$ commutes with any gauge invariant operator and should be a constant. By using the same argument as in App.~\ref{TriangulationofTQFT},  $\exp(in\oint_{T_{i_1\cdots i_{d-1}}}B)=1$. Similarly, we find that $\exp(in\oint_{T_i}A)=1$ as well. The explains that the charges $k$ and $l$ of the non-local operators $V^k_{T_{i_1\cdots i_{d-1}}}$ and $W^l_{T_{i_d}}$ only take $n$ different values. 

We can define the ground states $|u_1\cdots u_d\rangle$ to be the eigenstates of $W_i^l$, and choose $V^k_{T_{i_1\cdots i_{d-1}}}$ as the raising and lowering operators acting on the ground states. Since $W_i^n=1$, the eigenvalues of $W_i$ should be $n$-th root of unity, i.e., $e^{-(-1)^d i2\pi u_i/n}$, where $u_i\in \{0, 1, \cdots, n-1\}$. Specifically, 
\begin{eqnarray}
\begin{split}
W_i^l|u_1\cdots u_d\rangle&=e^{-(-1)^di2\pi l u_i/n}|u_1\cdots u_d\rangle,\\
V^k_{T_{12\cdots (i-1)(i+1)\cdots d}}|u_1\cdots u_d\rangle&=|u_1\cdots u_{i-1}(u_i+1)u_{i+1}\cdots u_d\rangle,
\end{split}
\end{eqnarray}
where $u_i\in\{0, 1, \cdots, n-1\}$ for all $i$. Therefore, there are $n^d$ ground states on the $d$-dimensional spatial torus, $\mathrm{GSD}[T^d]=n^d$. 
\end{widetext}

To obtain the EE, we generalize the calculations of Sec.~\ref{EntanglementEntropyOfTQFT}. Since most of the calculations are similar, we will only present the crucial steps. 

We start by formulating the theory on the higher dimensional triangulated spacetime lattice $\mathcal{M}_{d+1}$. The ground state wavefunction is still the equal weight superposition of loop configurations in the dual of the spatial lattice,
\begin{eqnarray}
|\psi\rangle=\mathfrak{C}\sum_{\mathcal{C}\in \mathcal{L}}|\mathcal{C}\rangle,
\end{eqnarray}
where the sum is taken over the set $\mathcal{L}$ of all possible loop configurations $\mathcal{C}$ at the dual lattice of spatial slice $S^d=\partial\mathcal{M}_{d+1}$. We choose the entanglement surface to be a $(d-1)$-dimensional torus, separating the space into two regions A and $\mathrm{A}^{\mathrm{c}}$. The wavefunction is 
\begin{eqnarray}
|\psi\rangle=\mathfrak{C}\sum_{\mathcal{C}_{\mathrm{E}}}\sum_{a=1}^{N_{\mathrm{A}}(\mathcal{C}_{\mathrm{E}})}\sum_{b=1}^{N_{\mathrm{A}^{\mathrm{c}}}(\mathcal{C}_{\mathrm{E}})}|\mathrm{A}_a^{\mathcal{C}_{\mathrm{E}}}\rangle|\mathrm{A}_b^{\mathrm{c}\mathcal{C}_{\mathrm{E}}}\rangle,
\end{eqnarray}
from which one can obtain the reduced density matrix by tracing over the degrees of freedom in region $\mathrm{A}^\mathrm{c}$, 
\begin{eqnarray}
\rho_{\mathrm{A}}=|\mathfrak{C}|^2\sum_{\mathcal{C}_{\mathrm{E}}}N_{\mathrm{A}^{\mathrm{c}}}(\mathcal{C}_{\mathrm{E}})\sum_{a,a'=1}^{N_{\mathrm{A}}(\mathcal{C}_{\mathrm{E}})}|\mathrm{A}_{a}^{\mathcal{C}_{\mathrm{E}}}\rangle\langle \mathrm{A}_{a'}^{\mathcal{C}_{\mathrm{E}}}|.
\end{eqnarray}
The normalization constant $\mathfrak{C}$ is determined by $\Tr_{\ch_\mathrm{A}}\rho_{\mathrm{A}}=|\mathfrak{C}|^2N_{\mathrm{A}}(\mathcal{C}_{\mathrm{E}})N_{\mathrm{A}^{\mathrm{c}}}(\mathcal{C}_{\mathrm{E}})n^{|\Sigma|-1}=1$, where $|\Sigma|$ is the number of $(d-1)$-simplices on the entanglement surface. The EE is
\begin{equation}
\begin{split}
S(\mathrm{A})&=-\Tr_{\ch_\mathrm{A}}\rho_{\mathrm{A}}\log \rho_{\mathrm{A}}=\frac{d}{dN}\Bigg(-\frac{\Tr_{\ch_\mathrm{A}}\rho_{\mathrm{A}}^N}{(\Tr_{\ch_\mathrm{A}}\rho_{\mathrm{A}})^N}\Bigg)\bigg|_{N=1}\\
&= -\frac{d}{dN}\bigg(|\mathfrak{C}|^{2N}\sum_{\mathcal{C}_{\mathrm{E}}}N_{\mathrm{A}^{\mathrm{c}}}(\mathcal{C}_{\mathrm{E}})^NN_{\mathrm{A}}(\mathcal{C}_{\mathrm{E}})^N\bigg)\bigg|_{N=1}\\
&=-\frac{d}{dN}\bigg(\sum_{\mathcal{C}_{\mathrm{E}}}n^{-(|\Sigma|-1)N}\bigg)\bigg|_{N=1}\\
&= -\frac{d}{dN}\bigg(n^{-(|\Sigma|-1)(N-1)}\bigg)\bigg|_{N=1}\\
&=|\Sigma|\log n-\log n.
\end{split}
\end{equation}\\
In the second line, we used the normalization $\Tr_{\ch_\mathrm{A}}\rho_{\mathrm{A}}=1$, $\Tr_{\ch_\mathrm{A}}\rho_{\mathrm{A}}^N=|\mathfrak{C}|^{2N}\sum_{\mathcal{C}_{\mathrm{E}}}N_{\mathrm{A}^{\mathrm{c}}}(\mathcal{C}_{\mathrm{E}})^NN_{\mathrm{A}}(\mathcal{C}_{\mathrm{E}})^N$. In the third line, we used $|\mathfrak{C}|^2N_{\mathrm{A}}(\mathcal{C}_{\mathrm{E}})N_{\mathrm{A}^{\mathrm{c}}}(\mathcal{C}_{\mathrm{E}})=n^{-(|\Sigma|-1)}$. In the fourth line, since the summand does not depend on $\mathcal{C}_{\mathrm{E}}$, we just multiply the summand by the number of $\mathcal{C}_{\mathrm{E}}$ $n^{|\Sigma|-1}$. In the last line, we take the differential with respect to $N$ and take $N=1$. 
Therefore, the constant part of the EE across $T^{d-1}$ is $-\log n$, which we conjecture to be the TEE across $T^{d-1}$.  Combining the results $\mathrm{GSD}[T^d]=n^d$ and $S_{\mathrm{topo}}[T^{d-1}]=-\log n$, we expect that the conjecture $\exp(-dS_{\mathrm{topo}}[T^{d-1}])=\mathrm{GSD}[T^d]$ of Eq.~\eqref{conjecture} holds for the $(d+1)$-dimensional BF theory.

\bibliography{3dTEE}

\end{document}